# Random Geometries in Quantum Gravity

Ph.D. thesis

Charlotte Følge Kristjansen

The Niels Bohr Institute
University of Copenhagen
September 1993

# Acknowledgements

It is a pleasure to thank Jan Ambjørn for always inspiring and stimulating supervision and collaboration. I am also grateful for the valuable and fruitful collaboration that I have had with Jerzy Jurkiewicz and Yuri Makeenko as well as Zdzisek Burda and Leonid Chekhov. Finally the positive atmosphere and the support that I have experienced in the high energy theory group at NBI is greatly appreciated.

# Contents

# 1  Introduction

Since the beginning of this century we have had two remarkably successful theories, quantum theory and general relativity. Each theory describes its own domain with an astonishing accuracy. No experiment has ever been in contradiction with either theory. However, history provides several examples that attempts to describe apparently unrelated phenomena within a common theoretical framework have led to great progress in our understanding of nature. A unification of quantum theory and general relativity might shed light on fundamental questions concerning the nature and origin of space time.

During the past decades the problem of quantizing gravity has attracted the interest of an increasingly number of physicists from different fields. Numerous approaches to the problem have been developed. Here we will report of some results of approaching the problem keeping to the conventional quantum field theoretical framework. It is well known that by following this line of action one immediately comes up against severe difficulties. General relativity based on the Einstein Hilbert action is not perturbatively renormalizable in dimensions higher than two. Adding higher derivative terms to the interaction does not improve the situation. Renormalizability can be obtained but then unitarity is lost [1]. In 1979 Weinberg pointed out that under special circumstances one can construct a reasonable quantum field theory out of a theory apparently needing an infinite number of counter terms [2]. The special circumstances required are that the original theory has an ultraviolet fixed point for which the associated critical surface is finite dimensional. In that case only a finite number of parameters is needed to describe the ultraviolet behaviour of the theory and the situation resembles the one encountered for a renormalizable theory. This scenario is known as asymptotic safety. Weinberg and others have studied in perturbation theory the renormalization group flow of Newtons constant using dimensional continuation [2]. The outcome of these studies is that gravity appears to be asymptotically safe in $2 + \epsilon$ dimensions ($\epsilon > 0$). Since the analyses are based on perturbation theory and $\epsilon$ is assumed to be infinitesimal we can not really take these results as an indicator of what happens in the ultraviolet regime of 4-dimensional gravity. Nevertheless Weinbergs ideas are sill viable and the work we will report on is greatly influenced by them.

An approach to quantizing field theories which has made possible addressing non perturbative questions is the Euclidean path integral method. At the same time as Weinberg put forward his ideas, Hawking suggested to apply the path integral approach to the problem of quantizing gravity [3]. In particular he suggested that the rotation from Lorentzian to Euclidean space in the case of ordinary field theories should be generalized to a rotation from Minkowskian to Riemannian space. Writing down a Riemannian path integral for quantum gravity is straightforward but the possibility of recovering Minkowskian space times from Riemannian ones by some kind of analytical continuation remains an assumption. Even though one chooses to postpone the clarification of this point, ones troubles are not over. The Riemannian action is unbounded from below due to the conformal model. Several suggestions for a solution of this problem exists. Hawking et. al. have suggested that one might rotate the contour of integration for the conformal factor so that it becomes parallel to the imaginary axis thereby making the path integral convergent [4]. Greensite has advocated that stochastic stabilization provides a way of attributing meaning to Riemannian quantum gravity [5]. Here we will analyse the Riemannian path integral from what one might call a statistical mechanical point of view. By discretizing the path integral we construct a lattice model of quantum gravity. We then search the coupling constant space of the discretized model for critical points, the hope being that a second order phase transition point where physical observables show appropriate scaling behaviour exists. If such a situation is indeed encountered we might attempt to define a continuum limit. We deal with the problem of the unboundedness of the Riemannian action by choosing a discretization which for fixed space time volume and topology provides a regularization of the path integral. Of course this means that in the first place we just postpone the problem untill the infinite volume limit is taken. Several other open questions of the Riemannian approach to quantum gravity turn up also in the discretized models. One example is the question of how to choose the path integral measure. Another example is the question of whether or not one should sum over space time manifolds of different topology. In two dimensions the sum over topologies is known to be badly divergent and the situation only gets worse in higher dimensions, let alone the problem that no classification of manifolds according to topology is known in higher dimensions. There is of course also still the question if we can extract any information about Minkowskian gravity from the Riemannian model. Finally in the lattice approach a new problem turns up. General relativity is reparametrization invariant and we want a quantum theory of gravity to possess the same symmetry. By discretizing the Einstein Hilbert action we break explicitly reparametrization invariance. Therefore it is of outmost importance in case the lattice theory offers to us the possibility of defining a continuum limit to check whether reparametrization invariance is recovered in this limit.

Despite the non negligible number of unclarified points it is still the hope that something interesting can be learnt from studying discretized quantum gravity. Needless to say that predictions of experimental relevance are not to be expected in the immediate future. It is tempting the repeat the statement made by A. Salam at the first Oxford conference on quantum gravity in 1975 [6] and repeated by C. J. Isham at the second Oxford conference on quantum gravity in 1980 [7]

> "In particle physics we have become conditioned never to ask what the theory can do for us; instead we humbly try to see what we can do for the theory."

In the following sections we will describe some examples of what has been done for the theory. In section 2 we introduce the discretization scheme known as dynamical triangulations. The remaining sections describe the application of this discretization scheme in two, three and four space time dimensions, focusing on aspects addressed in the work of the author.

# 2 Dynamical Triangulations

## 2.1 The Discretization Programme

The first step in the discretization programme consists in replacing the continuous $d$-dimensional space time manifold by a special type of lattice known as a $d$-dimensional simplicial manifold. The building blocks of such a lattice are so-called $d$-simplexes. A 2-simplex is a triangle, a 3-simplex is a tetrahedron. The $d$-dimensional analogue can be thought of as an object consisting of $(d+1)$ vertices connected via $\frac{(d+1)d}{2}$ links. A $d$-dimensional simplicial manifold is a collection of $d$-simplexes which are glued together along their $(d-1)$-dimensional sub-simplexes in such a way that the set of $d$-simplexes surrounding any given vertex is homeomorphic to the unit ball, $B^d$, in $R^d$. Hence per definition a simplicial manifold does not have any boundary. In the following we will in addition assume that our simplicial manifolds are connected. Let us introduce the notation that $N_i$ denotes the total number of $i$-simplexes in such a $d$-dimensional manifold while $n_i$ refers to a given $i$-simplex. Furthermore let us denote by $o(n_i)$ the order of $n_i$, i.e. the number of $d$-simplexes to which $n_i$ belongs. Then the Euler characteristic of a $d$-dimensional simplicial manifold, $\chi_d$, is given by

$$\chi_d = \sum_{i=0}^{d} (-1)^i N_i. \tag{2.1}$$

Furthermore from the definition of a simplicial manifold it follows that in $d$ dimensions the following relations hold between the $N_i$'s [8]

$$N_i = \sum_{k=i}^{d} (-1)^{k+d} \binom{k+1}{i+1} N_k. \tag{2.2}$$

For $d = 3$ and $d = 4$, (2.2) gives rise to two independent relations between the $N_i$'s. For $d = 2$ only one constraint arises. In the case $d = 3$ one of the constraints is that the Euler characteristic has to equal zero. This holds always for simplicial manifolds of odd dimension.

$$\chi_{d=2n+1} = 0. \tag{2.3}$$

One can introduce the notion of distance on a simplicial manifold, i.e. equip the manifold with a metric by specifying the lengths of all its links and assuming the metric to be flat in the interior of its $d$- and $(d-1)$-dimensional sub-simplexes and continuous across faces. In this way one obtains a so-called piecewise linear space. We note that the number of edges of a $d$-simplex equals the number of independent components of the metric tensor in $d$ dimensions. To obtain the discrete analogue of a Riemannian manifold restrictions must be imposed on the edge lengths in order to ensure that the metric is positive definite. The construction of piecewise linear spaces is due to Regge who also showed how one can assign curvature to such spaces [9]. Since the geometry of a $d$-dimensional piecewise linear space is Euclidean on its $d$- and $(d-1)$-dimensional sub-simplexes its Riemann tensor is $\delta$-function like having support only on the $(d-2)$-dimensional sub-simplexes, the hinges. We will not enter into a discussion of how to define the full Riemann tensor for a piecewise linear space. Such a discussion can be found in reference [10].

Our aim here is only to find the discretized version of the Riemannian Einstein Hilbert action given in the continuum by

$$S_{EH}[\mathcal{M}] = \Lambda \int_{\mathcal{M}} d^d x \sqrt{g} - \frac{1}{16\pi G_N} \int_{\mathcal{M}} d^d x \sqrt{g} R \qquad (2.4)$$

where $\mathcal{M}$ is the space time manifold, $\Lambda$ the cosmological constant and $G_N$ Newtons constant. Hence as far as curvature is concerned all we need to know is how to calculate the integrated scalar curvature. The integrated scalar curvature for a given piecewise linear space is a sum of contributions from each of its hinges, namely

$$\int_{\mathcal{M}} d^d x \sqrt{g} R \longrightarrow \sum_{hinges, h} V(h) \delta_h \qquad (2.5)$$

where $V(h)$ is the volume of the hinge, $h$, and $\delta_h$ the deficit angle associated with the hinge. The deficit angle at a hinge, $h$, is $2\pi$ minus the sum over $d$-simplexes containing $h$, $n_d^{(i)}(h)$, of the angle between the two faces of $n_d^{(i)}(h)$ which have $h$ in common. In general the deficit angle $\delta_h$ is a complicated function of the edge lengths of all $d$-simplexes meeting at $h$. A discussion of the general case can be found for instance in references [11, 12]. However we will advocate a discretization scheme where one only considers piecewise linear spaces built from regular simplexes. For a regular $d$-simplex the angle, $\theta_d$, between any two adjacent faces is given by

$$\cos \theta_d = \frac{1}{d} \qquad (2.6)$$

and hence the integrated scalar curvature for a $d$-dimensional Riemannian piecewise linear space built from regular simplexes can be written as

$$\int_{\mathcal{M}} d^d x \sqrt{g} R \longrightarrow \sum_{n_{d-2}} (c_d - o(n_{d-2})) = c_d N_{d-2} - K_{d+1,2} N_d \qquad (2.7)$$

where

$$c_d = \frac{2\pi}{\theta_d} \quad \text{and} \quad K_{d+1,2} = \binom{d+1}{2}. \qquad (2.8)$$

Only for $d = 2$, $c_d$ is an integer and hence only in two dimensions we can have a regular tessellation of flat space. Furthermore in two dimensions the integrated scalar curvature is a topological invariant, namely

$$\int_{\mathcal{M}} d^2 x \sqrt{g} R = 4\pi \chi = 8\pi (1-g), \qquad (d=2) \qquad (2.9)$$

where $g$ is the genus of manifold $\mathcal{M}$. The discrete analogue of the cosmological term reads

$$\int_{\mathcal{M}} d^d x \sqrt{g} \longrightarrow N_d. \qquad (2.10)$$

It is also possible to construct discrete versions of integrals involving higher derivative terms. However this requires some care since higher derivative terms involve higher powers of the Riemann tensor which as mentioned above is $\delta$-function like. We will return to this problem in section 4.4.3. From (2.7) and (2.10) it appears that for a triangulation, $T$, consisting of $N_i$ $i$-simplexes, $i = 0, \ldots, d$, the Einstein Hilbert action turns into

$$S_{EH}[T] = \kappa_d N_d - \kappa_{d-2} N_{d-2} \qquad (2.11)$$

where $\kappa_{d-2}$ and $\kappa_d$ are dimensionless coupling constants, $\kappa_d$ being a linear combination of the cosmological constant and the inverse gravitational constant and $\kappa_{d-2}$ being proportional to the latter. In two dimensions we can also write

$$S_{EH}[T] = \kappa_2 N_2 - \kappa \chi, \qquad (d = 2). \qquad (2.12)$$

Having arrived at a lattice version of the action we have completed the first step of the discretization procedure. We come now to the second and more subtle step, namely the discretization of the path integral measure. In the continuum formulation in order to calculate the partition function we are supposed to integrate over all Riemannian metrics of our space time manifold modulo diffeomorphisms. This integration we will replace by a summation over all simplicial manifolds built from regular simplexes, a prescription known as dynamical triangulation [1]

$$\int_{\mathcal{M}} \frac{d[g_{\mathcal{M}}]}{\text{diff}_{\mathcal{M}}} e^{-S_{EH}[\mathcal{M}]} \longrightarrow \sum_T \frac{1}{C(T)} e^{-S_{EH}[T]}. \qquad (2.13)$$

Here we have divided the contribution from a given triangulation by the order of the symmetry group of the triangulation, $C(T)$. This can be thought of as eliminating the last remainings of the diffeomorphism group. Apart from the symmetry factor we choose the same weight for all simplicial manifolds. There are no indications that other choices should be made. On the contrary results obtained in two dimensions speak in favour of a uniform weight (cf. to section 2.3). One may also argue that some kind of universality should hold so that the detailed form of the measure should not be of importance. There have, however, been studies of dynamical triangulations with a non uniform measure [13].

So far we have not made precise what class of manifolds should be included in the functional integral. In a theory of quantum gravity one might ultimately want to include manifolds of all possible topologies. However at the present stage of studies a restriction of topology is necessary. This is evident already in two dimensions. Let us write the partition function for dynamically triangulated 2-dimensional gravity as

$$Z(\kappa, \kappa_2) = \sum_\chi e^{\kappa \chi} \left\{ \sum_{N_2} e^{-\kappa_2 N_2} N_\chi(N_2) \right\} \qquad (2.14)$$

---

[1]Since we have restricted ourselves to considering only connected simplicial manifolds it is actually the free energy rather than the partition function we are calculating by the prescription (2.13). However we will carry on the historically conditioned misuse of notation and denote the right hand side of (2.13) as the partition function of dynamically triangulated gravity.

where
$$N_\chi(N_2) = \sum_{T \in \mathcal{T}(N_2,\chi)} \frac{1}{C(T)}. \qquad (2.15)$$

Here $\sum_{T \in \mathcal{T}(N_2,\chi)}$ means summation over all triangulations corresponding to manifolds of volume $N_2$ and Euler characteristic $\chi$. Two-dimensional dynamically triangulated gravity can be solved by analytical means (cf. to section 3.1) and one finds in the limit $N_2 \to \infty$ [14, 15]
$$N_\chi(N_2) \sim N_2^{\gamma_\chi - 3} e^{\kappa_2^c N_2}, \qquad (2.16)$$
where
$$\gamma_\chi - 2 = (\gamma_0 - 2)(1 - g) \quad \text{and} \quad \gamma_0 = -\frac{1}{2}. \qquad (2.17)$$

It is important to note that $\kappa_2^c$ does not depend on topology. If we restrict the topology the sum (2.14) will be well behaved for $\kappa_2 > \kappa_2^c$. However, for unrestricted $\chi$ the sum in equation (2.14) will diverge. We can not even hope for Borel summability. In higher dimensions the need for a restriction of topology can only be more pronounced. In addition we would not even know how to attribute a meaning to the phrase "summing over all topologies" since a topological classification of manifolds is not known for $d > 2$. We will take as the partition function for 3- and 4-dimensional dynamically triangulated gravity

$$Z(\kappa_{d-2}, \kappa_d) = \sum_{T \sim S^d} \frac{1}{C(T)} e^{-\kappa_d N_d + \kappa_{d-2} N_{d-2}} \qquad (2.18)$$

where the sum is over simplicial manifolds homeomorphic to $S^d$. Considering only manifolds homeomorphic to a sphere does not impose further restrictions on the $N_i$'s for $d = 3$ while for $d = 2$ and $d = 4$ it dictates that the Euler characteristic must equal two and hence imposes one additional constraint on the $N_i$'s. In conclusion we see that for $d$-dimensional simplicial manifolds homeomorphic to $S^d$ we have for $d = 2$ only one independent $N_i$ and for $d = 3$ and $d = 4$ only two independent $N_i$'s. In particular for $d = 3$ and $d = 4$ the action (2.11) is the most general one we can have involving only $N_i$'s.

Yet no powerful analytical tools for studying dynamical triangulated gravity in three and four dimensions have been developed. However, numerical simulations have been performed. A priori it is not clear that the number of manifolds with spherical topology and a given volume is exponentially bounded by the volume in higher dimensions as it is in two dimensions (cf. to equation (2.16)). However it has now been proven numerically both for $d = 3$ [16, 17] and for $d = 4$ [18, 19, 20, 21]. Since for $d = 3$ and $d = 4$, $N_{d-2} \leq const \times N_d$ there exists a line $\kappa_d = \kappa_d^c(\kappa_{d-2})$ in the $(\kappa_{d-2}, \kappa_d)$ coupling constant plane with the property that for $\kappa_d > \kappa_d^c(\kappa_{d-2})$ the partition function (2.18) is well defined. The question of whether the subleading correction to the exponential growth is power like in three and four dimensions has also been addressed numerically. In three dimensions one finds the subleading correction to be power like only in some part of the coupling constant space. The situation might very well be the same in 4 dimensions [22, 23]. We will treat the numerical simulations of quantum gravity in three and four dimensions in more detail in sections 4.3 and 4.4. The dynamical triangulation prescription for discretizing the path integral in quantum gravity has many attractive features. For a given volume of the space time manifold the Einstein Hilbert action is automatically bounded from below as well as from above. Furthermore we have no problem with overcounting. Two different triangulations lead to different curvature assignment and hence correspond to two truly different Riemannian metrics. Dynamical triangulation as an approach to discretizing the path integral in quantum gravity was originally introduced for gravity in two dimensions in references [24, 25, 26, 27]. Let us mention that there exists another approach which one could denote as "static triangulation" which has been advocated by Hamber and Williams. For an introduction see for instance [11]. In this approach one considers a simplicial manifold with fixed connectivity and the path integral is calculated by summing over all edge lengths of the given triangulation. For static triangulations the path integral is not automatically regulated. A cut off on the edge lengths has to be introduced. Furthermore one does not automatically have a Riemannian metric. To ensure this further constraints on the edge lengths are needed. The question of the choice of the measure is also more complicated for static triangulations since many choices of edge lengths correspond to the same metric. To divide out the diffeomorphism group is not practically doable. Finally one might fear that a fixed triangulation does not allow one to explore completely the space of all possible Riemannian metrics. Of course one could also argue that not all Riemannian manifolds can be approximated with a simplicial manifold built from regular simplexes. However, our task is not to approximate a given Riemannian metric but to pick out with appropriate weights representatives from all regions of the space of such metrics. Whether we succeed in doing so should of course be tested. Another important test concerns reparametrization invariance. As pointed out in the introduction, in case the discretized theory offers to us the possibility of defining a continuum limit we must make sure that reparametrization invariance is recovered in this limit. We will return to these points in section 2.3. Let us just mention that dynamically triangulated gravity seems to pass the tests whereas the situation is more doubtful for the static triangulations.

## 2.2  Coupling to Matter

A quantum version of general relativity should of course include a description of the interaction between space time and matter. Matter fields can easily be coupled to dynamically triangulated gravity. The probably simplest type of matter one can introduce is an Ising spin system. The spin variables, $\sigma_i$, can be placed either on the vertices or in the center of the $d$-simplexes of the $d$-dimensional simplicial manifolds. One simply adds the Ising model action

$$S_{Ising} = \beta \sum_{<i,j>} \sigma_i \sigma_j \qquad (2.19)$$

where $\sum_{<i,j>}$ denotes the sum over neighbouring pairs of spins, to the Einstein Hilbert action and extends the summation over manifolds to include a summation over all possible spin configurations for a given manifold. One can couple any number of such spin systems to dynamically triangulated gravity. Other spin systems such as $q$-state Potts models can also be considered. By making use of rules of simplicial differential calculus one can couple fields with tensor properties to the geometrical system. For example the continuum action for a massless scalar field $\phi$

$$S_{cont} = \int d^d x \sqrt{g} g^{\mu\nu} \partial_\mu \phi \partial_\nu \phi \tag{2.20}$$

translates into

$$S_{lattice} = \sum_{<i,j>} (\phi_i - \phi_j)^2 . \tag{2.21}$$

Also in this case we can choose to place the field variables either on the vertices or in the center of our $d$-simplexes. For $d=2$ only the model consisting of one Ising spin system coupled to dynamically triangulated gravity can be solved analytically [28, 29]. Needless to say that for $d=3$ and $d=4$ no analytical tools for studying dynamically triangulated gravity coupled to matter exist.

Let us try to describe what effects one might expect to encounter when matter is coupled to the geometrical system. In case the matter field variables are independent as is the case for instance in the disordered phase of the Ising model the geometry can not be affected by their presence. However, if the matter field variables are correlated they might be able to influence the properties of the geometrical system. An example of such behaviour is seen in the model consisting of one Ising spin system coupled to 2-dimensional dynamically triangulated gravity. At the critical point of the spin system the critical exponent, $\gamma_0$, describing the fractal structure of space time is changed from $-\frac{1}{2}$ to $-\frac{1}{3}$. In dimensions higher than two one could imagine that the presence of matter could alter the nature of a possible phase transition in the geometrical system (cf. to section (2.3)). The interaction between geometry and matter might also cause a change in the critical properties of the matter fields. This effect is seen in the above mentioned model as well. Whereas the phase transition of the Ising model on a regular 2-dimensional lattice is of second order it becomes of third order when the Ising model is coupled to two-dimensional dynamically triangulated gravity. In section 4.3.2 and 4.4.2 we will describe what effects are observed when matter is coupled to dynamically triangulated gravity in higher dimensions.

The prescription given above for the coupling of matter to dynamically triangulated gravity is a very direct one. In two dimensions matter can be introduced in a less direct way. By considering triangulations consisting of different polygons some of which appear with negative weights one can obtain models describing non unitary conformal matter with $(p,q) = (2, 2m-1)$ coupled to two-dimensional gravity [30, 31, 32]. These models, the Kazakov multi-critical models can be studied by analytical means and are the subject of sections 3.2 and 3.3. Recently it has furthermore been shown that by coupling an Ising spin system to triangulations of the type just mentioned one can construct models describing two-dimensional quantum gravity coupled to any minimal conformal model and that by a slight generalization of this idea all $(p,q)$ rational matter fields can be reached [33, 34]. Many of the analytical tools developed for the $(2, 2m-1)$ models are probably applicable also in the generic case. We will comment on this new development in section 3.4. The matter fields which can be coupled to gravity in this indirect way all have central charge, $c < 1$, namely

$$c = 1 - \frac{6(p-q)^2}{pq}. \tag{2.22}$$

For matter fields with $c \geq 1$ the direct approach must be used and only numerical investigations are possible. Results of such investigations can be found in references [35, 36, 37].

## 2.3 The Continuum Limit

The ultimate goal for our studies of dynamically triangulated gravity is of course the construction of a continuum theory of quantum gravity. Let us try to describe what possibilities for defining a continuum limit the discretized models might offer us. There is an important difference between the two-dimensional and the higher dimensional cases which can be traced back to the fact that in two dimensions the curvature term in the Einstein Hilbert action is purely topological. Let us consider the partition function of dynamically triangulated gravity in two dimensions for spherical topology. From the expressions (2.14), (2.15) and (2.16) it follows that in the limit $N_2 \to \infty$

$$Z(\kappa_2)\big|_{g=0} \sim \sum_{N_2} N_2^{\gamma_0 - 3} e^{-(\kappa_2 - \kappa_2^c)N_2} \tag{2.23}$$

where we have left out the trivial factor $e^{2\kappa}$. We see that for $\kappa_2 < \kappa_2^c$ the sum will be totally dominated by contributions from surfaces with large areas and will not converge. On the other hand for $\kappa_2 > \kappa_2^c$ surfaces with large areas will be heavily suppressed. Only by letting $\kappa_2$ approach $\kappa_2^c$ from above we can arrange that surfaces of all sizes contribute to the sum and still the sum is well behaved. Considering the limit $\kappa_2 \to (\kappa_2^c)^+$ is hence our only hope for arriving at a continuum theory. In this limit we have[2]

$$Z(\kappa_2)\big|_{g=0} \sim (\kappa_2 - \kappa_2^c)^{2-\gamma_0}. \tag{2.24}$$

Hence the continuum limit we can obtain in two dimensions is not one to which a divergent correlation length is associated. Only moments of the volume of order higher than three diverge. In $d = 3$ and $d = 4$ we have the possibility of another situation. In these cases after having restricted the topology of our space time manifold we still have two coupling constants $\kappa_{d-2}$ and $\kappa_d$ at our hand (cf. to equation (2.18)). It follows from the discussion in section 2.1 that the partition functions for 3- and 4-dimensional dynamically triangulated gravity behave as

$$Z(\kappa_{d-2}, \kappa_d) = \sum_{N_d} f(N_d, \kappa_{d-2}) e^{-(\kappa_d - \kappa_d^c(\kappa_{d-2}))N_d} \tag{2.25}$$

---

[2] Actually also some analytic terms of the type $(\kappa_2 - \kappa_2^c)^n$, $0 < n \leq 2 - \gamma_0$ appear on the right hand side of (2.24). This will be evident from the discussion in section 3.2.

where $f(N_d, \kappa_{d-2})$ is a subleading correction. By the same arguments as used in the two-dimensional case one reaches the conclusion that only by approaching the line $\kappa_d = \kappa_d^c(\kappa_{d-2})$ from the region where $\kappa_d > \kappa_d^c(\kappa_{d-2})$ one has the hope of being able to address continuum physics. However in this case we still have one free parameter left and our hope is that when we move along the line $\kappa_d = \kappa_d^c(\kappa_{d-2})$ we will encounter a divergence in an appropriately defined susceptibility for quantum gravity, i.e. a second order phase transition. This signals the existence of a continuum theory having a massless excitation. In practise one does not perform the sum over $N_d$ in (2.18). One investigates the behaviour of the partition function for a fixed $N_d$. However, by considering a sequence of increasing $N_d$'s one can extract information about the infinite volume limit.

In case a second order phase transition point is encountered one must as in ordinary statistical mechanics to define a continuum theory prescribe how the critical point should be approached. In particular such a prescription must imply a meaningful scaling of physical observables. In two dimensions an interesting recipe for defining a continuum theory exists [38, 39, 40]. This recipe applies not only to the genus zero partition function but to the complete partition function involving all possible topologies. It consists in defining a renormalized cosmological constant $\Lambda_R$ by $(\kappa_2 - \kappa_2^c) = a^2 \Lambda_R$ and sending $a$ to zero and $e^{2\kappa}$ to infinity while keeping fixed the parameter

$$G^{-1} = a^5 \, e^{2\kappa}. \tag{2.26}$$

With this prescription known as the double scaling prescription one can write down a power series representation of the all genus continuum partition function where contributions from surfaces of genus $g$ appear with the factor $(G^{-1}\Lambda_R^{5/2})^{(1-g)}$. The problem of summing over topologies is not solved by this procedure, though, only reformulated. The series is still not even Borel summable. However the double scaling idea has revived the hope that we might eventually be able to address questions in 2-dimensional quantum gravity reaching beyond the perturbative expansion. We will return to this later in section 3.3 and 3.5. As pointed out in the previous section for $d = 3$ and $d = 4$ we have at the present stage no possibility of addressing the question of summing over all topologies. We will describe the present state of the art as regards the possibility of defining a continuum theory for space times homeomorphic to the sphere in section 4.3 and 4.4.

In case we eventually arrive at some continuum theory we must of course provide evidence that it qualifies as a quantum theory of gravity. As mentioned earlier one very important requirement to be imposed on such a theory is that of reparametrization invariance. In two dimensions we have the possibility of testing whether reparametrization invariance is recovered in the continuum limit. A continuum formalism, Liouville theory, allows us to study two-dimensional quantum gravity coupled to conformal matter fields with central charge $c < 1$ without breaking reparametrization invariance. Liouville theory predicts a set of scaling laws for physical observables, referred to as KPZ scaling [41, 42, 43]. One of them is that the contribution to the partition function from surfaces of genus $g$ behaves as

$$Z_g \sim \Lambda_R^{(2-\gamma_{str})(1-g)}. \tag{2.27}$$

For unitary matter fields $\gamma_{str}$ is given by

$$\gamma_{str} = \frac{1}{12}\left(c - 1 - \sqrt{(25-c)(1-c)}\right) \tag{2.28}$$

where $c$ is the central charge of the matter field. For non unitary $(p,q)$ matter (2.28) is replaced by [32]

$$\gamma_{str} = -\frac{2}{p+q-1}. \tag{2.29}$$

Liouville theory hence predicts a value for $\gamma_0$ (cf. to (2.24)). For pure gravity, $c = 0$, and the predicted value of $\gamma_0$ is $-\frac{1}{2}$. This is exactly what is found using the framework of dynamical triangulations. Dynamically triangulated gravity coupled to one Ising spin system is the unitary $c = \frac{1}{2}$ case. Here the prediction of KPZ scaling is $\gamma_0 = -\frac{1}{3}$ and as appears from the previous section this prediction is also confirmed. No results obtained by dynamical triangulation have ever been in conflict with KPZ scaling. Hence it seems that dynamical triangulations *do* allow us to explore the complete space of Riemannian metrics and reparametrization invariance *is* recovered in the continuum limit. Furthermore our choice of the uniform measure is justified. We can of course not take the two-dimensional results as a guarantee that reparametrization invariance is also recovered in the possible continuum limit of dynamically triangulated gravity in higher dimensions. However, since the continuum formalism can not be generalized to $d = 3$ and $d = 4$ while dynamically triangulated gravity can, we consider this a sound way of continuing the analysis. (We note that dynamical triangulations also provide another route for continuing the analysis. While the Liouville approach breaks down at $c = 1$ the coupling of matter fields with $c \geq 1$ to gravity can be handled in a straightforward way within the framework of dynamical triangulations (cf. to section 2.2).)

Let us close this section by mentioning that the static triangulations have not been quite as successful with respect to reproducing the continuum results. In this approach coupling an Ising spin system to gravity does not lead to a modification of $\gamma_0$ as predicted by the formula (2.28) [44].

# 3  D=2. Matrix Models

## 3.1  The Complete Perturbative Solution

### 3.1.1  The Hermitian 1-Matrix Model

It is well known that the partition function of 2-dimensional dynamically triangulated gravity given by (2.12) and (2.13) can be identified with the free energy, $F$, of the following matrix model

$$Z[\lambda, N] = e^{F[\lambda,N]} = \int_{N\times N} d\phi \exp\left\{ N \operatorname{Tr}\left(-\frac{1}{2}\phi^2 + \lambda\phi^3\right)\right\} \tag{3.1}$$

where the integration is over $N \times N$ hermitian matrices and where the gaussian measure is assumed to be normalized, if we set

$$\lambda = e^{-\kappa_2}, \qquad N = e^{\kappa}. \tag{3.2}$$

This can be shown in the following very straightforward manner [45]. The building blocks of our two-dimensional simplicial manifolds are equilateral, oriented triangles. We can consider the term $\lambda \operatorname{Tr} \phi^3 = \lambda\, \phi_{ij}\, \phi_{jk}\, \phi_{ki}$ in the matrix model action as being associated with such a building block, $t$, as indicated in figure 1.

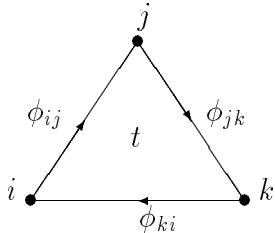

Figure 1. Labels of a triangle.

Here we have labeled the vertices of $t$ by $i, j$ and $k$ and equipped its links with matrix variables $\phi_{ij}, \phi_{jk}, \phi_{ki}$ where the order of the matrix indices reflects the orientation of the links. The term $\lambda \operatorname{Tr} \phi^3$ is now the product of matrix elements along the boundary of $t$. With this assignment it is easy to see that one generates by the perturbative expansion of $F[\lambda, N]$ all connected, closed, orientable 2-dimensional simplicial manifolds since Wick contractions correspond to gluing together triangles along their links always identifying only links with opposite orientation. A given triangulation, $T$, appears with the weight, $W(T)$

$$W(T) = \frac{1}{C(T)} \lambda^{N_2} N^{N_2 - N_1 + N_0} \tag{3.3}$$

which follows easily from the usual Feynman rules. A comparison of (3.3) with (2.12) and (2.13) leads to the identification in (3.2). In the matrix model language the topological expansion (2.14) reads [46]

$$F[\lambda, N] = \sum_{g=0}^{\infty} N^{2-2g} F_g[\lambda] \tag{3.4}$$

where the term $N^{2-2g}F_g[\lambda]$ is the contribution to the free energy from surfaces of genus $g$. Each $F_g$ in (3.4) is a power series in $\lambda$ which as explained in section 2.1 is convergent for $|\lambda| < \lambda_c = e^{-\kappa_2^c}$. Hence each $F_g$ is an analytic function of $\lambda$ in a region around the origin. As also explained in section 2.1 singular behaviour arises when $\lambda \to (\lambda_c)^-$. In this limit one has (cf. to (2.16))

$$F_g \sim (\lambda_c - \lambda)^{(2-\gamma_0)(1-g)}, \qquad \gamma_0 = -\frac{1}{2}. \tag{3.5}$$

It is the singularity at $\lambda = \lambda_c$ which is interesting from the point of view of continuum physics.

In the following we will consider a completely general hermitian 1-matrix model given by

$$Z[\{g_i\}, N] = e^{F[\{g_i\}, N]} = \int_{N \times N} d\phi \exp(-N \operatorname{Tr} V(\phi)) \tag{3.6}$$

where

$$V(\phi) = \sum_{j=1}^{\infty} \frac{g_j}{j} \phi^j. \tag{3.7}$$

For the generic model we still have a genus expansion like (3.4). However, the surfaces which are generated by Wick contractions are now built not only from triangles but from all types of polygons, and different polygons typically appear with different weights. Apart from the partition function we will be interested in expectation values, $\langle \mathcal{Q}(\phi) \rangle$, of operators of the following type

$$\mathcal{Q}(\phi) = \operatorname{Tr} \phi^{k_1} \ldots \operatorname{Tr} \phi^{k_s} \tag{3.8}$$

where we by $\langle \mathcal{Q}(\phi) \rangle$ mean

$$\langle \mathcal{Q}(\phi) \rangle = \frac{1}{Z} \int d\phi \exp(-N \operatorname{Tr} V(\phi)) \mathcal{Q}(\phi). \tag{3.9}$$

The surfaces which appear by the perturbative expansion of (3.9) have $s$ polygonal boundaries of length $k_1, \ldots, k_s$. As usual we will only be interested in contributions from connected surfaces. In stead of working with the expectation values themselves we will work with their generating functionals, the $s$-loop correlators, defined by

$$W(p_1, \ldots, p_s) = \left\langle \operatorname{Tr} \frac{1}{p_1 - \phi} \ldots \operatorname{Tr} \frac{1}{p_s - \phi} \right\rangle_{conn}, \qquad s \geq 1 \tag{3.10}$$

where *conn* refers to the connected part. With the normalization chosen in (3.10), the genus expansion of the correlators reads

$$W(p_1, \ldots, p_s) = \sum_{g=0}^{\infty} \frac{1}{N^{2g-2+s}} W_g(p_1, \ldots, p_s), \qquad s \geq 1. \tag{3.11}$$

In the following we will describe an iterative procedure which allows us to solve the model (3.6) completely in perturbation theory, i.e. to calculate $W_g(p_1, \ldots, p_s)$ and $F_g$ for (in principle) any $g$ and any $s$. In particular we will see how the singular behaviour (3.5) emerges in a certain region of the coupling constant space. Likewise in other regions of the coupling constant space we will see the emergence of Kazakovs $m$'th multi-critical models characterized by $\gamma_0$ taking the value $-\frac{1}{m}$ [30]. Traditionally matrix model are studied by the technique of orthogonal polynomials [15]. However as regards explicit calculations away from the double scaling limit this technique is in practise only applicable to matrix models with a small number of interaction terms. As opposed to this our technique works for a completely general potential. The result is expressed without any reference to the original matrix model coupling constants and $W_g(p_1, \ldots, p_s)$ as well as $F_g$ depend only on a *finite* number of parameters, namely at most $2 \times (3g - 2 + s)$ and $2 \times (3g - 2)$ respectively. The version of the iterative procedure that we will present here was developed in [47]. An earlier version can be found in [48]. The procedure is based on three ingredients, the loop equations, a loop insertion operator, and a set of variables denoted as moments. The loop equations and the loop insertion operator are the subjects of the next section while the moment description is introduced in section 3.1.3. There exists a similar iterative technique for the complex matrix model [49]. Here we will consider only the hermitian case.

### 3.1.2 The Loop Equations

The loop equations are nothing but the Dyson Schwinger equations of the hermitian 1-matrix model [51, 52]. They consist of a set of relations between expectation values which express the invariance of the model under field redefinitions. To derive a version of the loop equations appropriate for our iterative scheme it is convenient to consider the following transformation of the field, $\phi$

$$\phi \to \phi + \epsilon \sum_{n \geq 0} \frac{\phi^n}{p^{n+1}} = \phi + \epsilon \frac{1}{p - \phi}. \tag{3.12}$$

Under a transformation of this type the measure changes as

$$d\phi \to d\phi \left(1 + \epsilon \left(\operatorname{Tr} \frac{1}{p - \phi}\right)^2\right) \tag{3.13}$$

and we get to first order in $\epsilon$

$$\int d\phi \left\{ \left(\operatorname{Tr} \left(\frac{1}{p - \phi}\right)\right)^2 - N \operatorname{Tr} \left(V'(\phi) \frac{1}{p - \phi}\right) \right\} e^{-N \operatorname{Tr} V(\phi)} = 0. \tag{3.14}$$

Introducing corresponding to the matrix, $\phi$, an eigenvalues density, $u(\lambda)$, with support on the real axis [14] this equation can be written as [50]

$$\oint_C \frac{d\omega}{2\pi i} \frac{V'(\omega)}{p - \omega} W(\omega) = (W(p))^2 + \frac{1}{N^2} W(p, p) \tag{3.15}$$

where $V(\omega) = \sum_{j=1}^{\infty} g_j \omega^j / j$ and where $C$ is a curve which encloses the support of the eigenvalue distribution but not the point $\omega = p$. To obtain (3.15) we have also made use of the fact that the expectation value of a product of traces factorizes in the following way

$$\langle \text{Tr } \phi^{k_1} \text{Tr } \phi^{k_2} \ldots \text{Tr } \phi^{k_s}\rangle = \sum_{\{P\}} \prod_{I=1}^{N_p} \langle \prod_{i\in P_I} \text{Tr } \phi^{k_i}\rangle_{conn} N^{2-2N_p-s} \qquad (3.16)$$

where the sum is over all partitions, $\{P\}$, of the set $\{1,\ldots,s\}$, $P_1,\ldots, P_{N_p}$ being the components of the partition $P$. It is easy to convince oneself that (3.16) holds by considering the effect of gluing together two disconnected surfaces along some polygon. In the following we will always assume that $u(\lambda)$ has support only on one interval, $[y,x]$, on the real axis and that

$$1 = \int_y^x u(\lambda) d\lambda. \qquad (3.17)$$

This implies that the 1-loop correlator, $W(p)$, is analytic in the complex plane except from a square root branch cut, $[y,x]$, and that it behaves as $\frac{1}{p}$ as $p\to\infty$ [14]. This is the only assumption we need to find the complete perturbative solution of the model. Neglecting the last term on the right hand side of (3.15) and making use of the assumption about the analyticity structure of $W(p)$ we find the following expression for $W_0(p)$

$$W_0(p) = \frac{1}{2} \oint_C \frac{d\omega}{2\pi i} \frac{V'(\omega)}{p-\omega} \left\{\frac{(p-x)(p-y)}{(\omega-x)(\omega-y)}\right\}^{1/2} \qquad (3.18)$$

where $x$ and $y$ are determined by the matrix model potential in the following way

$$B_1(x,y) = \oint_C \frac{d\omega}{2\pi i} \frac{V'(\omega)}{\sqrt{(\omega-x)(\omega-y)}} = 0, \qquad (3.19)$$

$$B_2(x,y) = \oint_C \frac{d\omega}{2\pi i} \frac{\omega V'(\omega)}{\sqrt{(\omega-x)(\omega-y)}} = 2. \qquad (3.20)$$

One strategy for calculating higher genera contributions to $W(p)$ using loop equations was outlined by Migdal [52] and later elaborated by David [53]. Equation (3.15) can be considered as only the first in an entire chain of loop equations. The n'th equation in the chain involves the $s$-loop correlators for $s=1,2,\ldots,n+1$ and can be derived by exploiting the invariance of the $(n-1)$-loop correlator under field redefinitions of the type (3.12). As in equation (3.15) the term involving the correlator with the largest number of loops is suppressed by a factor $1/N^2$ when compared to the rest. This makes possible an iterative scheme for calculating $W_g(p)$. However, in each step of iteration a new loop equation must be taken into use and calculations quickly become rather involved even for simple matrix model potentials. By means of this method $W_1(p)$ was calculated for even potentials of degree four and six in reference [54].

The iterative scheme referred to in the previous section only involves the first in the chain of loop equations, i.e. equation (3.15). We avoid the use of other loop equations by introducing a so-called loop insertion operator, $\frac{d}{dV(p)}$, defined by

$$\frac{d}{dV(p)} \equiv -\sum_{j=1}^{\infty} \frac{j}{p^{j+1}} \frac{d}{dg_j}. \qquad (3.21)$$

As indicated by its name this operator allows us to step from the $n$-loop correlator to the $(n+1)$-loop correlator. More generally we have

$$W(p_1,\ldots,p_s) = \frac{d}{dV(p_s)}\frac{d}{dV(p_{s-1})}\cdots\frac{dF}{dV(p_1)}. \qquad (3.22)$$

At first sight it seems as if nothing is gained by introducing an operator having an infinite number of terms. However, as will appear from the next sections, in the moment description the loop insertion operator in all cases of interest has effectively only a finite number of terms. We note that in order to have a loop insertion operator it is necessary to work with a generic potential. A symmetrical potential will not do. Now writing $W(p,p)$ as $\frac{d}{dV(p)}W(p)$ and inserting the genus expansion (3.11) in (3.15) we get

$$\left\{\hat{K} - 2W_0(p)\right\} W_g(p) = \sum_{g'=1}^{g-1} W_{g'}(p)\, W_{g-g'}(p) + \frac{d}{dV(p)} W_{g-1}(p), \qquad g \geq 1 \qquad (3.23)$$

where $\hat{K}$ is a linear operator, namely

$$\hat{K} f(p) \equiv \oint_C \frac{d\omega}{2\pi i}\, \frac{V'(\omega)}{p-\omega} f(\omega). \qquad (3.24)$$

Equation (3.23) expresses $W_g(p)$ in terms of $W_{g_i}(p)$ with $g_i < g$. This equation allows us to determine $W_g(p)$ for any value of $g$ starting from the expression (3.18) for $W_0(p)$. From $W(p)$ all correlators involving more loops can be found by application of the loop insertion operator. Furthermore we can find the free energy, $F$, from $W(p)$ by, in a sense to be explained later, applying the inverse loop insertion operator.

### 3.1.3 The Moment Description

To characterize the matrix model potential it proves convenient to introduce in stead of the coupling constants, $\{g_j\}$, a new set of variables, $\{J_i, M_i\}$, denoted as moments and defined by

$$M_k(x,y,\{g_i\}) = \oint_C \frac{d\omega}{2\pi i}\, \frac{V'(\omega)}{(\omega-x)^{k+1/2}\,(\omega-y)^{1/2}}, \qquad k \geq -1, \qquad (3.25)$$

$$J_k(x,y,\{g_i\}) = \oint_C \frac{d\omega}{2\pi i}\, \frac{V'(\omega)}{(\omega-x)^{1/2}\,(\omega-y)^{k+1/2}}, \qquad k \geq -1. \qquad (3.26)$$

We note that the moments $J_k$ and $M_k$ depend only explicitly on coupling constants, $g_j$, with $j \geq k+1$. The moments facilitate the study of the 1-matrix model in several ways. First for each term in the genus expansion of the correlators and the free energy the dependence of the entire set of coupling constants, $\{g_i\}$, arranges into a simple function of a *finte* number of the moments. (Cf. to section 3.1.4.) Furthermore the moments reflect more directly than the couplings the possible critical behaviour of the matrix model. This can be seen from the following rewriting of $W_0(p)$

$$W_0(p) = \frac{1}{2}V'(p) - \frac{1}{4}(p-x)^{1/2}(p-y)^{1/2}\sum_{q=1}^{\infty}\left\{(p-x)^{q-1}M_q + (p-y)^{q-1}J_q\right\}. \qquad (3.27)$$

To derive equation (3.27) one first deforms the the contour of integration in (3.18) into two, one which encloses the point $\omega = p$ and one which encircles infinity. In the contribution from the latter on then writes $(p - \omega)^{-1}$ as

$$\frac{1}{p - \omega} = \frac{1}{2}\left\{\frac{1}{(p-x) - (\omega - x)}\right\} + \frac{1}{2}\left\{\frac{1}{(p-y) - (\omega - y)}\right\} \qquad (3.28)$$

and expand in powers of $\left(\frac{p-x}{\omega-x}\right)$ and $\left(\frac{p-y}{\omega-y}\right)$ respectively. The expansion procedure is justified by the fact that the contour of integration encircles infinity which allows us to assume that $\omega \gg p$. From the classical work of Brezin et. al. [14] it follows that with $W_0(p)$ given as in (3.27) the eigenvalue distribution of the matrix model on spherical level reads

$$u_0(\lambda) = \frac{1}{4\pi} \sum_{q=1}^{\infty} \left\{(\lambda - x)^{q-1} M_q + (\lambda - y)^{q-1} J_q\right\} \sqrt{(x-\lambda)(\lambda - y)}, \qquad y \leq \lambda \leq x. \qquad (3.29)$$

Usually this function vanishes as a square root at both ends of its support. Critical behaviour arises when additional roots of $u_0(\lambda)$ approach $x$ or $y$. The type of critical behaviour is determined by the number, $n$, of extra zeros at the endpoint where most zeros accumulate. Kazakovs $m$'th multi-critical behaviour is found in the vicinity of points in the coupling constant space where $n = m - 1$. The simplest way of reaching a $m$'th multi- critical point is by demanding that $(m - 1)$ extra zeros accumulate at one endpoint, say $x$, while no extra zeros accumulate at $y$. The condition for being at such a point reads

$$M_0 = M_1 = \ldots = M_{m-1} = 0, \qquad M_m \neq 0, \quad J_1 \neq 0 \qquad (3.30)$$

which can easily be read off from (3.29) bearing in mind that the two bracketed terms are actually identical (cf. to equation (3.28)). The simplicity of (3.30) is an appealing feature of the moment description. An even more appealing feature is that the moments as opposed to the coupling constants have definite scaling properties in the vicinity of the $m$'th multi-critical points. This fact makes the moments well suited for addressing the continuum limit. We will analyse the continuum limit in sections 3.2 and 3.3. We note that the expression (3.27) allows us to determine very easily the inverse transformations $g_i = g_i(x, y, \{M_i\}, \{J_i\})$ since we know that $W_0(p)$ only contains negative powers of $p$. In the moment description the loop insertion operator reads

$$\frac{d}{dV(p)} = \sum_n \frac{dM_n}{dV(p)} \frac{\partial}{\partial M_n} + \sum_j \frac{dJ_j}{dV(p)} \frac{\partial}{\partial J_j} + \frac{dx}{dV(p)} \frac{\partial}{\partial x} + \frac{dy}{dV(p)} \frac{\partial}{\partial y} \qquad (3.31)$$

where

$$\begin{aligned}
\frac{dM_n}{dV(p)} &= -\frac{1}{2}(p-x)^{-n-1/2}(p-y)^{-3/2} - (n+1/2)(p-x)^{-n-3/2}(p-y)^{-1/2} \\
&\quad + \frac{1}{2}\left\{\sum_{i=1}^{n}(-1)^{n-i} M_i \left(\frac{1}{x-y}\right)^{n-i+1} + (-1)^n J_1 \left(\frac{1}{x-y}\right)^n\right\} \frac{dy}{dV(p)}
\end{aligned}$$

$$+(n+1/2)M_{n+1}\frac{dx}{dV(p)} \tag{3.32}$$

$$\frac{dx}{dV(p)} = \frac{1}{M_1}(p-x)^{-3/2}(p-y)^{-1/2}. \tag{3.33}$$

Of course the expressions for $dJ_n/dV(p)$ and $dy/dV(p)$ appear from (3.32) and (3.33) by the replacements $x \leftrightarrow y$, $J \leftrightarrow M$. To obtain $dM_n/dV(p)$ and $dx/dV(p)$ we have rewritten the loop insertion operator in the following way

$$\frac{d}{dV(p)} = \frac{\partial}{\partial V(p)} + \frac{dx}{dV(p)}\frac{\partial}{\partial x} + \frac{dy}{dV(p)}\frac{\partial}{\partial y} \tag{3.34}$$

where

$$\frac{\partial}{\partial V(p)} = -\sum_{j=1}^{\infty}\frac{j}{p^{j+1}}\frac{\partial}{\partial g_j} \tag{3.35}$$

and made use of the following relation

$$\frac{\partial}{\partial V(p)}V'(\omega) = \frac{d}{dp}\frac{1}{p-\omega} \tag{3.36}$$

as well as the boundary conditions (3.19) and (3.20).

As mentioned earlier to have a loop insertion operator it is necessary to work with a generic potential. However, the expressions obtained for correlators and the free energy away from the scaling limit can easily be translated to the symmetric case. One simply sets $x = -y = \sqrt{z}$ in the final expressions. This implies setting

$$J_k = (-1)^{k+1}M_k. \tag{3.37}$$

Of course results obtained under the assumption of being close to a critical point of the type given by (3.30) can not be taken over to the symmetric case since the possibility of having different behaviour at the two endpoints of the cut exists only in the generic case. For a symmetrical potential the $m$'th multi-critical points are characterized by the eigenvalue distribution acquiring $(m-1)$ additional zeros at both ends of its support and the condition for being at such a point is the vanishing of the first $(m-1)$ moments. In the vicinity of these points one effectively has a loop insertion operator in the symmetrical case too. We refer to [47] for a discussion of this point.

### 3.1.4 The Structure of the Solution

Below we will present the structure of $F_g$ and $W_g(p)$ for the generic hermitian 1-matrix model in the moment description. That the structure stated is actually correct to all orders in the genus expansion can be proven by induction by means of our iterative scheme. We will not go through the proof here. However the course of the proof will be evident from section 3.1.5 where we describe the iteration process.

The genus $g$ contribution to the free energy of the generic hermitian 1-matrix model takes the following form

$$F_g = \sum_{\substack{\alpha_j>1,\\ \beta_i>1}} \langle \alpha_1 \ldots \alpha_s; \beta_1 \ldots \beta_l | \alpha, \beta, \gamma \rangle_g \frac{M_{\alpha_1}\ldots M_{\alpha_s} J_{\beta_1}\ldots J_{\beta_l}}{M_1^\alpha J_1^\beta d^\gamma}, \qquad g \geq 2 \tag{3.38}$$

where $d = x - y$ is the distance between the endpoints of the support of the eigenvalue distribution, $\langle \ \rangle$ denotes a rational number and $\alpha$, $\beta$ and $\gamma$ are non-negative integers. The indices $\alpha_1, \ldots, \alpha_s, \beta_1, \ldots, \beta_l$ take values in the interval $[2, 3g-2]$ and the summation is over sets of indices. In particular $F_g$ depends on at most $2 \times (3g - 2)$ moments. Furthermore, the following relation holds

$$F: \quad \langle \alpha_1 \ldots \alpha_s; \beta_1 \ldots \beta_l | \alpha, \beta, \gamma \rangle = (-1)^\gamma \langle \beta_1 \ldots \beta_l; \alpha_1 \ldots \alpha_s | \beta, \alpha, \gamma \rangle \,. \tag{3.39}$$

This follows from the fact that $F_g$ must be invariant under the interchange of $x$ and $y$ since nothing allows us to distinguish between the two. In addition there are certain restrictions on the integers which enter equation(3.38), namely

$$s - \alpha \leq 0, \qquad l - \beta \leq 0 \tag{3.40}$$

and

$$F_g: \quad (s - \alpha) + (l - \beta) = 2 - 2g \,, \tag{3.41}$$

$$F_g: \quad \sum_{i=1}^{s}(\alpha_i - 1) + \sum_{j=1}^{l}(\beta_j - 1) + \gamma = 4g - 4 \,. \tag{3.42}$$

The relation (3.41) follows from the fact that the partition function $Z = e^{\sum_g N^{2-2g} F_g}$ is invariant under simultaneous rescalings of $N$ and the eigenvalue density, $u(\lambda)$; $N \to k \cdot N$, $u(\lambda) \to \frac{1}{k}u(\lambda)$. The relation (3.42) follows from the invariance of $Z$ under rescalings of the type $N \to k^2 \cdot N$, $g_j \to k^{j-2}g_j$. Finally the following inequality is fulfilled:

$$F_g: \quad \sum_{i=1}^{s}(\alpha_i - 1) + \sum_{j=1}^{l}(\beta_j - 1) \leq 3g - 3. \tag{3.43}$$

In combination with Eq. (3.42) it gives

$$g - 1 \leq \gamma \leq 4g - 4 \,. \tag{3.44}$$

The quantities $\langle \alpha_1 \ldots \alpha_s; \beta_1 \ldots \beta_l | \alpha, \beta, \gamma \rangle_g$ can be given a geometrical interpretation. We will return to this point in section 3.3.

To explain the structure of $W_g(p)$, let us introduce the basis vectors $\chi^{(n)}(p)$ and $\Psi^{(n)}(p)$:

$$\chi^{(n)}(p) = \frac{1}{M_1}\left\{\Phi_x^{(n)}(p) - \sum_{k=1}^{n-1}\chi^{(k)}(p)M_{n-k+1}\right\}, \qquad n \geq 1 \,, \tag{3.45}$$

$$\Psi^{(n)}(p) = \frac{1}{J_1}\left\{\Phi_y^{(n)}(p) - \sum_{k=1}^{n-1}\Psi^{(k)}(p)J_{n-k+1}\right\}, \qquad n \geq 1 \tag{3.46}$$

where

$$\Phi_x^{(n)}(p) = (p - x)^{-n}\left\{(p-x)(p-y)\right\}^{-1/2}, \qquad n \geq 0 \,, \tag{3.47}$$

$$\Phi_y^{(n)}(p) = (p - y)^{-n}\left\{(p-x)(p-y)\right\}^{-1/2}, \qquad n \geq 0 \,. \tag{3.48}$$

It is easy to show for the operator $\hat{K}$ defined by equation (3.24) that

$$\left\{\hat{K} - 2W_0(p)\right\} \chi^{(n)}(p) = \frac{1}{(p-x)^n}, \qquad n \geq 1, \tag{3.49}$$

$$\left\{\hat{K} - 2W_0(p)\right\} \Psi^{(n)}(p) = \frac{1}{(p-y)^n}, \qquad n \geq 1 \tag{3.50}$$

and that the kernel of $\left\{\hat{K} - 2W_0(p)\right\}$ is spanned by $\Phi_x^{(0)}(p) = \Phi_y^{(0)}(p)$.

In accordance with relation (3.22) and (3.38) the genus $g$ contribution to the 1-loop correlator can be written as

$$W_g(p) = \sum_{n=1}^{3g-1} \left\{ A_g^{(n)} \chi^{(n)}(p) + D_g^{(n)} \Psi^{(n)}(p) \right\}, \quad g \geq 1. \tag{3.51}$$

We note that $\Phi_x^{(0)}(p) = \Phi_y^{(0)}(p)$ does not appear in (3.51). The presence of such terms would contradict the boundary condition $W(p) \to \frac{1}{p}$ for $p \to \infty$ since this behaviour was already obtained for genus zero. Furthermore we note that the structure (3.51) of $W_g(p)$ is in agreement with the assumption that $W(p)$ is analytic in the complex plane except for a square root branch cut $[y, x]$ on the real axis.

The coefficients $A_g^{(n)}$ have the same structure as $F_g$ (cf. to equation (3.38)) and the relation (3.40) is valid also for $A_g^{(n)}$. However the indices $\alpha_1, \ldots, \alpha_s, \beta_1, \ldots, \beta_l$ now take values in the interval $[2, 3g - n]$. The invariance of the partition function under the rescalings described above has the following implications for the structure of $A_g^{(n)}$:

$$A_g^{(n)}: \quad (s - \alpha) + (l - \beta) = 2 - 2g, \tag{3.52}$$

$$A_g^{(n)}: \quad \sum_{i=1}^{s}(\alpha_i - 1) + \sum_{j=1}^{l}(\beta_j - 1) + \gamma = 4g - 2 - n. \tag{3.53}$$

We also have an analogue of (3.43) for $A_g^{(n)}$. It reads

$$A_g^{(n)}: \quad \sum_{i=1}^{s}(\alpha_i - 1) + \sum_{j=1}^{l}(\beta_j - 1) \leq 3g - n - 1. \tag{3.54}$$

Combining (3.53) and (3.54) one gets again the bound (3.44) on $\gamma$. The coefficient $D_g^{(n)}$ appears from $A_g^{(n)}$ by the replacements $d \to -d$, $J \leftrightarrow M$. This follows from the fact that $W_g(p)$ must be invariant under the interchange of $x$ and $y$. (We note that we do not have a relation like (3.39) for the $A_g^{(n)}$'s.) Summing up the information just given about $W_g(p)$ one finds that $W_g(p)$ depends on at most $2 \times (3g - 1)$ moments. Then taking a look at the loop insertion operator (3.31)–(3.33) bearing in mind the relation (3.22) one easily convinces oneself that $W_g(p_1, \ldots, p_s)$ depends on at most $2 \times (3g - 2 + s)$ parameters.

There are two cases in which the correlators and the free energy depend on less than the maximum number of parameters. One case is when one considers a matrix model potential with a finite number of terms, i.e. a potential for which $g_i = 0$ for $i \geq k$. Then $M_j$ and $J_j$ vanish for $j \geq k - 1$. The other case is when one considers a symmetrical potential. Then the two sets of moments become identical and $W_g(p_1, \ldots, p_s)$ depends on at most $(3g - 2 + s)$ parameters, the free energy on at most $(3g - 2)$ parameters.

### 3.1.5 The Iterative Procedure

According to Eq. (3.23), in order to start the iterative procedure we need to calculate $W_0(p,p)$. To do this we first find $W_0(p_1,p_2)$ by applying $\frac{d}{dV(p_2)}$ in the form (3.34) to $W_0(p)$ written as in equation (3.18). This gives [55]

$$W_0(p_1,p_2) = \frac{1}{2(p_1-p_2)^2}\left\{\frac{p_1 p_2 - \frac{1}{2}(p_1+p_2)(x+y)+xy}{\sqrt{(p_1-x)(p_1-y)(p_2-x)(p_2-y)}} - 1\right\}. \qquad (3.55)$$

Then taking the limit $p_1 \to p_2$ we obtain

$$W_0(p,p) = \frac{(x-y)^2}{16(p-x)^2(p-y)^2}. \qquad (3.56)$$

Now we can determine $W_1(p)$ and we see that it is of the form (3.51) with

$$A_1^{(1)} = -\frac{1}{8d}, \qquad A_1^{(2)} = \frac{1}{16}, \qquad (3.57)$$

$$D_1^{(1)} = \frac{1}{8d}, \qquad D_1^{(2)} = \frac{1}{16}. \qquad (3.58)$$

Carrying on the iteration process is straightforward. In each step one must calculate the right hand side of the loop equation (3.23). Decomposing the result obtained into fractions of the type $(p-x)^{-n}$, $(p-y)^{-n}$ allows one to identify immediately the coefficients $A_g^{(n)}$ and $D_g^{(n)}$ (cf. to equations (3.49), (3.50) and (3.51)). To calculate $W_g(p,p)$ one makes use of the expression (3.31)–(3.33) for the loop insertion operator. We note that there is no simplification of the algorithm in the case of a symmetric potential. We can only put $x = -y = \sqrt{z}$ at the end of the calculation. This complication of course stems from the fact that we have to keep the coupling constants with odd indices in the loop insertion operator until all differentiations have been performed. Only hereafter can they be put equal to zero. The same is not true in the double scaling limit however. This point is explained in reference [47].

By taking a closer look at the loop insertion operator (3.31)–(3.33) and bearing in mind the results (3.56), (3.57) and (3.58), it is easy to convince oneself that $A_g^{(n)}$ and $D_g^{(n)}$ depend only on $x$ and $y$ via $(x-y)$ and have the structure shown in equation (3.38). The results for $g=2$ obtained with the aid of *Mathematica* read

$$\begin{aligned}A_2^{(1)} &= \frac{201}{256\, d^5\, J_1{}^2} - \frac{67\, J_2}{128\, d^4\, J_1{}^3} - \frac{5\, J_3}{32\, d^3\, J_1{}^3} + \frac{49\, J_2{}^2}{256\, d^3\, J_1{}^4} \\ &\quad + \frac{57}{64\, d^5\, J_1\, M_1} - \frac{11\, J_2}{128\, d^4\, J_1{}^2\, M_1} + \frac{49\, M_2{}^2}{256\, d^3\, M_1{}^4} \\ &\quad + \frac{201}{256\, d^5\, M_1{}^2} + \frac{22\, M_2}{128\, d^4\, J_1\, M_1{}^2} - \frac{J_2\, M_2}{64\, d^3\, J_1{}^2\, M_1{}^2} \\ &\quad + \frac{67\, M_2}{128\, d^4\, M_1{}^3} - \frac{5\, M_3}{32\, d^3\, M_1{}^3}, \end{aligned}$$

$$\begin{aligned}
A_2^{(2)} &= -\frac{57}{128\, d^4\, J_1\, M_1} + \frac{8\, J_2}{128\, d^3\, J_1{}^2\, M_1} - \frac{49\, M_2{}^2}{256\, d^2\, M_1{}^4} \\
&\quad - \frac{201}{256\, d^4\, M_1{}^2} - \frac{3\, M_2}{128\, d^3\, J_1\, M_1{}^2} + \frac{J_2\, M_2}{128\, d^2\, J_1{}^2\, M_1{}^2} \\
&\quad - \frac{67\, M_2}{128\, d^3\, M_1{}^3} + \frac{5\, M_3}{32\, d^3\, M_1{}^3} \,, \\
A_2^{(3)} &= \frac{49\, M_2{}^2}{256\, d\, M_1{}^4} - \frac{5\, M_3}{32\, d\, M_1{}^3} + \frac{67\, M_2}{128\, d^2\, M_1{}^3} \\
&\quad + \frac{201}{256\, d^3\, M_1{}^2} + \frac{15}{128\, d^3\, J_1\, M_1} - \frac{5\, J_2}{128\, d^2\, J_1{}^2\, M_1} \,, \\
A_2^{(4)} &= -\frac{49\, M_2}{128\, d\, M_1{}^3} - \frac{189}{256\, d^2\, M_1{}^2} \,, \\
A_2^{(5)} &= \frac{105}{256\, d\, M_1{}^2} \,; \\
D_2^{(1)} &= A_2^{(1)}\ (M \longleftrightarrow J,\ d \longrightarrow -d)\,, \\
D_2^{(2)} &= A_2^{(2)}\ (M \longleftrightarrow J,\ d \longrightarrow -d)\,, \\
D_2^{(3)} &= A_2^{(3)}\ (M \longleftrightarrow J,\ d \longrightarrow -d)\,, \\
D_2^{(4)} &= A_2^{(4)}\ (M \longleftrightarrow J,\ d \longrightarrow -d)\,, \\
D_2^{(5)} &= A_2^{(5)}\ (M \longleftrightarrow J,\ d \longrightarrow -d)\,.
\end{aligned}$$

The genus two contribution to $W(p)$ is now determined by equation (3.51).

Determining $F_g$ when $W_g(p)$ is known is straightforward. The strategy consists in writing the basis vectors $\chi^{(n)}(p)$ and $\Psi^{(n)}(p)$ as derivatives with respect to $V(p)$. It is easy to verify that the following relations hold

$$\chi^{(1)}(p) = \frac{dx}{dV(p)}\,, \tag{3.59}$$

$$\Psi^{(1)}(p) = \frac{dy}{dV(p)}\,, \tag{3.60}$$

$$\chi^{(2)}(p) = \frac{d}{dV(p)}\left\{-\frac{2}{3}\ln M_1 - \frac{1}{3}\ln d\right\}\,, \tag{3.61}$$

$$\Psi^{(2)}(p) = \frac{d}{dV(p)}\left\{-\frac{2}{3}\ln J_1 - \frac{1}{3}\ln d\right\}\,. \tag{3.62}$$

Combining this with the results (3.57) and (3.58) one immediately finds

$$F_1 = -\frac{1}{24}\ln M_1 - \frac{1}{24}\ln J_1 - \frac{1}{6}\ln d. \tag{3.63}$$

For $g > 1$ things are not quite as simple. The basis vectors can not be written as total derivatives. This is of course in accordance with the fact that the $A$ and $D$ coefficients now have a more complicated dependence on the potential. However, a rewriting of the basis vectors allows one to identify relatively simply $W_g(p)$ as a total derivative. In the case of $\chi^{(n)}(p)$ this rewriting reads

$$\chi^{(n)}(p) = \frac{1}{M_1}\left\{-\frac{1}{2n-1}\sum_{i=1}^{n-1}(-1)^{n-i-1}\left\{\Phi_x^{(i)} - M_i\frac{dy}{dV(p)}\right\}\left(\frac{1}{x-y}\right)^{n-i}\right.$$
$$\left.-\frac{2}{2n-1}\frac{dM_{n-1}}{dV(p)} - \sum_{k=2}^{n-1}\chi^{(k)}M_{n-k+1}\right\}, \qquad n \geq 2 \qquad (3.64)$$

where $\Phi_x^{(n)}$ should be written as

$$\Phi_x^{(n)} = -\frac{1}{2n-1}\sum_{i=1}^{n-1}(-1)^{n-i-1}\left\{\Phi_x^{(i)} - M_i\frac{dy}{dV(p)}\right\}\left(\frac{1}{x-y}\right)^{n-i}$$
$$+M_n\frac{dx}{dV(p)} - \frac{2}{2n-1}\frac{dM_{n-1}}{dV(p)}, \qquad n \geq 2, \qquad (3.65)$$
$$\Phi_x^{(1)} = M_1\frac{dx}{dV(p)}. \qquad (3.66)$$

The basis vector $\chi^{(1)}(p)$ should of course still be written as in (3.59). The rewriting of the $\Psi^{(n)}$'s is analogous to that of the $\chi^{(n)}$'s. It can be obtained by performing the replacements $J \leftrightarrow M$ and $x \leftrightarrow y$ in the formulas above.

By means of these rewritings we have been able to determine $F_2$. The result reads

$$F_2 = -\frac{119}{7680\,J_1^2\,d^4} - \frac{119}{7680\,M_1^2\,d^4} + \frac{181\,J_2}{480\,J_1^3\,d^3} - \frac{181\,M_2}{480\,M_1^3\,d^3}$$
$$+\frac{3\,J_2}{64\,J_1^2\,M_1\,d^3} - \frac{3\,M_2}{64\,J_1\,M_1^2\,d^3} - \frac{11\,J_2^2}{40\,J_1^4\,d^2} - \frac{11\,M_2^2}{40\,M_1^4\,d^2}$$
$$+\frac{43\,M_3}{192\,M_1^3\,d^2} + \frac{43\,J_3}{192\,J_1^3\,d^2} + \frac{J_2\,M_2}{64\,J_1^2\,M_1^2\,d^2} - \frac{17}{128\,J_1\,M_1\,d^4}$$
$$+\frac{21\,J_2^3}{160\,J_1^5\,d} - \frac{29\,J_2\,J_3}{128\,J_1^4\,d} + \frac{35\,J_4}{384\,J_1^3\,d} - \frac{21\,M_2^3}{160\,M_1^5\,d}$$
$$+\frac{29\,M_2\,M_3}{128\,M_1^4\,d} - \frac{35\,M_4}{384\,M_1^3\,d}. \qquad (3.67)$$

It is obvious from the formulas above that $F_g$ depends for a non-symmetric potential on at most $2 \times (3g - 2)$ moments, and for a symmetrical potential on at most $(3g - 2)$ moments.

## 3.2 The Double Scaling Limit

### 3.2.1 The Partition Function

The regions in the coupling constant space which are interesting from the point of view of continuum physics are those where the free energy becomes singular. From equation (3.38) this is seen to take place when $M_1(x, y, \{g_i\})$ or $J_1(x, y, \{g_i\})$ acquires a zero of some order. If $M_1$ has a zero in $x$ of order $(m-1)$ for $(x, y, \{g_i\}) = (x_c, y_c, \{g_i^c\})$, $M_i$ for $i < m$ will at this point have a zero in $x$ of order $(m - i)$. This corresponds to the situation where $(m-1)$ additional zeros have accumulated at the endpoint $x$ of the eigenvalue distribution (cf. to section 3.1.3). If in addition we assume that $J_1(x_c, y_c, \{g_i^c\}) \neq 0$ we have exactly the situation described in equation (3.30). Let us try to analyse the behaviour of $F_g$ in the vicinity of such a singularity. If we perturb the coupling constants, $\{g_i\}$, away from their critical values, $\{g_i^c\}$, by an amount, $\delta g_i \sim O(a^m)$, in general $[y_c, x_c]$ will move to $[y, x]$. If we assume that

$$x - x_c \sim a \tag{3.68}$$

we will have

$$M_k \sim a^{m-k}, \qquad k \in [0, m] \tag{3.69}$$

while the $J$-moments do not scale. From the boundary condition (3.19) it follows that

$$y - y_c \sim a^m. \tag{3.70}$$

(An explanation of this point will be given shortly.) A given term in the expansion (3.38) of $F_g$ ($g \geq 2$) under these circumstances scales with a negative power of $a$, $P_-$, given by

$$P_- = \alpha(m-1) - \sum_{i=1}^{s}(m - \alpha_i) = (\alpha - s)(m-1) + \sum_{i=1}^{s}(\alpha_i - 1). \tag{3.71}$$

From equation (3.40), (3.41) and (3.43) it follows that the terms which dominate in the limit $a \to 0$ are terms for which

$$(\alpha - s) = 2g - 2, \qquad \sum_{i=1}^{s}(\alpha_i - 1) = 3g - 3. \tag{3.72}$$

These dominant terms hence do not depend on any $J$-moments and have

$$\gamma = g - 1. \tag{3.73}$$

In particular we find that

$$P_-^{max}(F_g) = (g-1)(2m+1). \tag{3.74}$$

In deriving (3.71) we implicitly assumed scaling for all $M$-moments involved. By a slight modification of the argument given above it is easy to convince oneself that terms which contain a moment $M_k$ with $k > m$ will be subdominant in the limit $a \to 0$. Hence in this limit $F_g$ can be written as [3]

$$F_g^{(NS)} = \sum_{1 < \alpha_j \leq m} \langle \alpha_1 \ldots \alpha_s | \alpha \rangle_g \frac{M_{\alpha_1} \ldots M_{\alpha_s}}{M_1^\alpha d_c^{g-1}}, \qquad g \geq 2 \tag{3.75}$$

---

[3] Here and in the following we use the notation that a superscript $(NS)$ refers to a non symmetrical model where the critical behaviour is associated with the endpoint $x$ of the eigenvalue distribution while a superscript $(S)$ refers to a symmetrical model.

where the indices fulfill the requirements (3.72). Each term in (3.75) is $O(a^{(1-g)(2m+1)})$. We now see the possibility of the double scaling limit emerging (cf. to section (2.3)). Namely, if we send $a$ to zero and $N$ to infinity keeping fixed the product

$$G^{-1} = a^{2m+1} N^2 \tag{3.76}$$

we can rewrite the contributions from surfaces of genus $g \geq 2$ to the free energy as a power series in $G$. Genus zero and genus one are special. It appears that in the double scaling limit $F_g$ for $g \geq 1$ depends on at most $(3g-2)$ parameters.

It is easy to convince oneself that if, in addition to the $m$ zeros at the endpoint $x$, $n$ zeros accumulate at the endpoint $y$, the scaling behaviour of $F_g$ is still given by (3.72)–(3.75), as long as $n < m$. The case $n = m$ is special, however. In this case $F_g$ is a sum of two terms of the type (3.75), one involving $M$-moments and one involving $J$-moments. All terms which mix $M$- and $J$-moments are subdominant in the double scaling limit. This is easily seen to be true for $g = 2$ from equation (3.67) and is proven in the general case in reference [47]. If the potential is not symmetric the double scaling limit of $F_g$ for $n = m$ will depend on $2 \times (3g-2)$ parameters. If the potential is symmetric the two sets of moments are identical and only $(3g-2)$ parameters appear.

What is traditionally referred to as $m$'th multi-critical behaviour is obtained when the critical point given by (3.30) is approached along the straight line $g_i = g \cdot g_i^c$ in the coupling constant space. At a given point on this line, assuming the support of the eigenvalue distribution to be $[y, x]$ the boundary conditions (3.19) and (3.20) reduce to

$$c_m(x-x_c)^m g\, M_m^c + c_1(y-y_c)\, g\, J_1^c = 0, \tag{3.77}$$
$$c_m(x-x_c)^m x_c\, g\, M_m^c + c_1(y-y_c)\, y_c\, g\, J_1^c = 2(1-g) \tag{3.78}$$

where we have kept only leading order terms and where

$$M_i^c = M_i(x_c, y_c, \{g_i^c\}) \qquad J_i^c = J_i(x_c, y_c, \{g_i^c\}) \tag{3.79}$$

and

$$c_k = \frac{(2k-1)!!}{k!\, 2^k}. \tag{3.80}$$

From equation (3.77) we get

$$(y - y_c) = -\frac{2 c_m M_m^c}{J_1^c}(x - x_c)^m \tag{3.81}$$

which we note explains the statement (3.70). Inserting (3.81) in (3.78) one finds

$$(1-g) = \frac{1}{2} c_m\, d_c\, g\, M_m^c\, (x - x_c)^m \tag{3.82}$$

and with the scaling (3.68) of $x$ we are led to introduce a renormalized cosmological constant $\Lambda_R$ by

$$(1-g) = a^m \Lambda_R. \tag{3.83}$$

Then we obviously have that the $M$-moments scale as

$$M_k \sim (a\Lambda_R^{1/m})^{m-k} M_m^c, \qquad k \in [1, m] \tag{3.84}$$

and we see from (3.75) the emergence of the formula (2.27) with

$$\gamma_{str} = -\frac{1}{m}. \tag{3.85}$$

It is when the $m$'th multi-critical point is approached as just described that the specific heat (as a function of the renormalized cosmological constant) obeys the celebrated string equation [39, 40, 56, 57]. One can consider a more general way of approaching the critical point, however. In stead of having only one scaling variable, $\Lambda_R$, one can introduce $m$ different scaling variables, $\mu_1, \ldots, \mu_m$ corresponding to the $m$ different moments of relevance — for example in the following way

$$M_k = a^{m-k}\mu_k, k \in [1, m] \tag{3.86}$$

With this prescription one finds

$$F_g^{(NS)} = \sum_{1 < \alpha_j \leq m} \langle \alpha_1 \ldots \alpha_s | \alpha \rangle_g \frac{\mu_{\alpha_1} \ldots \mu_{\alpha_s}}{\mu_1^\alpha} \frac{1}{[a^{2m+1} d_c]^{g-1}}, \qquad g \geq 2. \tag{3.87}$$

The idea of approaching the $m$'th multi-critical point in a more general way will allow us in section 3.3 to establish in a very direct way the connection between the double scaling limit of the partition function of the hermitian 1-matrix model and the $\tau$-function of the kdV hierarchy.

### 3.2.2 The Correlators

In order to be able to speak about the double scaling limit in the context of correlators we must introduce a scaling variable $\pi$ to replace $p$. In the case of a $m$'th multi-critical model where the critical behaviour of the matrix model is determined by the conditions at only one endpoint of the eigenvalue distribution, $x$, we set

$$p_i = x_c + a\pi_i, \tag{3.88}$$
$$x = x_c + a\Lambda_R^{1/m}. \tag{3.89}$$

The scaling appropriate for $y$ depends on the conditions at the other end of the eigenvalue distribution as well as of the details of the prescription for approaching the critical point. It must be determined from the boundary equations in each specific case. However it always holds that $(y - y_c) = o(x - x_c)$. By inserting the scaling relations for $p$, $x$, $y$ and $M_j$ in the general expression (3.51) for $W_g(p)$ it is possible by means of the equations (3.40) and (3.52)–(3.54) to determine $P_-^{max}(W_g(p))$ and to derive a set of requirements like (3.72) that a given term of $W_g(p)$ must fulfill in order not to be subdominant in the limit $a \to 0$. Likewise one can study the behaviour of the multi-loop correlators by inserting the various scaling relations in the explicit expressions for these correlators. However, by analyzing in some detail the loop insertion operator a more direct way of studying the properties of the dominant contributions to the correlators can be found. From the point of view of the limit $a \to 0$ the effect of a given operator in $d/dV(p)$ when acting on an expression like (3.38) is to lower the power of $a$ by some amount. Examining carefully each term in $d/dV(p)$ shows that the power of $a$ is maximally lowered by $m + 3/2$. All operators which do not lower the power of $a$ by this amount hence only give rise to terms which are subdominant in the limit $a \to 0$. One finds that in the this limit the loop insertion operator reduces to

$$\frac{d}{dV(p)}\bigg|_x = \sum_n \frac{dM_n}{dV(p)} \frac{\partial}{\partial M_n} + \frac{dx}{dV(p)} \frac{\partial}{\partial x} \quad (d.s.l.) \qquad (3.90)$$

where

$$\frac{dM_n}{dV(p)} = -(n+1/2)\left\{\Phi_x^{(n+1)}(p) - \frac{M_{n+1}}{M_1}\Phi_x^{(1)}(p)\right\} \quad (d.s.l.), \qquad (3.91)$$

$$\frac{dx}{dV(p)} = \frac{1}{M_1}\Phi_x^1(p) \quad (d.s.l.) \qquad (3.92)$$

and

$$\Phi_x^{(n)}(p) = (p-x)^{-n}\left\{d_c(p-x)\right\}^{-1/2} \quad (d.s.l.). \qquad (3.93)$$

By applying this version of the loop insertion operator to the double scaling relevant part of $F_g$, i.e. the expression (3.75) (with $d_c$ replaced by $d$) we get the directly the dominant contributions to the multi-loop correlators. It appears that in the limit $a \to 0$ the deviation of $y$ from $y_c$ will never be of importance, so we actually do not need to know the scaling behaviour of $y$. Furthermore we see that all dependence of the $J$-moments disappears in the limit $a \to 0$ and that $W_g(p_1, \ldots, p_s)$, $g \geq 1$ depends on at most $(3g-2+s)$ moments. From the expression (3.55) it follows that this statement is also true for $g = 0$ when $s \geq 2$. We note that in the derivation of the expression above for the loop insertion operator we implicitly assumed scaling for all moments involved. As in the case of $F_g$ it holds for the correlators that for a $m$'th critical model any term which contain $M_k$ with $k > m$ will be subdominant in the limit $a \to 0$ and can be left out. From the scaling behaviour of the loop insertion operator it follows that

$$P_-^{max}(W_g(p_1, \ldots, p_s)) = (m+1/2)(2g-2+s) + s, \qquad g \geq 1. \qquad (3.94)$$

This shows that for the contributions to the $s$-loop correlators from surfaces of $g \geq 1$ we need a multiplicative renormalization by $a^s$ in order to have a double scaling expansion (cf. to equation (3.11)). For $g = 0$ additional renormalization is needed. From (3.18) we see that $W_0(p)$ has a part that does not scale as in equation (3.94), namely $\frac{1}{2}V'(p)$. This part must hence be subtracted. So far we have deliberately not used the word "continuum correlators". The precise definition of this concept will be given in section 3.3. As we shall see, when we want to make contact with continuum physics we must also subtract the term $-\frac{1}{2}\frac{1}{(p_1-p_2)^2}$ from $W_0(p_1,p_2)$ even though it has the right scaling behaviour. This is not surprising, though, since $-\frac{1}{2}\frac{1}{(p_1-p_2)^2}$ is exactly $\frac{d}{dV(p_2)} \frac{1}{2} V'(p_1)$, i.e. the remnant of the term which is subtracted from $W_0(p_1)$. No additional renormalization is needed in the case of $W_0(p_1, \ldots, p_s)$ for $s \geq 3$.

The scaling prescription (3.88) must be modified if at the critical point the same number of zeros accumulate at the two endpoints, $x$ and $y$, of the eigenvalue distribution. In the case of a symmetrical potential we have $x = -y = \sqrt{z}$ and it follows from the assumption about the analyticity structure of $W_g(p)$ that all correlators depend only on $p$ via $p^2$. For a symmetrical potential we set

$$p_i^2 = z_c + 2a\pi_i \sqrt{z_c}, \qquad z = z_c + 2a\Lambda_R^{1/m} \sqrt{z_c}. \tag{3.95}$$

(The normalization of (3.95) is chosen in order to facilitate comparison with the non symmetric case.) As mentioned earlier, in the double scaling limit one effectively has a loop insertion operator in the case of the symmetrical potential also. We refer to [47] for the explanation. Here we just note that as in the asymmetric case $W_g(p_1, \ldots, p_s)$ depends on at most $(3g - 2 + s)$ moments except when $g = 0$ and $s = 1$. Similarly, what was stated above about the scaling behaviour of the correlators and the need for renormalization applies to the symmetrical case as well. In the very special case where at the critical point the same number of zeros accumulate at $x$ and $y$ and yet the potential is not symmetric, it is not possible to devise any sensible scaling for $p$. In the following we will refrain from considering this situation.

### 3.2.3 The Perturbative Solution

Having calculated $W_g(p)$ and $F_g$ using the iterative procedure described in section 3.1.5 one can easily by means of the scaling relations for $x$, $y$, $p$ and the moments determine which terms survive in the double scaling limit. However, this is a rather uneconomic method for obtaining the double scaling relevant terms. For example from the long list of complicated expressions for the $A$ and $D$ coefficients of $W_2(p)$ (cf. to section 3.1.5) only $A_2^{(5)}$, the first term of $A_2^{(4)}$ and the two first terms of $A_2^{(3)}$ survive in the double scaling limit when we consider a situation described by (3.88) and (3.89). In reference [47] an algorithm which gives as output only double scaling relevant terms was described. This algorithm not only saves us the trouble of the scaling analysis but also allows us to probe higher genera. To develop the algorithm we carried out a simple but careful scaling analysis of the loop equation. We do not intend to repeat this analysis here. Let us just give a brief outline of the course of the iteration process.

Let us consider first the case where the critical behaviour of the matrix model is determined by the condition at only one endpoint, $x$, of the eigenvalue distribution. It appears from the expressions (3.90)–(3.93) and (3.75) for the double scaling relevant part of the loop insertion operator and the free energy respectively that in the double scaling relevant part of $W_g(p)$ only the basis vectors $\chi^{(n)}(p)$ will appear and only in a form where $(p - y)$ has been replaced by $d_c$. Therefore in the iteration process we must use the basis vectors $\chi^{(n)}(p)$ in this slightly modified form. Apart from this modification the recipe is simple. We do exactly as described in section (3.1.5) just starting from the double scaled version of $W_0(p,p)$

$$W_0^{(NS)}(p,p) = \frac{1}{16}\frac{1}{(p-x)^2} \qquad (d.s.l.) \qquad (3.96)$$

and using the double scaled version of the loop insertion operator. Following this recipe we have calculated $W_g(p)$ as it looks in the double scaling limit for $g = 2, 3$ and $4$. The results can be found in reference [47].

The procedure for calculating $F_g$ when $W_g(p)$ is known given in section (3.1.5) can easily be adjusted to the double scaling limit as well. The starting point is of course the double scaling relevant version of $W_g(p)$ — and the strategy consists as before in rewriting the basis vectors in a form which allows one to identify $W_g(p)$ as a total derivative. However, this time the rewriting in the case of $\chi^{(n)}$ is made with the aid of (3.91) instead of (3.32). For genus one we immediately find

$$F_1^{(NS)} = -\frac{1}{24}\ln M_1 \qquad (d.s.l.). \qquad (3.97)$$

The outcome of the iteration process for $g = 2, 3$ and $4$ is

$$F_2^{(NS)} = \frac{-21\,M_2^3}{160\,d_c\,M_1^5} + \frac{29\,M_2\,M_3}{128\,d_c\,M_1^4} - \frac{35\,M_4}{384\,d_c\,M_1^3}. \qquad (3.98)$$

$$\begin{aligned}F_3^{(NS)} = &\;\frac{2205\,M_2^6}{256\,d_c^2\,M_1^{10}} - \frac{8685\,M_2^4\,M_3}{256\,d_c^2\,M_1^9} + \frac{15375\,M_2^2\,M_3^2}{512\,d_c^2\,M_1^8} + \frac{5565\,M_2^3\,M_4}{256\,d_c^2\,M_1^8} \\ &\;- \frac{5605\,M_2\,M_3\,M_4}{256\,d_c^2\,M_1^7} - \frac{72875\,M_3^3}{21504\,d_c^2\,M_1^7} - \frac{3213\,M_2^2\,M_5}{256\,d_c^2\,M_1^7} + \frac{2515\,M_3\,M_5}{512\,d_c^2\,M_1^6} \\ &\;+ \frac{21245\,M_4^2}{9216\,d_c^2\,M_1^6} + \frac{5929\,M_2\,M_6}{1024\,d_c^2\,M_1^6} - \frac{5005\,M_7}{3072\,d_c^2\,M_1^5}\end{aligned} \qquad (3.99)$$

$$\begin{aligned}F_4^{(NS)} = &\;-\frac{21023793\,M_2^9}{10240\,d_c^3\,M_1^{15}} + \frac{12829887\,M_2^7\,M_3}{1024\,d_c^3\,M_1^{14}} - \frac{98342775\,M_2^5\,M_3^2}{4096\,d_c^3\,M_1^{13}} \\ &\;- \frac{4456305\,M_2^6\,M_4}{512\,d_c^3\,M_1^{13}} + \frac{16200375\,M_2^3\,M_3^3}{1024\,d_c^3\,M_1^{12}} + \frac{26413065\,M_2^4\,M_3\,M_4}{1024\,d_c^3\,M_1^{12}} \\ &\;+ \frac{12093543\,M_2^5\,M_5}{2048\,d_c^3\,M_1^{12}} - \frac{83895625\,M_2\,M_3^4}{32768\,d_c^3\,M_1^{11}} - \frac{68294625\,M_2^2\,M_3^2\,M_4}{4096\,d_c^3\,M_1^{11}} \\ &\;- \frac{12367845\,M_2^3\,M_4^2}{2048\,d_c^3\,M_1^{11}} - \frac{13024935\,M_2^3\,M_3\,M_5}{1024\,d_c^3\,M_1^{11}} - \frac{15411627\,M_2^4\,M_6}{4096\,d_c^3\,M_1^{11}}\end{aligned}$$

$$+\frac{32418925\,M_3{}^3\,M_4}{24576\,d_c^3\,M_1{}^{10}} + \frac{17562825\,M_2\,M_3\,M_4{}^2}{4096\,d_c^3\,M_1{}^{10}} + \frac{578655\,M_2\,M_3{}^2\,M_5}{128\,d_c^3\,M_1{}^{10}}$$

$$+\frac{10050831\,M_2{}^2\,M_4\,M_5}{2048\,d_c^3\,M_1{}^{10}} + \frac{5472621\,M_2{}^2\,M_3\,M_6}{1024\,d_c^3\,M_1{}^{10}} + \frac{44207163\,M_2{}^3\,M_7}{20480\,d_c^3\,M_1{}^{10}}$$

$$-\frac{1511055\,M_2\,M_5{}^2}{2048\,d_c^3\,M_1{}^9} - \frac{7503125\,M_4{}^3}{36864\,d_c^3\,M_1{}^9} - \frac{2642325\,M_3\,M_4\,M_5}{2048\,d_c^3\,M_1{}^9}$$

$$-\frac{11532675\,M_3{}^2\,M_6}{16384\,d_c^3\,M_1{}^9} - \frac{6242775\,M_2\,M_4\,M_6}{4096\,d_c^3\,M_1{}^9} - \frac{6968247\,M_2\,M_3\,M_7}{4096\,d_c^3\,M_1{}^9}$$

$$-\frac{4297293\,M_2{}^2\,M_8}{4096\,d_c^3\,M_1{}^9} + \frac{12677665\,M_2\,M_9}{32768\,d_c^3\,M_1{}^8} + \frac{8437275\,M_5\,M_6}{32768\,d_c^3\,M_1{}^8}$$

$$+\frac{8913905\,M_4\,M_7}{32768\,d_c^3\,M_1{}^8} + \frac{10156575\,M_3\,M_8}{32768\,d_c^3\,M_1{}^8} - \frac{8083075\,M_{10}}{98304\,d_c^3\,M_1{}^7}\,. \tag{3.100}$$

We note that the terms listed above are only potentially relevant. For a $m$'th multi-critical model all terms involving $M_k$, $k > m$ will be subdominant in the double scaling limit and can be ignored. Obviously the arguments and results presented above can be translated to the case where the critical behaviour of the matrix model is associated with the endpoint, $y$, of the eigenvalue distribution just be replacing $M$ by $J$ and $d_c$ by $-d_c$.

Let us turn now to the case of a symmetrical potential. In view of the scaling relations (3.95) it seems unnatural to work with terms like $(p - x)$ and $(p - y)$ and as shown in reference [47] there exists a way of avoiding this. However, an interesting relation can be derived if one analyses the scaling behaviour of $W_g(p)$ in the $x$, $y$ formalism. The details of this scaling analysis can be found in reference [47]. Here we just state the result. It turns out that in the case of a symmetrical potential the loop equation decouples completely into two independent equations. Each of these is a double scaled version of the loop equation for the asymmetric potential. One corresponds to the case where the critical behaviour is associated with the endpoint, $x$, of the eigenvalue distribution, the other to the case where the critical behaviour is associated with the endpoint, $y$. In particular the free energy of the symmetrical model can be written as a sum of two contributions one from each endpoint of the support of the eigenvalue distribution. The contribution from the endpoint, $x$ takes the form of equation (3.75) and the contribution from the endpoint $y$ appears from (3.75) when $M$ is replaced by $J$ and $d_c$ by $-d_c$. However, since for a symmetrical potential $J_k = (-1)^{k+1} M_k$ we find

$$F_g^{(S)} = 2 F_g^{(NS)} \qquad (d.s.l.). \tag{3.101}$$

A similar conclusion was reached in another approach in reference [58].

## 3.3 From 1-Matrix Model to (almost) $\tau$-function

### 3.3.1 To Square or not to Square

It is a well established fact that the partition function of the hermitian 1-matrix model in the double scaling limit is closely related to a $\tau$-function of the kdV hierarchy. In references [59, 60] a continuum version of the loop equations of the symmetric hermitian matrix model was derived and it was found that the corresponding continuum partition function was the square of a $\tau$-function of the kdV hierarchy. A similar conclusion was reached in reference [61] where discrete loop equations were used as the starting point— but it appeared that just taking the double scaling limit of the partition function of the symmetric hermitian matrix model does not leave one with the square of a $\tau$-function. In an attempt to overcome these difficulties a renomalization procedure was prescribed. Equation (3.101) indicates that the reason why it is the $\tau$-function squared which appears in the here mentioned analyses is that a symmetrical potential for the matrix model is assumed. For a generic potential the continuum partition function should be related to the $\tau$-function itself rather than the $\tau$-function squared. As we shall see this can indeed be proven to be the case.

After the analyses of references [59], [60] and [61] new development has taken place. A matrix model realization of a $\tau$-function of the kdV hierarchy—the Kontsevich model—has been found [62]. Since this discovery there have been attempts to relate the partition function of the Kontsevich model to the partition function of various 1-matrix models. In reference [63] a limiting procedure which allows one, on path integral level, to pass from the partition function of the reduced hermitian matrix model to the square of the partition function of the Kontsevich model was presented. In reference [47] a different limiting procedure provided a way of passing from the partition function of the generic hermitian 1-matrix model to the non-squared partition function of the Kontsevich model. However, the latter limiting procedure was rather unconventional involving an analytic continuation of the size of the matrices entering the original 1-matrix model from $N$ to $-\xi N$. Here we will show using the conventional double scaling prescription that it is possible to define continuum time variables for the generic hermitian 1-matrix model such that its correlators after appropriate renormalization—in compliance with the analysis of section 3.2.2—turn into the correlators of the Kontsevich model expressed in terms of kdV times. Furthermore it appears that the double scaling limit of the partition function of the generic hermitian 1-matrix model agrees with the partition function of the Kontsevich model (and not the square of the partition function) except for some complications at genus zero. The complications encountered for genus zero are of the same type as those encountered in reference [61]. However, we see no sensible way of removing these complications. The analysis which will be carried out below is based on reference [64].

### 3.3.2 The Kontsevich Model

The Kontsevich model is defined by the partition function

$$Z^{Kont}[N, M] = e^{F^{Kont}[N,M]} = \frac{\int dX \exp\left\{-N \operatorname{Tr}\left(\frac{MX^2}{2} - \frac{iX^3}{6}\right)\right\}}{\int dX \exp\left\{-N \operatorname{Tr}\left(\frac{MX^2}{2}\right)\right\}} \qquad (3.102)$$

where the integration is over $N \times N$ hermitian matrices. This partition function only depends on the parameters $t_k$

$$t_k = \frac{1}{N} \operatorname{Tr} M^{-(2k+1)} \tag{3.103}$$

and expressed in terms of these it is a $\tau$-function of the kdV hierarchy [62]. As is the case for 1-matrix models the free energy $F^{Kont}$ has a genus expansion (cf. to equation (3.4)). The genus zero contribution to $F^{Kont}$ reads

$$\begin{aligned} F_0^{Kont} &= \frac{1}{3}\frac{1}{N}\sum_{i=1}^N m_i^3 - \frac{1}{3}\frac{1}{N}\sum_{i=1}^N (m_i^2 - 2u_0)^{3/2} - u_0 \frac{1}{N}\sum_{i=1}^N (m_i^2 - 2u_0)^{1/2} \\ &+ \frac{u_0^3}{6} - \frac{1}{2}\frac{1}{N^2}\sum_{i,j=1}^N \ln\left\{\frac{(m_i^2 - 2u_0)^{1/2} + (m_j^2 - 2u_0)^{1/2}}{m_i + m_j}\right\} \end{aligned} \tag{3.104}$$

where the $m_i$'s are the eigenvalues of the matrix $M$ and the parameter $u_0$ is given by the boundary condition

$$u_0 + \frac{1}{N}\sum_i \frac{1}{\sqrt{m_i^2 - 2u_0}} = 0. \tag{3.105}$$

It can be derived by means of the Dyson Schwinger equations of the model as done in references [65, 66, 67]. Alternatively it can be found by exploiting the fact that $Z^{Kont}[N, \{t_k\}]$ is a $\tau$-function of the kdV hierarchy [68]. The higher genera contributions can be written in the following form [67, 68]

$$F_1^{Kont} = -\frac{1}{24}\ln(I_1 - 1) \tag{3.106}$$

$$F_g^{Kont} = \sum_{\alpha_j > 1} \langle \alpha_1 \ldots \alpha_s | \alpha \rangle_g^{Kont} \frac{I_{\alpha_1} \ldots I_{\alpha_s}}{(I_1 - 1)^\alpha}, \qquad g \geq 2 \tag{3.107}$$

where the sum is over sets of indices obeying the following restrictions

$$\sum_{j=1}^s (\alpha_j - 1) = 3g - 3, \qquad (\alpha - s) = 2g - 2 \tag{3.108}$$

and where the moments $I_k$ are defined by

$$I_k = \frac{1}{N}\sum_{j=1}^N \frac{1}{(m_j^2 - 2u_0)^{k+1/2}}, \qquad k \geq 0. \tag{3.109}$$

The quantities $\langle \alpha_1 \ldots \alpha_s | \alpha \rangle_g^{Kont}$ have a geometrical interpretation in terms of intersection indices on the moduli space $\mathcal{M}_{g,s}$ of Riemann surfaces with $g$ handles and $s$ marked points [62, 69]. Different types of singular behaviour are possible for (3.107). These different types of singular behaviour can be indexed by an integer, $m$. As shown in reference [68] to reach the m'th type of critical behaviour one should neglect $I_k$ with $k > m$, keep $I_m$ constant, send $I_1 - 1, I_2, \ldots, I_{m-1}$ to zero and introduce a scaling parameter, $s$, such that

$$v_q = \frac{I_q(1-I_1)^{q-2}}{I_2^{q-1}}, \qquad 3 \leq q \leq m \tag{3.110}$$

remains finite while

$$\frac{1-I_1}{I_2^{3/5}} = \frac{v_1}{v_2^{3/5}} \cdot s \tag{3.111}$$

tends to zero. In this limit one has

$$F_g^{Kont} = \sum_{1 < \alpha_j \leq m} \langle \alpha_1 \ldots \alpha_s | \alpha \rangle_g^{Kont} \frac{v_{\alpha_1} \ldots v_{\alpha_s}}{v_1^\alpha} \frac{1}{s^{5(g-1)}}, \qquad g \geq 2. \tag{3.112}$$

We see that this prescription for fine tuning the $I$'s is completely equivalent to the fine tuning of the $M$'s in the vicinity of a $m$'th multi-critical point, $s^5$ playing the role of $a^{(2m+1)}$ (cf. to equation (3.86) and (3.87)). However in the case of the hermitian 1-matrix model our starting point is the expression (3.38) for the free energy which as opposed to (3.107) contains terms which will never be of importance for any type of critical behaviour. Our conjecture is that with an appropriate definition of continuum times for the hermitian 1-matrix model the potentially relevant terms of (3.38) are exactly given by (3.107).

### 3.3.3 Continuum Time Variables, Correlators and Loop Equations

To motivate our choice of continuum time variables $T_k$ for the non symmetrical hermitian 1-matrix model let us write the boundary equation for the Kontsevich model as

$$\sum_{k=0}^{\infty} c_k \left(t_k + \delta_{k,1}\right)(2u_0)^k = 0 \tag{3.113}$$

where $c_k$ was defined in (3.80).

The idea is now to define $T_k$ in such a way that by taking the double scaling limit of the boundary equations (3.19) and (3.20) we reproduce equation (3.113) with the $T_k$'s replacing the $t_k$'s. Here and in the following we will consider a 1-matrix model for which the eigenvalues, at the critical point, are confined to the interval $[y_c, x_c]$. As mentioned earlier, when we move away from a given $m$'th multi-critical point by a change of coupling constants $\delta g_i \sim O(a^m)$ in general both $x$ and $y$ will change. To keep the presentation as simple as possible we restrict ourselves to considering only the subclass of deformations for which $y$ is kept fixed at $y_c$ [4]. Expanding equation (3.19) in powers of $(x - x_c)$ we find

$$\sum_{p=0}^{\infty} c_p(x - x_c)^p M_p^c(\{g_i\}) = 0 \tag{3.114}$$

---

[4]We note that this subclass of deformations does not include the deformations that lead to the emergence of the matrix model string equations (cf. to page 25). However the arguments of this and all following sections can be generalized to the case where $y$ is not kept fixed. The expressions just become more involved.

where
$$M_p^c(\{g_i\}) = M_p(x_c, y_c, \{g_i\}). \tag{3.115}$$

Note that the coupling constants entering $M_p^c(\{g_i\})$ are here completely arbitrary. Rewriting the boundary equation (3.20) we find

$$\sum_{p=0}^{\infty} c_p(x - x_c)^p \left\{ x_c M_p^c(\{g_i\}) + M_{p-1}^c(\{g_i\}) \right\} = 2. \tag{3.116}$$

Since $M_p^c(\{g_i\}) \sim a^{m-p}$, for $0 \leq k \leq m$ the term $M_{p-1}^c(\{g_i\})$ is subleading when compared to $M_p^c(\{g_i\})$ except for $M_{-1}^c(\{g_i\})$ which is equal to 2 up to subleading terms of $O(a^{m+1})$. If we set

$$x - x_c = a(2u_0) \tag{3.117}$$

and define our continuum time variables by

$$T_k + \delta_{k,1} = a^{k+1/2} d_c^{1/2} M_k^c(\{g_i\}), \qquad k \geq 0 \tag{3.118}$$

both (3.114) and (3.116) turn into the boundary equation of the Kontsevich model (3.113). (The reason why we include an additional factor $\sqrt{a}\, d_c^{1/2}$ will become clear later.)

We define continuum correlators $W^{cont}(\pi_1, \ldots, \pi_s)$ by

$$W(p) - \frac{1}{2}V'(p) \xrightarrow{d.s.l.} \frac{1}{a}\left( W^{cont}(\pi) - \frac{1}{2}\sum_{k=0}^{\infty}(T_k + \delta_{k,1})\pi^{k-1/2} \right) \tag{3.119}$$

$$W(p_1, p_2) + \frac{1}{2}\frac{1}{(p_1 - p_2)^2} \xrightarrow{d.s.l.} \frac{1}{a^2}\left( W^{cont}(\pi_1, \pi_2) + \frac{\frac{1}{2}(\pi_1 + \pi_2)}{2(\pi_1 - \pi_2)^2 \sqrt{\pi_1 \pi_2}} \right) \tag{3.120}$$

$$W(p_1, \ldots, p_s) \xrightarrow{d.s.l.} \frac{1}{a^s} W^{cont}(\pi_1, \ldots, \pi_s) \tag{3.121}$$

where the scaling prescription (3.88) for $p_i$ is understood. Multiplicative renormalization is prescribed for all correlators. Additional subtractions appear in the case of the 1- and 2-loop correlators. This is exactly the type of renormalization that we found to be necessary in the analysis of section 3.2.2 We note that the subtractions appearing above only concern the *genus zero* contributions to the 1- and 2-loop correlators. For the 2-loop correlator in the case where the two momenta coincide we use the following renormalization prescription

$$W(p, p) \xrightarrow{d.s.l.} \frac{1}{a^2}\left( W^{cont}(\pi, \pi) + \frac{1}{16\pi^2} \right). \tag{3.122}$$

The rationale for the multiplicative renormalization can also be given without referring explicitly to the detailed analysis of section 3.2.2. By expanding the moment $M_p$ in powers of $x - x_c$ one gets

$$M_p = \sum_{l=p}^{\infty} a^{-p-1/2}(T_l + \delta_{1,l})(2u_0)^{l-p}\frac{\Gamma(l + 1/2)}{\Gamma(p + 1/2)(l - p)!}, \qquad p \geq 0. \tag{3.123}$$

Using equation (3.123) the scaling relation (3.117) and (3.88) for $x$ and $p$ respectively and the boundary equation expressed in terms of $T_k$'s one can by means of the chain rule show that the following relation holds

$$\frac{d}{dV(p)} \xrightarrow{d.s.l.} \frac{1}{a} \frac{d}{dV^{cont}(\pi)} = \frac{1}{a} \left\{ -\sum_{k=0}^{\infty} (k+1/2) \frac{1}{\pi^{k+3/2}} \frac{d}{dT_k} \right\}. \tag{3.124}$$

Bearing in mind the relation (3.22) it appears natural to extract one power of $a^{-1}$ for each loop in a given correlator. The relation (3.124) will be essential for the proof of our conjecture. (We note that from the equation (3.123) one can read off the continuum scaling behaviour of the moments for a given $m$'th multi-critical model and the relation (3.69) is easily reproduced.) Due to the peculiarities of the genus zero contributions to the 1- and 2-loop correlators it is convenient to use the loop equations in the genus expanded version (3.23). Let us introduce in equation (3.23) the continuum correlators. First we note that since $W_0(p)$ itself does not scale we have in the double scaling limit

$$\begin{aligned} a^2 \cdot \left\{ \hat{K} - 2W_0(p) \right\} W_g(p) &\xrightarrow{d.s.l.} \oint_C \frac{d\omega}{2\pi i} \frac{V'(\omega) - 2W_0(\omega)}{p - \omega} W_g^{cont}(\omega) \\ &= -\left\{ 2W_0^{cont}(\pi) - \sum_k (T_k + \delta_{k,1}) \pi^{k-1/2} \right\} W_g^{cont}(\pi) \\ &\quad - \oint_\infty \frac{d\omega}{2\pi i} \left\{ \frac{2W_0^{cont}(\omega) - \sum_k (T_k + \delta_{k,1}) \omega^{k-1/2}}{\pi - \omega} \right\} W_g^{cont}(\omega) \end{aligned} \tag{3.125}$$

where $\oint_\infty$ denotes a contour integral where the contour encircles infinity and where we have used the definition (3.119). To proceed let us write $W_g^{cont}(\omega)$ as

$$W_g^{cont}(\omega) = \sum_{q=0}^{\infty} \omega^{-q-3/2} W_g^{cont,q}. \tag{3.126}$$

That $W_g^{cont}$ allows an expansion of this type is obvious for $g \geq 1$ since for $g \geq 1$ we have $W_g^{cont}(\omega) = \frac{d}{dV^{cont}(\omega)} F_g$, where $\frac{d}{dV^{cont}(\omega)}$ is given by equation (3.124). (It is also evident from the explicit expression for $W_g(p)$ given in equation (3.51).) That the same is true for $g = 0$ will become clear in section (3.3.5). Performing the contour integral in (3.125) and making use of the definition (3.120) one obtains the following continuum loop equation

$$-\left[ 2W_0^{cont}(\pi) - \sum_k (T_k + \delta_{k,1}) \pi^{k-1/2} \right] W_g^{cont}(\pi) - \sum_{q,a} (T_{2+a+q} + \delta_{2+q+a,1}) W_g^{cont,q} \pi^a$$
$$= \sum_{g'=1}^{g-1} W_{g'}^{cont}(\pi) W_{g-g'}^{cont}(\pi) + W_{g-1}^{cont}(\pi, \pi) + \delta_{g,1} \cdot \frac{1}{16\pi^2}. \tag{3.127}$$

### 3.3.4 Loop Equations and Correlators for the Kontsevich Model

Inspired by the equation (3.124) let us introduce a loop insertion operator for the Kontsevich model by

$$\frac{d}{dV^{Kont}(\pi)} \equiv -\sum_k (k+1/2)\frac{1}{\pi^{k+3/2}}\frac{d}{dt_k} \qquad (3.128)$$

and multi-loop correlators by

$$W^{Kont}(\pi_1,\ldots,\pi_s) = \frac{d}{dV^{Kont}(\pi_s)}\cdots\frac{d}{dV^{Kont}(\pi_1)}F^{Kont}, \qquad s \geq 1. \qquad (3.129)$$

Using the relation (3.103) it is easy to show by means of the chain rule that

$$\frac{d}{dV^{Kont}(\pi)} = N \left.\frac{\partial}{\partial m_i^2}\right|_{m_i^2=\pi}. \qquad (3.130)$$

In this form the loop insertion operator can readily be applied to (3.104) to yield

$$W_0^{Kont}(\pi) = \frac{1}{2}\left[-\sqrt{\pi-2u_0} + \sqrt{\pi} + \frac{1}{N}\sum_{j=1}^N \frac{1}{\pi-m_j^2}\left(\frac{\sqrt{\pi-2u_0}}{\sqrt{m_j^2-2u_0}} - \frac{m_j}{\sqrt{\pi}}\right)\right] \qquad (3.131)$$

and

$$W_0^{Kont}(\pi_1,\pi_2) = \frac{1}{4(\pi_1-\pi_2)^2}\left\{\frac{\pi_1+\pi_2-4u_0}{\sqrt{\pi_1-2u_0}\sqrt{\pi_2-2u_0}} - \frac{\pi_1+\pi_2}{\sqrt{\pi_1\pi_2}}\right\}. \qquad (3.132)$$

Furthermore taking the limit $\pi_1 \to \pi_2$ in (3.132) we find

$$W_0^{Kont}(\pi,\pi) = \frac{u_0}{4(\pi-2u_0)\pi^2}. \qquad (3.133)$$

We note that the 1-loop correlator of the Kontsevich model is analytic in the complex plane except for a square root branch cut $[-\infty, 2u_0]$ on the real axis. This is exactly the same analyticity structure as one obtains for the 1-loop correlator of the hermitian 1-matrix model by the scaling prescriptions (3.117) and (3.88). With the definitions given above the master equation of the Kontsevich model can be written as

$$\frac{1}{N^2}\left\{W^{Kont}(\pi,\pi) + \frac{1}{16\pi^2}\right\} + \left(\tilde{W}^{Kont}(\pi)\right)^2 + \frac{1}{N}\sum_{j=1}^N \frac{\tilde{W}^{Kont}(m_j^2)}{m_j^2-\pi} = \frac{\pi}{4} \qquad (3.134)$$

where

$$\tilde{W}^{Kont}(\pi) = W^{Kont}(\pi) - \frac{1}{2}\sum_k (t_k+\delta_{k,1})\pi^{k-1/2}. \qquad (3.135)$$

To make contact with the previous section we will rewrite this equation in a genus expanded version. To deal with the sum appearing on the left hand side of equation (3.134) let us note that the genus $g$ contribution to the 1-loop correlator can be expanded in the following way

$$W_g^{Kont}(\pi) = \sum_{q=0}^\infty \pi^{-q-3/2} W_g^{Kont,q}. \qquad (3.136)$$

Then making use of the definition of the $t_k$'s, (3.103), we arrive at the following form of the loop equation.

$$-\left[2W_0^{Kont}(\pi) - \sum_k (t_k + \delta_{k,1})\pi^{k-1/2}\right] W_g^{Kont}(\pi) - \sum_{q,a}(t_{2+a+q} + \delta_{2+q+a,1}) W_g^{Kont,q} \pi^a$$

$$= \sum_{g'=1}^{g-1} W_{g'}^{Kont}(\pi) W_{g-g'}^{Kont}(\pi) + W_{g-1}^{Kont}(\pi,\pi) + \delta_{g,1} \cdot \frac{1}{16\pi^2}. \tag{3.137}$$

### 3.3.5 Proof of our Conjecture

The task of proving statement made in section (3.3.1) amounts to proving the following two identities

$$W_0^{cont}(\pi) = W_0^{Kont}(\pi), \tag{3.138}$$
$$W_0^{cont}(\pi_1, \pi_2) = W_0^{Kont}(\pi_1, \pi_2) \tag{3.139}$$

where it is understood that the quantities on the left hand side should be expressed in terms of $T_k$'s whereas those on the right hand side should be expressed in terms of $t_k$'s. By comparing equation (3.127) and (3.137) it is easily seen that once this task has been fulfilled it follows by induction that

$$W_g^{cont}(\pi, \{T_k\}) = W_g^{Kont}(\pi, \{t_k\}), \qquad g \geq 1 \tag{3.140}$$

since now the two sets of loop equations and corresponding boundary equations only differ by $\{T_k\}$ appearing in one case and $\{t_k\}$ in the other. Furthermore by taking a glance at equation (3.124) and (3.128) bearing in mind the relations (3.22) and (3.119) – (3.121) one easily convinces oneself that

$$W_g^{cont}(\pi_1, \ldots, \pi_s, \{T_k\}) = W_g^{Kont}(\pi_1, \ldots, \pi_s, \{t_k\}), \qquad g, s \geq 1. \tag{3.141}$$

From equation (3.119) and (3.140) it follows that

$$W_g(p) \xrightarrow{d.s.l.} \frac{1}{a} W_g^{Kont}(\pi), \qquad g \geq 1 \tag{3.142}$$

since both the second term on the left hand side and the second term on the right hand side of (3.119) are of zeroth order in genus. Now due to the similarity between $\frac{d}{dV^{cont}(\pi)}$ and $\frac{d}{dV^{Kont}(\pi)}$ we immediately find

$$F_g^{(NS)} \xrightarrow{d.s.l.} F_g^{Kont}, \qquad g \geq 1. \tag{3.143}$$

To address the $g = 0$ case we note that equation (3.120) and (3.139) imply that the double scaling limit of $W_0(p_1, p_2)$ differs from $W_0^{Kont}(\pi_1, \pi_2)$ only by a term which does not depend on any couplings. Therefore we have

$$W_0^{cont}(\pi_1, \ldots, \pi_s) = W_0^{Kont}(\pi_1, \ldots, \pi_s), \qquad s \geq 3. \tag{3.144}$$

However, equation (3.138) does not allow us to conclude anything about the relation between the genus zero contributions to the partition functions of the two models because of the subtractions appearing in equation (3.119).

Let us now turn to the proof of the relations (3.138) and (3.139). The proof of the latter is by far the most straightforward since $W_0(p_1, p_2)$ is universal, i.e. it does not contain any explicit reference to the coupling constants. Taking the double scaling limit of (3.55) following the prescriptions (3.117) and (3.88) one easily reproduces the genus zero part of (3.120) with $W_0^{Kont}(\pi_1, \pi_2)$ replacing $W_0^{cont}(\pi_1, \pi_2)$. Likewise taking the double scaling limit of (3.56) one gets exactly the genus zero part of the relation (3.122) with $W_0^{cont}(\pi, \pi)$ replaced by $W_0^{Kont}(\pi, \pi)$. To prove the relation (3.138) we will prove that

$$W_0(p) - \frac{1}{2}V'(p) + \frac{1}{a}\frac{1}{2}\sum_{k=0}^{\infty}(T_k + \delta_{k,1})\pi^{k-1/2} \xrightarrow{d.s.l.} \frac{1}{a}W_0^{Kont}(\pi). \qquad (3.145)$$

First we note that due to equation (3.18) we can write the two first terms as

$$W_0(p) - \frac{1}{2}V'(p) = \frac{1}{2}\oint_{\infty}\frac{d\omega}{2\pi i}\frac{V'(\omega)}{p-\omega}\left\{\frac{(p-x)(p-y)}{(\omega-x)(\omega-y)}\right\}^{1/2}. \qquad (3.146)$$

Furthermore inserting our definition of continuum times (3.118) into the remaining term on the left hand side of (3.145) we get

$$\begin{aligned}
\frac{1}{a}\frac{1}{2}\sum_{k=0}^{\infty}(T_k + \delta_{k,1})\pi^{k-1/2} &= \frac{1}{2}\sum_{k=0}^{\infty}(a\pi)^{k-1/2}d_c^{1/2}M_k^c(\{g_i\}) \\
&= -\left\{W_0(p) - \frac{1}{2}V'(p)\right\}\bigg|_{x=x_c, y=y_c} + \frac{1}{2}(p-x_c)^{-1/2}(p-y_c)^{1/2}M_0^c(\{g_i\}) \\
&= -\frac{1}{2}\oint_{\infty}\frac{d\omega}{2\pi i}\frac{V'(\omega)}{p-\omega}\left\{\frac{(p-y_c)(\omega-x_c)}{(p-x_c)(\omega-y_c)}\right\}^{1/2}. 
\end{aligned} \qquad (3.147)$$

To obtain the second equality sign we have made use of the rewriting of $W_0(p)$ given in (3.27) and the scaling relation for $p$, (3.88). (In particular we have used that $(p-y)$ in the continuum limit can be replaced by $d_c$.) We note that it was in order to be able to carry out this step that we had to multiply our boundary equation by $d_c^{1/2}\sqrt{a}$. To obtain the third equality sign we have made use of the relation (3.146). So our statement is now the following

$$W_0^{Kont} = \lim_{d.s.l} a \cdot \left[\frac{1}{2}\oint_{\infty}\frac{d\omega}{2\pi i}\frac{V'(\omega)}{p-\omega}\left\{\frac{\sqrt{(p-x)(p-y_c)}}{\sqrt{(\omega-x)(\omega-y_c)}} - \frac{\sqrt{(\omega-x_c)(p-y_c)}}{\sqrt{(p-x_c)(\omega-y_c)}}\right\}\right]. \qquad (3.148)$$

The similarity with equation (3.131) is striking and the equality is straightforward to prove. To do so one expands $(p-\omega)^{-1}$ in powers of $\left(\frac{p-x_c}{\omega-x_c}\right)$ and the quantities $(p-x)$ and $(\omega - x)$ in powers of $\left(\frac{x-x_c}{p-x_c}\right)$ and $\left(\frac{x-x_c}{\omega-x_c}\right)$ respectively. The factor $(p-y_c)$ can simply be replaced by $d_c$. The $\frac{1}{\sqrt{\pi}}$ term of (3.148) vanishes as it should. This is actually ensured by the boundary equation. In the process of expanding the integrand it proves convenient to pull out a factor $(p-x)^{1/2}$. The result of the expansion procedure is the following expression for the right hand side of (3.148)

$$\begin{align}
rhs &= \lim_{d.s.l.} \left\{ \frac{1}{2}(p-x)^{1/2}(p-y_c)^{1/2} \sum_{b=1}^{\infty} \sum_{m=1}^{b-1} c_b \frac{(x-x_c)^b}{(p-x_c)^{b-m+1}} M_m^c(\{g_i\}) \right\} \tag{3.149} \\
&= \frac{1}{2}(\pi - 2u_0)^{1/2} \sum_{b=1}^{\infty} \sum_{m=1}^{b-1} c_b \frac{(2u_0)^b}{\pi^{b-m+1}} (T_m + \delta_{m,1}). \tag{3.150}
\end{align}$$

By rewriting (3.131) using the definition (3.103) of the time variables it is easy to show that the expression (3.150) is exactly $W_0^{Kont}(\pi)$.

Now comparing the expression for $M_p$, (3.123), with the expression for $I_p$, (3.109), bearing in the mind the relation (3.103) we see that $\langle \alpha_1 \ldots \alpha_s | \alpha \rangle_g^{Kont} = \langle \alpha_1 \ldots \alpha_s | \alpha \rangle_g$ and hence that we can carry over the geometrical interpretation of the former to the latter.

### 3.3.6 Comparison with other Approaches

Although it is already clear from equation (3.101) let us take the opportunity to show explicitly that there is no contradiction between our result and the results of references [59, 60, 61, 63]. The strategy applied to the generic hermitian matrix model in sections 3.3.3–3.3.5 allows us to address the symmetrical model as well. However one has to use the complex matrix model as an intermediate step. This is of course due to the fact that the symmetrical hermitian matrix model does not contain a complete set of operators. The complex matrix model does — and in reference [47] it was shown that in the double scaling limit the partition function of the complex matrix model involving matrices of size $N \times N$ equals the partition function of the symmetrical hermitian matrix model involving matrices of size $2N \times 2N$. Namely, it holds that

$$F_g^C = \frac{1}{4^{g-1}} F_g^{(S)} \qquad (d.s.l.) \tag{3.151}$$

where $C$ refers to the complex matrix model. The complex matrix model is in many respects very similar to the generic hermitian matrix model. It has a set of loop equations which can be written in the same form as that of equation (3.15). The appropriate requirement concerning the analyticity structure of its 1-loop correlator is that it has only one square root branch cut $[-\sqrt{z}, \sqrt{z}]$ on the real axis. This corresponds to the eigenvalues of the matrix $(\phi^\dagger \phi)$ being confined to the interval $[0, z]$. With this requirement one can solve the loop equations genus by genus. The solution of course depends on the parameter $z$ which is determined by a boundary condition similar to (3.20). As before the multi-loop correlators can be found by applying a loop insertion operator to the 1-loop correlator and as before expressing the higher genera contributions to the correlators is facilitated by introducing a moment description. The $m$'th multicritical point is reached when the distribution of eigenvalues of $(\phi^\dagger \phi)$ acquires $(m-1)$ extra zeros at the endpoint $z$. To relate the double scaling of the partition function of the complex matrix model to the one of the Kontsevich model one takes the same line of action as for the generic hermitian matrix model. Appropriate continuum time variables are defined by the requirement that the boundary equation of the complex matrix model reproduces the boundary equation of the Kontsevich model when the double scaling limit is taken. The resulting time variables turn out to be related to the moments of the complex matrix model by an equation similar to (3.118). For the correlators the appropriate renormalization prescriptions are very similar to (3.119)–(3.121) and furthermore for the loop insertion operator one has again a relation like (3.124). However, a closer analysis of the loop equations shows that in stead of (3.143) we have

$$F_g^C \xrightarrow{d.s.l.} \frac{1}{4^{g-1}} \left(2 F_g^{Kont}\right), \qquad g \geq 1. \tag{3.152}$$

By comparing (3.152) with (3.151) we immediately see that except for some complications at genus zero we get in the double scaling limit from the partition function of the symmetrical hermitian matrix model the square of the partition function of the Kontsevich model. A similar statement of course holds for the complex matrix model, the only difference being that starting from complex matrices of size $N \times N$ one ends up with the Kontsevich model involving matrices of size $2N \times 2N$.

Let us close this section by briefly discussing the unconventional limiting procedure mentioned in section (3.3.1). This procedure consists in analytically continuing the size of the matrices of the generic hermitian 1-matrix model from $N$ to $-\xi \cdot N = -\frac{1}{\epsilon^3} N$, rescaling matrices and couplings appropriately and sending $\epsilon$ to zero. With this prescription one can turn the path integral defining the generic hermitian 1-matrix model into the path integral defining the Kontsevich model. However, solving the boundary equations (3.19) and (3.20) in the limit $\epsilon \to 0$ assuming a non symmetric support of the eigenvalue distribution also allows us to apply the $\epsilon \to 0$ procedure to the representation (3.38) of the free energy of the generic hermitian 1-matrix model. Following this line of action one finds that only terms of (3.38) which fulfill the requirements (3.72) survive in the limit $\epsilon \to 0$. Hence the $\epsilon \to 0$ procedure leaves us with exactly the terms which are potentially relevant for the double scaling limit! In addition the limiting procedure sends a given of the surviving terms of (3.38) into the term of (3.107) which appears from it when $(M_k + \delta_{k,1})$ is replaced by $I_k$ and $d_c$ removed. We refer to [47] for details. The $\epsilon \to 0$ prescription is an amazing limiting procedure. It picks out exactly the terms which are relevant for the double scaling limit and yet we can not relate $\epsilon$ in any way to the double scaling parameter, $a$. In addition it allows us to access less explored regions of the moduli spaces $\mathcal{M}_{g,s}$ of Riemann surfaces with $g$ handles and $s$ marked points. In the same way as the model arising in the limit $\epsilon \to 0$, i.e. the Kontsevich model, can be viewed as a generating functional for intersection indices on moduli spaces $\mathcal{M}_{g,s}$, the model corresponding to a finite $\epsilon$ can be viewed as a generating functional for intersection indices on discretized versions of moduli spaces, $\epsilon$ being the step of discretization. This gives rise to a geometrical interpretation of the coefficients of those terms in (3.38) which are subdominant in the limit $\epsilon \to 0$ and hence in the double scaling limit. For a discretized moduli space one must take into consideration contributions from the boundary when calculating intersection indices. The boundary of the moduli space $\mathcal{M}_{g,s}$ appears by the so-called Deligne-Mumford compactification [70] and consists of (products of) lower dimensional moduli spaces. While the coefficients of the leading order terms of $F_g$ are related to intersection indices on moduli spaces $\mathcal{M}_{g,s}$, $s \leq 3g - 3$, the coefficients of the subleading terms are related to intersection indices on the boundary of these spaces. For a description of the idea of the discretized moduli space we refer to [71, 72].

## 3.4 A Remark on the Two-Matrix Model

The two-matrix model has turned out to encode a vast amount of information about the interaction of matter fields with 2-dimensional quantum gravity. This property is most clearly exposed by the following two-matrix integral [28]

$$Z[b, \lambda, N] = e^{F[b,\lambda,N]}$$
$$= \int d\phi_+ d\phi_- \exp\left\{-N \operatorname{Tr} \left(\frac{1}{2}\phi_+^2 + \frac{1}{2}\phi_-^2 - b\phi_+\phi_- - \lambda\left(\phi_+^3 + \phi_-^3\right)\right)\right\}. \quad (3.153)$$

By the perturbative expansion of $F[b, \lambda, N]$ one generates the same simplicial manifolds as by the perturbative expansion of $F[\lambda, N]$ appearing in (3.1). However each triangle is now equipped with a label which is either $+$ or $-$, and the weight of a given triangulation is modified by a factor $\frac{1}{1-b^2}$ for each pair of neighbouring triangles with identical labels and a factor $\frac{b}{1-b^2}$ for each pair of neighbouring triangles with different labels. This clearly corresponds to a situation where we have added to the Einstein Hilbert action (2.12) the Ising model action (2.19) with

$$e^{-2\beta} = b, \quad (3.154)$$

the spins being placed in the centres of the triangles. The genus zero contribution to the free energy of the model (3.153), $F_0[b, \lambda]$ can be calculated using the technique of orthogonal polynomials [73, 74]. Analyzing the singularity structure of $F_0[b, \lambda]$ one finds that there exists a point in the coupling constant plane for which the spin system has a third order phase transition and the geometrical system becomes critical in the sense described in section 2.3 [28, 29]. In the vicinity of this point the model describes unitary conformal matter with $c = \frac{1}{2}$ coupled to 2-dimensional quantum gravity.

The generic hermitian two-matrix model takes the following form

$$Z[c, \{g_i\}, \{\hat{g}_i\}, N] = e^{F[c,\{g_i\},\{\hat{g}_i\},N]}$$
$$= \int dX dY \exp\left\{-N \operatorname{Tr} \left(V(X) + \hat{V}(Y) - cXY\right)\right\} \quad (3.155)$$

where

$$V(X) = \sum_{j=1}^{\infty} \frac{g_j}{j} X^j, \qquad \hat{V}(Y) = \sum_{j=1}^{\infty} \frac{\hat{g}_j}{j} Y^j. \quad (3.156)$$

Recently it has been shown that in the coupling constant space of this model one can find critical points describing the interaction of any rational conformal matter field with quantum gravity [34]. For rational $(p, q)$ matter the critical potentials $V_c$ and $\hat{V}_c$ can be taken to be of degree $p$ and $q$ respectively. For unitary conformal matter with $(p, q) = (m, m + 1)$ one also has the possibility of choosing a critical point for which the corresponding potentials are identical and of order $m$. Hence all the unitary models can be viewed as Ising models living on random surfaces.

In the light of this recent development it would be interesting to try to generalize the techniques described in the previous sections to the two-matrix case. As mentioned above the genus zero contribution to the free energy for the model given by (3.153) has been calculated using the method of orthogonal polynomials. The study of loop equations for this model has also been initiated [75, 76, 77] but explicit results for correlators are few and limited to genus zero. We are of the opinion that a more detailed analysis along the lines of the previous sections should indeed be possible and might lead to a better understanding of the structure of two-matrix models both away from and in the continuum. Let us try to describe our present understanding of how such an analysis should be carried out [79]. The starting point would of course be the loop equations of the generic hermitian two-matrix model. To derive these in an appropriate form it is convenient to consider the following redefinition of the field, $X$

$$X \to X + \varepsilon \frac{1}{p - X} \frac{1}{q - Y}. \tag{3.157}$$

Under a transformation of this type the measure changes as

$$dX \to dX \left\{ 1 + \varepsilon \operatorname{Tr}\left(\frac{1}{p - X}\right) \operatorname{Tr}\left(\frac{1}{p - X} \frac{1}{q - Y}\right) \right\} \tag{3.158}$$

and we get to the first order in $\epsilon$

$$\begin{aligned}
0 &= \int dX dY \left\{ \operatorname{Tr}\left(\frac{1}{p - X}\right) \operatorname{Tr}\left(\frac{1}{p - X} \frac{1}{q - Y}\right) - N \operatorname{Tr}\left(V'(X) \frac{1}{p - X} \frac{1}{q - Y}\right) \right. \\
&\quad \left. + cN \operatorname{Tr}\left(Y \frac{1}{p - X} \frac{1}{q - Y}\right) \right\} \exp\left\{ -N \operatorname{Tr}\left(V(X) + \hat{V}(Y) - cXY\right) \right\}. \tag{3.159}
\end{aligned}$$

For the two-matrix model we have of course a new type of correlators, namely correlators involving both $X$ and $Y$ fields. Let us introduce the following definitions

$$G_{pq} = \langle \operatorname{Tr} \frac{1}{p - X} \frac{1}{q - Y} \rangle \tag{3.160}$$

and

$$G^m(p) = \langle \operatorname{Tr} \frac{1}{p - X} Y^m \rangle, \qquad \hat{G}^m(q) = \langle \operatorname{Tr} X^m \frac{1}{q - Y} \rangle. \tag{3.161}$$

Then it obviously holds that

$$G_{pq} = \sum_{n=0}^{\infty} \frac{G^n(p)}{q^{n+1}} = \sum_{m=0}^{\infty} \frac{\hat{G}^m(q)}{p^{m+1}}. \tag{3.162}$$

Furthermore let us define loop insertion operators by

$$\frac{d}{dV(p)} = -\sum_{j=1}^{\infty} \frac{j}{p^{j+1}} \frac{d}{dg_j}, \qquad \frac{d}{d\hat{V}(q)} = -\sum_{j=1}^{\infty} \frac{j}{q^{j+1}} \frac{d}{d\hat{g}_j}. \tag{3.163}$$

Assuming that eigenvalue distributions for $X$ and $Y$ have been introduced we can now write the equation (3.159) as

$$\oint_C \frac{d\omega}{2\pi i} \frac{V'(\omega)}{p-\omega} G_{\omega q} = G^0(p) G_{pq} + c(qG_{pq} - G^0(p)) + \frac{d}{dV(p)} G_{pq} \qquad (3.164)$$

where $C$ is a curve which encircles the supports of the eigenvalue distributions and where we have made use of a factorization condition similar to (3.16). As in equation (3.15) the term $\frac{d}{dV(p)} G_{pq}$ is suppressed by a factor $\frac{1}{N^2}$ when compared to the rest. Let us describe how one can in principle solve equation (3.164) step by step. First we take the coefficient of $\frac{1}{p}$ in the expansion of (3.164) in powers of $\frac{1}{p}$. This gives

$$\sum_{j=1}^{\infty} g_j \hat{G}_{j-1}(q) = c(q \hat{G}^0(q) - 1). \qquad (3.165)$$

Of course (3.164) has a twin equation where the roles of $X$ and $Y$ are interchanged. Expanding this twin equation in powers of $\frac{1}{q}$ and taking the coefficient of $\frac{1}{q^n}$ we find the following recursion relation for $\hat{G}^n(q)$

$$c\hat{G}^n(q) = \oint_C \frac{d\omega}{2\pi i} \frac{\hat{V}(\omega)}{q-\omega} \hat{G}^{n-1}(\omega) - \hat{G}^0(q) \hat{G}^{n-1}(q) - \frac{d}{d\hat{V}(q)} \hat{G}^{n-1}(q). \qquad (3.166)$$

By means of this relation we can express $\hat{G}^n(q)$ entirely in terms of $\hat{G}^0(q)$. Neglecting the subleading term in (3.166) we that see that this equation allows us to obtain from (3.165) an algebraic equation for the genus zero contribution to the 1-loop correlator for the field $Y$. The degree of the algebraic equation is the same as the degree of the potential $V(X)$. This is in full compliance with the situation in the 1-matrix case where the genus zero contribution to the 1-loop correlator was determined by an algebraic equation of the second degree since a 1-matrix model can be considered a two matrix model where one potential is gaussian and has been integrated out. The equation that determines $\hat{G}^0(q)$ contains a number of unknown integration constants. These constants must be determined by imposing appropriate requirements on the analyticity structure of $\hat{G}^0(q)$. A similar situation was encountered in the 1-matrix case. Of course, unless special circumstances occur we can only solve the equation for $\hat{G}^0(q)$ when the degree of the potential $V(X)$ is less than or equal to four. However as long as the degree of $V(X)$ does not exceed four we can probably find and expression analogous to (3.18) giving $\hat{G}^0(q)$ for arbitrary $\hat{V}(Y)$. In case we succeed in arriving at such a formula we have the possibility of relating the various types of critical behaviour of the 2-matrix model to the behaviour of its eigenvalue distributions in analogy with what was done for the 1-matrix model. Furthermore such a formula might provide us with an important clue as to what are the fundamental variables of the two-matrix model. Maybe a set of generalized moments will reveal themselves when the 1-loop correlator is analysed in detail. This was what happened in the 1-matrix case (cf. to equation (3.27)). Finally a formula for $\hat{G}^0(q)$ for generic $\hat{V}(Y)$ might also allow us to develop an iterative procedure for calculating higher genera contributions to $\hat{G}^0(q)$ similar to the one described for the 1-matrix model. We can derive a genus expanded version of the equation for $\hat{G}^0(q)$ combining (3.165) and (3.166). With a generic $\hat{V}(Y)$ we have a loop insertion operator $\frac{d}{d\hat{V}(q)}$ and the equation closes. Probably the iterative solution is again conveniently expressed in terms of the eigenvectors of some linear operator. When we have calculated the genus zero contribution to $\hat{G}^0(q)$ we can in principle determine $G_{pq}$ from the twin equation of equation (3.164). While this is certainly doable when the potential $\hat{V}(Y)$ has a finite number of terms it might be practically impossible for a generic $\hat{V}(Y)$. However, if it is not, an iterative procedure for calculating higher genera contributions to $G_{pq}$ is very likely to exist since the structure of (3.164) is very similar to that of (3.15). From $\hat{G}^0(q)$ and $G_{pq}$ a whole series of correlators can be calculated by application of the loop insertion operator $\frac{d}{d\hat{V}(q)}$ and the operator $\frac{d}{dc}$. The $s$-loop correlators for the $X$ field seem difficult to access, however.

To our opinion one should not be deterred by the seemingly large degree of complexity of the loop equations of the two-matrix model. It might very well happen that a relatively simple structure of correlators and the free energy emerges as it was seen in the case of the 1-matrix model. Studying the loop equations of the two matrix model might lead to new insight concerning rational conformal theories coupled to 2-dimensional quantum gravity. Furthermore the same loop equations have been found to turn up in the Kazakov-Migdal model of induced QCD [78].

## 3.5 Non Perturbative 2D Quantum Gravity ?

### 3.5.1 Non Perturbative Effects in Matrix Models

The wish to understand in detail the the simple model (3.1) has brought us far. We have found the complete perturbative solution of the generic hermitian 1-matrix model (3.6), i.e. we know (at least in principle) all terms in the genus expansion of its free energy as well as of any of its correlators. We have localized and classified all critical points in the coupling constant space of the generic model and devised a prescription, namely the double scaling prescription for how to obtain continuum theories in the vicinity of these points. In the same sense as for the discrete models we have determined the complete perturbative solution of these continuum theories, i.e. we have access to (in principle) all terms in the genus or double scaling expansion of their free energy and their correlators. However, as regards the understanding of 2-dimensional quantum gravity beyond the topological expansion our achievements are less impressive. Most efforts have naturally been concentrated on studying the pure gravity, $c = 0$, case. Let us briefly summarize the results of these efforts. For convenience we will refer to a specific $m = 2$ multi-critical point, namely the one given by

$$g_1^c = \frac{3}{2^{2/3}}, \quad g_3^c = -\frac{1}{3}, \quad g_i^c = 0, \quad \text{for} \quad i > 3, \, i = 2. \qquad (3.167)$$

This critical point can be reached starting from the following matrix model potential

$$V(\phi) = g\phi - \frac{1}{3}\phi^3 \qquad (3.168)$$

which appears from the one in (3.1) by a simple rescaling. Each term in the topological expansion of the model with the potential (3.168) is a power series in $(1/g)^{3/4}$ which is convergent when $g \geq g_c$, divergent otherwise. Let us (for $g > g_c$) set $g - g_c = a^2 \Lambda_R$. Then the double scaling limit is defined as the limit $a \to 0$, $N \to \infty$, $G^{-1} = a^5 N^2$ fixed. The topological expansion in this limit becomes an expansion in the parameter $G \Lambda_R^{-5/2}$. All attempts to perform the sum over topologies for this expansion have failed. The series is not even Borel summable. (We note that the divergencies appear only when we try to sum over topologies and are entirely due to the rapid increase of the entropy of manifolds with genus.) The Borel non summability of the perturbation series for the model given by the potential (3.168) can be understood as a consequence of the action not being bounded from below and hence the original matrix integral being ill defined. (Let us mention that it *is* actually possible to reach a $m = 2$ multi-critical point starting from a potential which is bounded from below i.e. a potential of the type $\sum_{i=1}^{2J} g_i \phi^i$ with $g_{2J} > 0$. We will not enter into a detailed discussion of these more complicated $m = 2$ multi-critical models. Let us just note that having a matrix model coupling constant $g_i > 0$ corresponds to considering space time manifolds where $i$-gons appear with negative weight.) For some time it was the hope that the connection between the double scaling limit of the 1-matrix model and the kdV hierarchy would lead to a non perturbative definition of 2D quantum gravity. For the model given by the potential (3.168) this connection implies that in the double scaling limit its specific heat, $C(\Lambda_R) = -\frac{d^2 F}{d\Lambda_R^2}$ obeys the following differential equation — the Painlevé equation

$$-\frac{G}{6} C'' + C^2 = \Lambda_R. \tag{3.169}$$

However this observation does not lead to very much progress as regards our search for a consistent non perturbative definition of 2D quantum gravity. There is an infinite number of solutions to the Painlevé equation which have the correct asymptotic behaviour, namely $C(\Lambda_R) \to \Lambda_R^{1/2}$ for $\Lambda_R \to \infty$ (cf. to equation (3.5)). The difference, $\delta C$, between two such solutions is invisible in perturbation theory. Its asymptotic behaviour can be found by linearizing (3.169). The result reads

$$\delta C \sim \Lambda_R^{-1/8} \exp\left(-\frac{12\sqrt{3}}{5} G^{-1/2} \Lambda_R^{5/4}\right). \tag{3.170}$$

Bearing in mind that our expansion parameter in the double scaling limit is $G \Lambda_R^{-5/2}$ it is obvious that the topological expansion can not provide us with any information which would allow us to choose between the different solutions. Additional physical constraints are needed. The search for such constraints has so far been unsuccessful. The above mentioned ambiguity is not the only problem, however. All real solutions to the Painlevé equation have an infinite number of poles on the negative real axis. The existence of these poles makes it impossible to define continuum $s$-loop correlators without violating the continuum loop equations [53].

The argument of the exponential in (3.170) can be recognized as (minus) the action of a real instanton present in the original matrix model [53, 80]. The presence of this real instanton explains the non Borel summability of the perturbation series. In the large-$N$ limit the eigenvalues of the matrix $\phi$ are confined to an interval $[y, x]$ on the real axis. The instanton configuration is one where the largest eigenvalue has been moved from the point $x$ to some point $\tilde{x} > x$ while the others remain at their original positions. This change of configuration can be described by the following change in the eigenvalue distribution

$$\triangle(u_0(\lambda)) = \frac{1}{N}\left[\delta(\lambda - \tilde{x}) - \delta(\lambda - x)\right]. \tag{3.171}$$

Under a deformation of $u_0(\lambda)$ of this type the effective action, $S$, of the matrix model, i.e. $-F_0$ changes (to the leading order in $N$) by

$$\begin{aligned}
\triangle S &= N^2 \int d\lambda \triangle(u_0(\lambda)) \times \left(V(\lambda) - 2\int d\mu u_0(\mu) \ln|\lambda - \mu|\right) \\
&= N \int_x^{\tilde{x}} d\lambda \left(V'(\lambda) - 2\int d\mu u_0(\mu)\frac{1}{\lambda - \mu}\right) \\
&= N \int_x^{\tilde{x}} d\lambda \left(-2W_0(\lambda) + V'(\lambda)\right) \\
&= -2N \int_x^{\tilde{x}} d\lambda W_0^{scal}(\lambda). \tag{3.172}
\end{aligned}$$

To arrive at the last equality sign we have once again made use of reference [14]. We note that $W_0^{scal}(\lambda)$ is exactly the scaling part of $W_0(\lambda)$. For the potential (3.168) we immediately find using the relation (3.27)

$$W_0^{scal}(\lambda) = -\frac{1}{2}(\lambda - x)^{1/2}(\lambda - y)^{1/2}(\lambda + \frac{1}{2}(x + y)) \tag{3.173}$$

where $x$ and $y$ are given by (3.19) and (3.20). The instanton configuration referred to above is characterized by $0 = \frac{d(\triangle S)}{d\tilde{x}} \propto W_0^{scal}(\tilde{x})$ and corresponds to the situation

$$\tilde{x} = -\frac{1}{2}(x + y). \tag{3.174}$$

It is easily verified that $\tilde{x} = -\frac{1}{2}(x + y) > x$, that $\tilde{x}$ is a maximum for $\triangle S$ and that $\triangle S \to -\infty$ for $\tilde{x} \to \infty$. Hence there is a barrier of height $\triangle S(\tilde{x} = -\frac{1}{2}(x + y))$ which prevents the eigenvalues from escaping to infinity. The probability that the largest eigenvalue tunnels through this barrier is given by $\exp(-\triangle S(\tilde{x} = -\frac{1}{2}(x + y)))$. In the large-$N$ limit this probability is exponentially suppressed and so is the contribution to the free energy from the instanton configuration. However, in the double scaling limit the situation is different. In this limit one finds using the boundary equations (3.19) and (3.20)

$$\begin{aligned}
x - x_c &= -2x_c \left(\frac{g - g_c}{g_c}\right)^{1/2} \tag{3.175} \\
y - y_c &= O(x - x_c)^2 \tag{3.176}
\end{aligned}$$

where
$$x_c = \frac{1}{2^{1/3}}, \qquad y_c = -3x_c. \tag{3.177}$$

Performing the following change of variable
$$\lambda - x_c = x_c \left(\frac{g - g_c}{g_c}\right)^{1/2} z \tag{3.178}$$

and inserting (3.175), (3.176) and (3.177) in (3.172) leads to

$$\begin{aligned}
(\triangle S)_{d.s.l.} &= N \left(\frac{g - g_c}{g_c}\right)^{5/4} \int_{-2}^{1} dz \sqrt{z + 2}\,(z - 1) \\
&= G^{-1/2} \Lambda_R^{5/4} \left(\frac{12\sqrt{3}}{5}\right).
\end{aligned} \tag{3.179}$$

We see that in the double scaling limit the action *is* influenced by the presence of the instanton and that as advertised the amount by which the action is modified is exactly equal to minus the argument of the exponential entering the non perturbative ambiguity of the Painlevé equation. As noted earlier the presence of the real instanton is what obstructs the Borel summation of the perturbation series.

Analyses of the properties of the perturbation series have also been carried out for matrix models describing gravity coupled to various matter fields [81]. For the $m$'th multi-critical models corresponding to gravity coupled to conformal matter with $(p, q) = (2, 2m - 1)$ the perturbation series has been found to be Borel summable if $m$ is odd and not Borel summable if $m$ is even. The non Borel summability when $m$ is even can in all cases be explained by the presence of a real instanton. The corresponding instanton configuration is of the same type as described for $m = 2$ above. In the case of the $m$ odd models there are no real instantons. For a discussion of the general $(p, q)$ case we refer to [81]. Let us just mention that for all the unitary models the perturbation series is not Borel summable.

In the case of pure gravity there have been several attempts to overcome the above mentioned problems by modifying the matrix model construction, the idea being that the modified theory should give a non perturbative definition of 2D quantum gravity while containing the same perturbative information as the original theory. Since the problems with the model given by the potential (3.168) can be traced back to the fact that the corresponding matrix integral is ill defined one way to proceed is to try to modify the theory directly on path integral level. In references [53, 80, 82] it was pointed out that one can obtain a well defined matrix integral with the same perturbative expansion as the original one with the potential (3.168) by choosing as the contour for the integration over eigenvalues in stead of the real axis a path which goes from $-\infty$ to $e^{i\frac{\pi}{3}}\infty$. This procedure of course also ensures that all correlators are well defined but they now have a non vanishing imaginary part. This holds in particular for the specific heat of the model which can be shown to correspond to a complex solution of the Painlevé equation. This solution is unique and has no poles on the real axis. However, due to the appearance of imaginary parts in the partition function and the correlators the contour deformation strategy does not lead to an immediately physically acceptable theory of quantum gravity.

Another strategy for constructing a non perturbative extension of the original model has been suggested by Dalley et al. [83, 84]. It consists in replacing the Painlevé equation by a differential equation of order four. This differential equation can be derived along the lines of the original derivation of the Painlevé equation for the model given by (3.6) and (3.168) using as the starting point a complex matrix model with a potential, $V(\phi^\dagger \phi)$, of the third degree in $(\phi^\dagger \phi)$. For the eigenvalues $\lambda_i^2$ of $(\phi^\dagger \phi)$ there is a lower cutoff, namely $\lambda_i^2 \geq 0$. The critical point of interest is characterized by the eigenvalues of $(\phi^\dagger \phi)$ being confined to the interval $[0, z]$ and the eigenvalue distribution acquiring two zeros at the endpoint 0, i.e. at the cutoff, in addition to its usual square root singularity at the endpoint $z$. This is a critical point of which no analogue exists for the hermitian matrix model since for hermitian matrices there is no lower cutoff on the eigenvalues. The above mentioned fourth order differential equation can also be derived without referring to a specific model but using the properties of the kdV flow equations [84]. This differential equation has solutions which have the correct asymptotic expansion for $\Lambda_R \to \infty$. Numerical investigations indicate that there is only one *real* solution with this property and that this solution does not have any poles on the real axis.

In the following we will describe yet another proposal for defining 2D quantum gravity non perturbatively which we will denote stochastic stabilization. It is based on the prescription of Greensite and Halpern for stabilizing bottomless action theories [85]. This prescription results in the original ill defined $d$-dimensional quantum field theory being replaced by a well defined $(d+1)$ dimensional quantum field theory which contains the same perturbative information as the original one. In the case of the zero-dimensional matrix model given by (3.6) and (3.168) the corresponding 1-dimensional theory corresponds to a simple quantum mechanical system. Stochastic stabilization in the context of matrix models has been studied by various groups [86]–[98]. For the stabilized model the partition function as well as all correlators are automatically real. Hence the non perturbative definition of 2D quantum gravity provided by stochastic stabilization differs from the one provided by the contour deformation method. A detailed comparison of correlators obtained by the two methods was performed in reference [90]. In reference [91] stochastically stabilized 2D gravity was compared to the non perturbative definition of 2D gravity advocated by Dalley et. al. and it was found that these two approaches do not agree either. In both approaches the original geometrical picture is lost. With stochastic stabilization we loose in addition the kdV flow structure. However, stochastic stabilization gives rise to a simple quantum mechanical interpretation of several matrix model characteristica. We will mention some examples in the next section. In particular we will reinterpret the non perturbative ambiguity of the Painlevé equation. Stochastic stabilization in addition seems to present to us the possibility of a strong coupling expansion of 2D quantum gravity. This will be the subject of section 3.5.5.

### 3.5.2 Stochastic Stabilization

Stochastic stabilization is a prescription which assigns to a $d$-dimensional bottomless action theory a well defined $(d+1)$-dimensional quantum field theory with the same perturbative information as the original ill defined theory. It relies on the observation that the expectation value of an operator $Q$ in a $d$-dimensional field theory with an action, $S[\phi]$, bounded from below can be viewed as the expectation value of the same operator in the ground state of a $(d+1)$-dimensional quantum field theory.

$$<Q> = \frac{1}{Z} \int d\phi \; e^{-S[\phi]/\hbar} Q[\phi] = \int d\phi \; \Psi_0^2[\phi] \; Q[\phi]. \qquad (3.180)$$

The $(d+1)$-dimensional theory is described by the Fokker-Planck Hamiltonian

$$H_{FP} = \int d^d x \left[ -\frac{\delta^2}{\delta \phi^2} + \frac{1}{4\hbar^2} \left( \frac{\delta S}{\delta \phi} \right)^2 - \frac{1}{2\hbar} \frac{\delta^2 S}{\delta \phi^2} \right] \qquad (3.181)$$

which is a positive semi-definite operator

$$H_{FP} = \int d^d x \; R^\dagger(x) R(x), \qquad R = i \frac{\delta}{\delta \phi} + \frac{i}{2\hbar} \frac{\delta S}{\delta \phi}. \qquad (3.182)$$

It is easily verified that

$$\Psi_0[\phi] = \frac{e^{-S[\phi]/2\hbar}}{\sqrt{Z}} \qquad (3.183)$$

is a normalizable eigenvector of $H_{FP}$ corresponding to the eigenvalue $E_0 = 0$ and hence qualifies as the ground state of the $(d+1)$-dimensional theory. However, if the action $S[\phi]$ is not bounded from below the wave functional given by (3.183) is not normalizable and does not qualify as the ground state of $H_{FP}$. The Fokker-Planck Hamiltonian is nevertheless still a positive semi-definite operator. Hence it also in this case has a well defined ground state, $\Phi_0(\phi)$, now corresponding to a vacuum energy $E_0 > 0$. This ground state, $\Phi_0(\phi)$, allows us to define a stabilized version of the original ill defined theory, namely the theory with action $S_{eff}$ given by

$$\Phi_0[\phi] = \frac{e^{-S_{eff}[\phi]/2\hbar}}{\sqrt{Z_{eff}}}. \qquad (3.184)$$

For the stabilized theory expectation values are defined by

$$\langle Q \rangle = \frac{1}{Z_{eff}} \int d\phi \; Q[\phi] \; e^{-S_{eff}[\phi]/\hbar} = \langle \Phi_0 | Q | \Phi_0 \rangle. \qquad (3.185)$$

It can be shown that the stabilized theory has the same classical limit, the same perturbative expansion in coupling constants and the same $\frac{1}{N}$-expansion as the original theory. In addition the action of the stabilized theory is of course bounded from below. Studying the stabilized theory amounts to studying the $(d+1)$-dimensional theory described by the Hamiltonian $H_{FP}$.

It is obvious that stochastic stabilization offers to us the possibility of a non perturbative definition of 2D quantum gravity. One can easily convince oneself that for the zero-dimensional matrix model defined by (3.6) $H_{FP}$ takes the following form

$$H_{FP} = N(R^\dagger R), \qquad R_{ab} = i\left(\frac{1}{N}\frac{\delta}{\delta\phi_{ab}} + \frac{1}{2}\frac{\delta V}{\delta\phi_{ab}}\right). \qquad (3.186)$$

Since our vacuum must be $U(N)$ invariant and since we will only be interested in $U(N)$ invariant observables we can write (3.185) as

$$\langle Q \rangle = \int \prod_{k=1}^{N} d\lambda_k \prod_{i<j}(\lambda_i - \lambda_j)^2 \, \Phi_0^2(\{\lambda_k\}) Q(\{\lambda_k\}) \qquad (3.187)$$

where $\{\lambda_i\}$ are the eigenvalues of the matrix $\phi$. Likewise we can diagonalize the operator $H_{FP}$. In this picture the wavefunction we are interested in, namely

$$\tilde{\Phi}_0(\{\lambda_k\}) \equiv \prod_{i<j}(\lambda_i - \lambda_j)\,\Phi_0(\{\lambda_k\}) \qquad (3.188)$$

is totally antisymmetric. Hence the 1-dimensional well defined theory which by the stochastic stabilization scheme gets associated with our original zero-dimensional matrix model in fact describes a Fermi gas consisting of $N$ particles. Let us now specialize to the matrix model with the potential (3.168). In this case the Hamiltonian of the fermionic system reads

$$H_{FP} = N\sum_{i=1}^{N} H_{fp}[\lambda_i], \qquad (3.189)$$

$$H_{fp}[\lambda] = -\frac{1}{N^2}\frac{d^2}{d\lambda^2} + V_{fp}(\lambda), \qquad (3.190)$$

$$V_{fp}(\lambda) = \frac{1}{4}(g-\lambda^2)^2 + \lambda. \qquad (3.191)$$

The situation is simple. The fermions do not interact. The ground state of the system is hence the Slater determinant of the $N$ lowest eigenstates of the single particle Hamiltonian $H_{fp}$ and the vacuum energy the sum of the corresponding eigenvalues times $N$. For potentials of degree larger than three the operator $\frac{\delta^2 V}{\delta\phi_{ab}\delta\phi_{ba}}$ gives rise to interactions between eigenvalues and the situation is more complicated.

The model given by (3.189) – (3.191) allows us to attribute a meaning to the partition function and the correlators of pure 2D gravity for any value of the coupling constant $g$ and without making any restrictions on topology. Hence we have a truly non perturbative definition of 2D quantum gravity. We note that the stochastic stabilization scheme always results in real answers for correlators and the partition function. No exact solution is known for the quantum mechanical problem given by $H_{FP}$. In reference [89] a recipe for numerical calculations of correlators of 2D quantum gravity based on a numerical solution of the quantum mechanical problem was given and some explicit results obtained. However, there is also quite a lot which can be learnt by studying the model given by (3.189)– (3.191) by semi-classical techniques. In particular many of the characteristica of the underlying matrix model can be reproduced and given a simple quantum mechanical interpretation. We will present some examples in the following sections.

### 3.5.3 The Large-$N$ limit

Let us consider the large-$N$ limit of the fermionic system given by (3.189)–(3.191). From (3.190) it appears that $\frac{1}{N^2}$ plays the role of $\hbar^2$. Hence we see that the genus expansion of our original matrix model is nothing but the WKB expansion of the quantum mechanical system defined by (3.189)–(3.191). In reference [98] the effect of applying the WKB approximation to the quantum mechanical system was analysed in some detail. Here we will just mention the main results. Since we are dealing with a system consisting of $N$ fermions let us denote the energy of the $N$'th level of $H_{fp}$ by $E_F$. Furthermore let us assume that for $E \leq E_F$ there is only one set of turning points, $(y(E), x(E))$, for a classical particle with energy $E$ moving in the potential $V_{fp}$. In the WKB approximation (with $\hbar = \frac{1}{N}$) $E_F$ is determined by

$$1 = \frac{1}{\pi} \int_{y(E_F)}^{x(E_F)} d\lambda \sqrt{E_F - V_{fp}(\lambda)}. \tag{3.192}$$

Now going back to the expression (3.29) one finds that for the matrix model with the potential (3.168) the spherical level eigenvalue distribution is given by

$$u_0(\lambda) = \frac{1}{\pi} \sqrt{c(x,y) - V_{fp}(\lambda)}, \qquad y \leq \lambda \leq x, \quad g > g_c \tag{3.193}$$

where $c(x,y)$ is equal to $\langle \frac{1}{N} \operatorname{Tr} \phi \rangle$ and can by expressed in terms of $x$ and $y$ as

$$c(x,y) = \langle \frac{1}{N} \operatorname{Tr} \phi \rangle = \left(\frac{1}{4}(x-y)\right)^4 + \frac{1}{2}\left(\frac{1}{4}(x^2 - y^2)\right)^2. \tag{3.194}$$

Bearing in mind that $u_0(x) = u_0(y) = 0$ and $u_0(\lambda) > 0$ for $\lambda \in ]y, x[$ it follows from the normalization condition (3.17) that for $g > g_c$, $E_F$ can be identified with the constant $c(x,y)$. This implies that

$$V_{fp}(\lambda) - E_F = \left(W_0^{scal}(\lambda)\right)^2, \qquad g > g_c, \quad \lambda > x \tag{3.195}$$

where we have made use of the relation between the eigenvalue distribution and the 1-loop-correlator (cf. to equation (3.27) and (3.29)). From the expression (3.173) for $W_0^{scal}(\lambda)$ we see that for $g > g_c$ we have the situation depicted in figure 2a. As already mentioned several times the critical point of the matrix model is characterized by the eigenvalue distribution $u_0(\lambda)$ acquiring an extra zero at the endpoint $x$. This means that as we let $g$ approach $g_c$ from above the point $-\frac{1}{2}(x+y)$ will approach the point $x$ until at $g = g_c$ they merge into a single point $x_c$ (cf. to (3.173)). We see that in the quantum mechanical picture the merging of the two points corresponds to the local minimum of the potential disappearing and the potential acquiring a horizontal tangent at $x = x_c$.

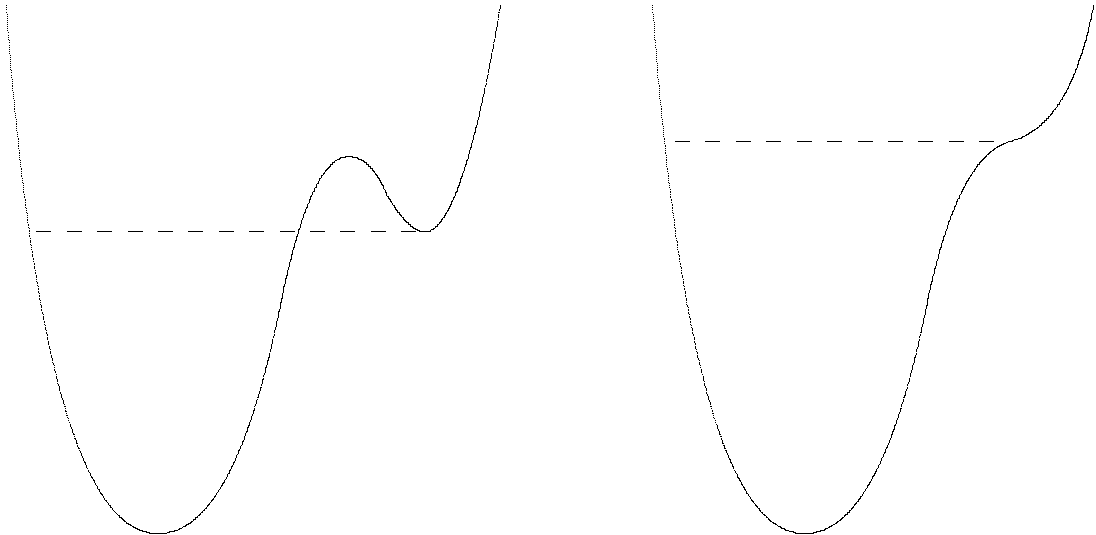

Figure 2. The figure to the left (fig.2a) represents the situation $g > g_c$. In the other figure (fig.2b) $g < g_c$. The dashed lines represent the Fermi energy in the two cases.

For $g < g_c$ we have the situation depicted in figure 2b. The potential now only has one stationary point, namely the point corresponding to the global minimum. It is evident that, as pointed out in the previous section, the quantum mechanical system still has a well defined ground state. As opposed to this our original matrix model makes no sense for $g < g_c$. Even the planar approximation breaks down. In the quantum mechanical picture this breakdown is signalled by the Fokker-Planck potential acquiring a horizontal tangent at $E = E_F$, a situation which conflicts with the standard WKB assumption that the potential can be approximated with a linear function $f(\lambda) = a\lambda + b$ with $a \neq 0$ in the vicinity of the turning points. However, the quantum mechanical system is of course perfectly well defined independently of the validity of the WKB approximation. Let us now try to translate the instanton considerations of section 3.5.1 to the quantum mechanical picture. These considerations of course refer to the situation where $g > g_c$. By comparison of (3.172) and (3.195) we see that

$$\triangle S\left(\tilde{x} = -\frac{1}{2}(x+y)\right) = \frac{2}{\hbar} \int_x^{-\frac{1}{2}(x+y)} \sqrt{V_{fp}(\lambda) - E_F}\, d\lambda, \qquad g > g_c \qquad (3.196)$$

and hence the matrix model probability that one eigenvalues tunnels through the barrier provided by the effective potential and escapes to infinity looks like the quantum mechanical probability that the fermion with energy $E_F$ tunnels from the large potential well to the smaller one. Strictly speaking there is of course no probability for tunneling when the Fermi energy exactly coincides with the value of the potential at the local minimum. But this situation only holds when $N$ is strictly infinite. For $N$ not infinite we should replace the left hand side of (3.192) by $1 + \frac{1}{2}\frac{1}{N}$. It is easy to show that this prescription causes the Fermi energy corresponding to the large well to be lifted by an amount $\triangle E_F \sim O\left(\frac{1}{N}\right)$ which makes it agree exactly with the lowest level in the small potential well. Furthermore it appears that the level spacing is the same in the two wells. We thus have a situation which is remarkably similar to that of a symmetric double well. As for this case the degenerate energy levels will split into two and the lowest of these will have to be identified with the Fermi level. This corrected Fermi level, $E_F^{cor}$, lies above the bottom of the small potential will and hence the fermion with energy $E_F^{cor}$ can tunnel between the two wells. The probability for this process is $e^{-\Gamma}$ where $\Gamma$ is given to leading order in $\frac{1}{N}$ by (3.196). We see that in accordance with the stabilizing effect of the stochastic scheme there is no possibility for the fermion to escape to infinity. Furthermore we note that as was the case in the matrix model picture the tunneling is exponentially suppressed in the large-$N$ limit. Likewise the correction to the Fermi energy $E_F^{cor} - E_F$ is exponentially small in $N$.

So far we have restricted ourselves to the simple matrix model given by the potential (3.168). As mentioned earlier for matrix models with potentials of degree larger than three the quantum mechanical system which arises by the stabilization procedure consists of $N$ *interacting* fermions. However, in the limit $N \to \infty$ these interacting systems are in fact equivalent to certain non interacting systems. A Hartree-Fock like approximation consistent with the large-$N$ limit can be used to decouple the fermions [94]. With this Hartree-Fock like approximation the Fokker-Planck Hamiltonian takes the same form as in (3.189) and (3.190). Of course the precise form of the Fokker-Planck potential $V_{fp}$ depends on the original matrix model potential. For a matrix model potential of degree $k$ the Fokker-Planck potential will have degree $k - 2$. It turns out that with the Hartree-Fock approximation the relations (3.193) and (3.195) still hold (in the region of the coupling constant space where the large-$N$ limit of the original matrix model is well defined). Hence we get a quantum mechanical description of all the matrix model $m$'th multi-critical points. We see that in the quantum mechanical picture the $m$'th multi-critical point is characterized by the $(m - 1)$ first derivatives of the Fokker-Planck potential vanishing at (at least) one of the turning points corresponding to the Fermi energy. We furthermore get a quantum mechanical description of the criterion for stability of the original 1-matrix model. As mentioned in section 3.5.1 a $m$'th multi-critical model is unstable if we when we approach the critical point encounter a real instanton, i.e. a real zero for $W_0^{scal}(\lambda)$ outside the support of the eigenvalue distribution. This situation in the quantum mechanical picture corresponds to the Fokker-Planck potential having a local minimum touching the Fermi level (for $N = \infty$).

### 3.5.4 The Double Scaling Limit

Let us consider the situation where $g > g_c$ (fig. 2a) and let us inspired by the calculations on page 46–47 introduce a scaled variable $z$ by

$$\lambda - x_c = x_c \left(\frac{g - g_c}{g_c}\right)^{1/2} z. \qquad (3.197)$$

Expressed in terms of $z$ the Fokker-Planck Hamiltonian (3.190) reads

$$H_{fp} = V_{fp}(0; g - g_c) + x_c^4 \frac{(g - g_c)^{3/2}}{g_c^{3/2}} h_{fp}(z), \tag{3.198}$$

$$h_{fp}(z) = -\hbar^2 \frac{d^2}{dz^2} + v_{fp}(z; \sqrt{g - g_c}) \tag{3.199}$$

where $V_{fp}(0; g - g_c)$ is a second order polynomial in $g - g_c$ and

$$v_{fp}(z; \sqrt{g - g_c}) = -3z + z^3 + \frac{\sqrt{g - g_c}}{\sqrt{g_c}} \left[ -\frac{3}{2} z^2 + \frac{1}{4} z^4 \right], \tag{3.200}$$

$$\hbar^2 = \frac{4 g_c^{5/2}}{N^2 (g - g_c)^{5/2}}. \tag{3.201}$$

We see that our problem of studying the Fokker-Planck Hamiltonian $H_{FP}$ given by (3.189)–(3.191) has now been reduced to studying the scaled Hamiltonian $h_{fp}(z)$. For $h_{fp}(z)$ the double scaling parameter plays the role of $\hbar^2$ i.e. the expansion parameter of the WKB approximation. Hence the double scaling expansion of our original matrix model is nothing but the WKB expansion of the quantum mechanical system given by $h_{fp}(z)$. We note that $v_{fp}(z; \sqrt{g - g_c})$ contains terms which are subdominant in the double scaling limit. This is not unexpected since we know that a lot of terms which appear in the full expression for the free energy and the correlators of the matrix model vanish in the double scaling limit (cf. to section 3.1.5 and 3.2.3). Likewise it comes as no surprise that we have in $H_{fp}$ a trivial analytic term and that a factor of $\left(\frac{g-g_c}{g_c}\right)^{3/2}$ should be extracted from the non analytic term in order to get a non trivial result in the double scaling limit. It follows from the conventional matrix model analysis bearing in mind that the Fermi energy of $H_{fp}$ equals $\langle \frac{1}{N} \text{Tr } \phi \rangle$ (cf. to page 51). The potential $v_{fp}(z; \sqrt{g - g_c})$ is of course of the same structure as $V_{fp}$. It has a global minimum as well as a local minimum. In the double scaling limit $g \to g_c$, $N \to \infty$, $\hbar$ fixed the local minimum coincides with the local minimum of the potential

$$v_{ds}(z) = -3z + z^3. \tag{3.202}$$

The local minimum of $v_{ds}(z)$ is located at $z = 1$ and $v_{ds}(z = 1) = -2$. In the WKB approximation the Fermi energy of the system given by $h_{fp}(z)$ is determined by

$$\hbar N = \int_{z_l(e_f)}^{z_r(e_f)} \sqrt{e_f - v_{fp}(z)} dz \tag{3.203}$$

where as usual $z_l(e_f)$ and $z_r(e_f)$ are the left and right turning points for a classical particle moving in the potential $v_{fp}(z)$ and where we have assumed that $N \gg 1$. The Fermi energy $e_f$ is related to $E_F$ by

$$E_F = V_{fp}(0; g - g_c) + x_c^4 \left(\frac{g - g_c}{g_c}\right)^{3/2} e_f. \tag{3.204}$$

Since the structure of the potential for the quantum mechanical system is unchanged by the scaling procedure it is obvious that $e_f$ coincides with the value of $v_{fp}$ at its local minimum. It can also be verified by direct computation. In particular one finds that in the limit $N \to \infty$, $g \to g_c$, $\hbar$ fixed, $e_f = -2$, i.e. the fermi energy equals the value of $v_{ds}$ at its local minimum. We note that $v_{ds}(z)$ also takes the value $-2$ for $z = -2$. We can carry over the symmetric double well analogy from the previous section to the scaled Hamiltonian as well. As before we should to be precise replace the $N$ entering equation (3.203) by $(N + \frac{1}{2})$ and as before this causes the Fermi energy corresponding to the large well of $v_{fp}$ to be lifted by an amount which makes it agree exactly with the lowest level in the small well. Furthermore one finds that the level spacing is the same in the two wells. Again the degenerate energy levels will split into two and the actual Fermi energy $e_f^{cor}$ will lie above the bottom of the small potential well. The probability that the fermion with energy $e_f^{cor}$ tunnels between the two wells is now given by $e^{-\Gamma_{scal}}$ where

$$\Gamma_{scal} = \frac{2}{\hbar} \int_{z_r(e_f^{cor})}^{z_l(e_f^{cor})} dz \sqrt{v_{fp}(z) - e_f}. \qquad (3.205)$$

Here $z_r(e_f^{cor})$ is the right turning point in the large well for a classical particle with energy $e_f^{cor}$ moving in the potential $v_{fp}$ and $z_l(e_f^{cor})$ the corresponding left turning point in the small well. Since the scaling prescription for $\lambda$, (3.197), is nothing but the usual double scaling prescription (cf. to equation (3.178)) the $\Gamma_{scal}$ appearing in equation (3.205) is nothing but the double scaled version of the $\Gamma$ appearing in section 3.5.3, i.e. the double scaled version of the instanton action, $\triangle S$

$$\Gamma_{scal} = (\triangle S)_{d.s.l.} \qquad (3.206)$$

We hence have arrived at a quantum mechanical interpretation of the non perturbative ambiguity of the Painlevé equation. In the matrix model picture we saw that the probability that the largest eigenvalue would tunnel through the barrier provided by the effective potential was suppressed in the large-$N$ limit but became of importance in the double scaling limit. The situation is very similar in the quantum mechanical picture. The probability that the most energetic fermion tunnels between the two potential wells of the Fokker-Planck potential is suppressed in the large-$N$ limit but becomes of importance in the double scaling limit. Needles to say that the tunneling probability is a non perturbative quantity.

Let us return for a moment to the expression (3.199) for the Hamiltonian of the quantum mechanical system. If we take the double scaling limit on operator level we are left with the following Hamiltonian

$$h_{ds} = -\hbar^2 \frac{d^2}{dz^2} + z^3 - 3z. \qquad (3.207)$$

It is thus tempting to state that 2D quantum gravity in the double scaling limit is equivalent to a fermionic system with Hamiltonian $h_{ds}$. However, some clarification is needed at this point. The operator $h_{ds}$ is not automatically self-adjoint. To ensure self-adjointness its domain must be carefully specified. As explained in references [97, 98] one can easily by means of the WKB approximation convince oneself that there exists a one parameter family of self-adjoint versions of $h_{ds}$ all of which have discrete spectra. Actually it can be proven (also without referring to the WKB approximation) that this is the situation for a large class of Hamiltonians with potentials unbounded from below, namely Hamiltonians, $h$, of the type

$$h = -\frac{d^2}{dy^2} + v(y) \qquad (3.208)$$

where $v(y) \to -\infty$ for $y \to -\infty$ so fast that

$$\int_{-\infty}^{0} \frac{dy}{\sqrt{|v(y)|}} < \infty \qquad (3.209)$$

and where $v(y)$ behaves in the same way when $y \to \infty$ or $v(y) \to \infty$ for $y \to \infty$. We refer to the mathematical literature, for instance [99] for a rigorous discussion of these points. Let us just note that the condition (3.209) implies that a classical particle, if it is not trapped in a local minimum of $v(y)$ will move to $-\infty$ or possibly $+\infty$ in a finite time. Furthermore let us note that the condition (3.209) hints to us the quantization of energies since in regions where the WKB approximation is valid we have for the density of states

$$\rho(e) = \int \frac{1}{2\pi\sqrt{e - v(y)}} \theta\left(e - v(y)\right) dy. \qquad (3.210)$$

As shown in reference [98] in the energy region below but close to the Fermi level of the full $v_{fp}$ the level spacing is the same for all the different self-adjoint extensions of $h_{ds}$. However, any two extensions will in this region differ by an overall displacement of the energy levels. This allows us to advocate exactly one of the self-adjoint extensions of $h_{ds}$ as the one relevant for 2D quantum gravity, namely the one having an energy level coinciding with the Fermi energy of the full Fokker-Planck potential, $v_{fp}$. Only with this Hamiltonian we will reproduce correctly the non perturbative effects of our original model.

Even though the quantum mechanical system given by (3.207) is hence perfectly well defined the calculations we were originally aiming at being able to carry out, namely calculations of expectation values of operators like $\langle \frac{1}{N} \text{Tr } \phi^n \rangle$ can not be performed without referring to the full Fokker-Planck potential. This is of course closely related to the fact that for these expectation values a multiplicative renormalization by some power of $(g - g_c)^{-1/2}$ was needed before a double scaling expansion could be extracted (cf. to section 3.2.2). With the Hamiltonian (3.207) we can only access quantities which can be expressed entirely in terms of the double scaling parameter. These aspects are discussed in more detail in reference [98] where it is also shown how one can easily extract the correct scaling behaviour of observables from $h_{ds}$ just by introducing an appropriate cutoff in the negative $z$.

### 3.5.5 A Strong Coupling Expansion ?

The considerations outlined in the previous section seem to hint to us the possibility of a convergent strong coupling expansion of 2D quantum gravity. To explain why let us recall a well understood problem from quantum mechanics, namely the anharmonic oscillator

$$H = -\frac{d^2}{dx^2} + x^2 + gx^4, \qquad g > 0. \tag{3.211}$$

For the operator $H$ the perturbative expansion in powers of $g$ is not a convergent power series expansion but only an asymptotic expansion. The term $gx^4$ is *not* a small perturbation for any $g > 0$ and clearly $g \to -g$ changes drastically the nature of $H$. However, the anharmonic oscillator has a convergent strong coupling expansion. This can be seen by a simple scaling argument. Setting

$$y = g^{1/6} x \tag{3.212}$$

we find

$$H(x) = \lambda^{-1/2} \tilde{H}(y) \tag{3.213}$$

where

$$\tilde{H}(y) = -\frac{d^2}{dy^2} + y^4 + \lambda y^2, \quad \lambda = \frac{1}{g^{2/3}}. \tag{3.214}$$

There exists a theorem stating that a term $v(y)$ is an analytic perturbation of an operator $H_0(y) = -\frac{d^2}{dy^2} + v_0(y)$ if

1. $D(v) \supseteq D(H_0)$

2. $||v\psi||_2 \leq a||H_0\psi||_2 + b||\psi||_2$ for some $a, b \in R$ and all $\psi \in D(H_0)$.

The potential $v(y) = y^2$ clearly satisfies the criteria above for being an analytic perturbation of $H_0(y) = -\frac{d^2}{dy^2} + y^4$. For an eigenvalue $E_n$ of the original operator $H$ we can therefore write

$$E_n(g(\lambda)) = \lambda^{-1/2} \sum_{k=0}^{\infty} c_{nk} \lambda^k \tag{3.215}$$

where the power series has a finite radius of convergence — and similarly for the eigenfunctions. The situation is very similar in the case of the Hamiltonian $h_{fp}(z)$. The expansion in powers of $\hbar^2$ is only an asymptotic expansion. However, we can by means of a simple scaling prescription transform the Hamiltonian $h_{fp}(z)$ into one for which the natural expansion parameter involves negative powers of $\hbar$. If we introduce scaled variables by

$$z = \hbar^{2/5} y, \qquad \lambda = \frac{1}{\hbar^{4/5}} = \frac{g - g_c}{g_c} \left(\frac{N}{2}\right)^{4/5} \tag{3.216}$$

we can write

$$h_{fp}(z) = \lambda^{-3/2} \tilde{h}_{fp}(y), \tag{3.217}$$

$$\tilde{h}_{fp}(y) = -\frac{d^2}{dy^2} + \tilde{v}_{fp}(y,\lambda,N) \tag{3.218}$$

where

$$\begin{aligned}\tilde{v}_{fp}(y,\lambda,N) &= y^3 - 3\lambda y + \left(\frac{2}{N}\right)^{2/5}\left[-\frac{3\lambda}{2}y^2 + \frac{1}{4}y^4\right] \\ &= \tilde{v}_{ds}(y,\lambda) + \left(\frac{2}{N}\right)^{2/5}\left[-\frac{3\lambda}{2}y^2 + \frac{1}{4}y^4\right].\end{aligned} \tag{3.219}$$

and the double scaling limit is obtained as before by letting $N \to \infty$, $g \to g_c$, but $\lambda$ fixed. The non-trivial information is contained in the Hamilton function $\tilde{h}_{fp}(y)$ which in the double scaling limit goes to

$$\tilde{h}_{ds}(y) = -\frac{d^2}{dy^2} + \tilde{v}_{ds} = -\frac{d^2}{dy^2} + y^3 - 3\lambda y \tag{3.220}$$

and we have formally the same situation as for the anharmonic oscillator: $-3\lambda y$ looks like a small perturbation with respect to $y^3$. It is thus natural to expect a *strong coupling* expansion of the energy eigenvalues of $h_{ds}(z)$ of the form

$$e_n(\hbar) = \lambda^{-3/2}\sum_{k=0}^{\infty} c_{nk}\lambda^k = \hbar^{6/5}\sum_{k=0}^{\infty} c_{nk}\hbar^{-4k/5} \tag{3.221}$$

where the series is convergent — and similarly for the eigenfunctions. The domain of the perturbation $v(y) = -3\lambda y$ does not include the domain of $h_0(y) = -d^2/dy^2 + y^3$ so we have no rigorous proof of this conjecture, but one should keep in mind that the requirements (1) and (2) mentioned above are only sufficient conditions, not necessary conditions, for analyticity.

# 4 D>2

## 4.1 Analytical Approaches

What caused the progress in our understanding of 2-dimensional quantum gravity was the discovery that the matrix model (3.1) exactly generates 2D orientable simplicial manifolds weighted with the Einstein Hilbert action. It is natural to look for a similar mapping of the partition function of 3D and 4D simplicial gravity given by (2.11) and (2.13) onto an exactly solvable model. In two dimensions one automatically gets a simplicial manifold if one glues together $d$-simplexes along their $(d-1)$-dimensional sub-simplexes until one has a closed complex. This is not the case in 3 and 4 dimensions. Constructing a 3- or 4-dimensional simplicial complex, $T$, following the recipe just given one only obtains a so-called pseudo-manifold. A pseudo-manifold, $T$, is a simplicial manifold if and only if $\chi(T) = 0$. In general $\chi(T) \geq 0$. This complication implies that one needs more parameters than the two required by the Einstein Hilbert action in order to construct an acceptable model of simplicial gravity in 3 and 4 dimensions. One way to search for viable theories of 3D and 4D quantum gravity is by trying to generalize the construction described on page 11 [45]. The most naive generalization of the construction does not work since only two parameters appear. There exist other generalizations involving more parameters [100, 101, 102]. However, none of these models are capable of at the same time eliminating complexes that are not simplicial manifolds and restricting the topology, let alone providing us with a topological expansion. Lacking a topological classification of 3D and 4D manifolds we do of course not really know what kind of expansion possibilities we can expect to encounter, if any. It is not an unreasonable line of action to try to generalize the two-dimensional models. However, since 3- and 4-dimensional geometry *is* much more complicated than two-dimensional geometry it might be necessary to consider models involving more complex structures.

A model of three-dimensional simplicial gravity involving the structure of quantum groups has been suggested by D. Boulatov [103, 104]. To define the model one uses as the starting point a real function, $g$, of three variables defined on the quantum group $SU_q(2)$ where $q = e^{2\pi i/(k+2)}$. The dynamical variables of the model are the fourier coefficients of the function $g$. The free energy of the model can be written as a sum over all connected orientable 3-dimensional pseudo-manifolds. A given pseudo-manifold, $T$, appears with a weight which in addition to the Einstein Hilbert weight factor involves the factor $I_q(T)\Lambda_q^{\chi(T)}$ where

$$\Lambda_q = -\frac{2(k+2)}{(q^{1/2} - q^{-1/2})^2} \qquad (4.222)$$

and where $I_q(T)$ is a topological invariant of $T$, the so-called Turaev Viro invariant associated with the quantum group $SU_q(2)$ [105]. If $T$ is a triangulation of an orientable 3-dimensional manifold, $M$, it holds that [106]

$$I_q(T) \leq \Lambda_q^{h(M)} \qquad (4.223)$$

where $h(M)$ is the Heegaard genus of the manifold $M$. The Heegaard genus, $h(M)$, equals zero if and only if $M$ is homeomorphic to $S^3$. Otherwise $h(M) \geq 1$. For the three-sphere one has $I_q(S^3) = 1$. Hence if one could send $\Lambda_q$ to zero keeping valid the inequality (4.223) one would be able to obtain a situation where only simplicial manifolds homeomorphic to $S^3$ would contribute to the free energy of the model. However, such a limiting procedure does not immediately make sense, since $\Lambda_q \to 0$ implies $q \to 0$ and $q$ is supposed to be a root of unity.

Ooguri has suggested another line of action for how to obtain from Boulatovs model the free energy of 3-dimensional simplicial Einstein Hilbert gravity where only simplicial manifolds homeomorphic to the sphere are taken into account [107]. This line of action is very similar to the line of action followed in section 3.1.5 where the free energy of the hermitian 1-matrix model was calculated. There we first calculated the 1-loop correlator using the Dyson Schwinger equations and afterwards integrated the result to obtain the free energy. Similarly Ooguri suggests that one should go via the Dyson Schwinger equations to obtain the free energy of 3-dimensional simplicial quantum gravity. Ooguri defines a 3-dimensional analogue of the matrix model expectation value $\langle \operatorname{Tr} \phi^n \rangle$ and derives a Dyson Schwinger equation analogous to equation (3.14) for this quantity. Just as the expectation value $\langle \operatorname{Tr} \phi^n \rangle$ in the matrix model case can be written as a sum over 2-dimensional simplicial manifolds with a loop of length $n$ inserted Ooguris expectation value can be written as a sum over 3-dimensional pseudo-manifolds with a 2-dimensional surface of a certain area and a certain number of handles inserted. And as in the matrix model case one can derive the factorization condition (3.16) which makes it possible to arrange that only 2-dimensional manifolds of genus zero enter the Dyson Schwinger equation Ooguri can write down a set of factorization rules by means of which one can arrange that the Dyson Schwinger equations of the 3-dimensional model involve only 3-dimensional manifolds homeomorphic to $S^3$. For this version of the Dyson Schwinger equations the solution related to the insertion of the 2-dimensional surface constituted by the boundary of a 3-simplex is a simple derivative of the free energy of 3-dimensional Einstein Hilbert gravity for simplicial manifolds homeomorphic to $S^3$. In principle it should be possible to calculate explicitly the observables of this model by means of the Dyson Schwinger equations and the factorization rules. However, such calculations have not yet been carried out. In the next section we will describe an approach to 3-dimensional simplicial quantum gravity where observables *have* been calculated, namely numerical simulations. Let us close this section by mentioning that Boulatovs model has also been generalized to 4 dimensions [108].

## 4.2 Numerical Simulations based on DT

### 4.2.1 The Numerical Method

As appears from the previous section so far no powerful analytical techniques for studying the partition function (2.18) of dynamically triangulated gravity for $d > 2$ have been developed. However, the 3- and 4-dimensional models can be studied by numerical means. The strategy when studying these models numerically consists in using

Monte Carlo techniques to generate $d$-dimensional simplicial manifolds homeomorphic to $S^d$ with weights given by the Einstein Hilbert action. For the generated simplicial manifolds one measures certain quantities and averaging over a large number of measurements one obtains the quantum averages of the quantities considered. To generate the simplicial manifolds one starts out with a given triangulation of $S^d$ and moves around in the space of combinatorially equivalent triangulations with a set of so-called moves, i.e. local changes of the triangulation[5]. These moves must fulfill the requirement of ergodicity, i.e. they must allow us to reach any simplicial manifold combinatorially equivalent to the initial one. Furthermore they should of course not change the topology of the manifold. In the grand canonical ensemble, i.e. the ensemble of manifolds where the volume is allowed to vary an ergodic set of moves has been known since 1930. These are the so-called Alexander moves [109]. The Alexander moves are, however, not convenient from a numerical point of view. More convenient are the so-called $(p, q)$ moves. In $d$ dimensions there are $(d + 1)$ such moves. They can be described in the following way. Let us consider a $(d + 1)$ simplex, $n_{d+1}$. Its boundary, $\partial n_{d+1}$ consists of $(d+2)$ $d$-simplexes. The recipe for performing a $(p, q)$ move in a $d$-dimensional simplicial manifold reads: Pick out a collection of $p$ $d$-simplexes which can be mapped onto a subset $S$ of $\partial n_{d+1}$ and replace it by the complement $\bar{S}$ in $\partial n_{d+1}$ consisting of $(d+2-q)$ $d$-simplexes. The $(p, q)$ moves were shown to be equivalent to the Alexander moves for $d = 3$ in reference [17] and for $d = 4$ in reference [110]. As opposed to what is the case in two dimensions in 3 and 4 dimensions no set of moves which are ergodic in the canonical ensemble, i.e. the ensemble of manifolds with a fixed volume, is known. One is hence forced to perform simulations in the grand canonical ensemble. However, there exists a method by means of which one can imitate a fixed volume simulation while still ensuring the ergodicity of ones algorithm [111]. As explained in section 2.1 a necessary condition for the partition function given by (2.18) to make sense is that the number of triangulations with the topology of $S^d$ and a given volume, $N_d$, must be exponentially bounded by the volume. Let us for a moment assume that such a bound exists, i.e that the partition function behaves as in (2.25). Normalizing the partition function such that the entropy factor for manifolds consisting of $N_d^0$ $d$-simplexes equals one we can write

$$Z(\kappa_d, \kappa_{d-2}) \sim \sum_{N_d} e^{-(\kappa_d - \kappa_d^c(\kappa_{d-2})^{eff})(N_d - N_d^0)} \qquad (4.224)$$

where

$$\kappa_d^c(\kappa_{d-2})^{eff} = \kappa_d^c(\kappa_{d-2}) + \frac{f'(N_d^0, \kappa_{d-2})}{f(N_d^0, \kappa_{d-2})}. \qquad (4.225)$$

We see that fixing $\kappa_d$ at the value $\kappa_d^c(\kappa_{d-2})^{eff}$ corresponding to some not too small volume $N_d^0$ and adding a term $\triangle\kappa_d |N_d - N_d^0|$, $0 < \triangle\kappa_d << 1$ to our original action results in the volume of the manifolds generated being peaked at $N_d^0$ and distributed around this value according to

$$P(N_d) \sim \exp(-\triangle\kappa_d |N_d - N_d^0|). \qquad (4.226)$$

---

[5]Two triangulations are combinatorially equivalent it they can be subdivided into the same finer triangulation.

This observation is very useful. First it makes it possible to check whether the exponential bound referred to above exists or not and if so to investigate further the nature of its subleading correction. Let us briefly describe how this works in practise. To check whether an exponential bound exists or not we start by choosing a given volume, $N_d^0$, a value of $\kappa_{d-2}$ and a value of $\triangle \kappa_d$. Then we search for a value of $\kappa_d$ for which the actual volume of the generated manifolds fluctuates in an interval around $N_d^0$. If one succeeds in finding a value of $\kappa_d$ for which such a situation is encountered one has a first estimate of $\kappa_d^c(\kappa_{d-2})^{eff}$. By examining the distribution of $N_d$ around $N_d^0$ one can improve the estimate for $\kappa_d^c(\kappa_{d-2})^{eff}$. (If $\kappa_d$ does not equal $\kappa_d^c(\kappa_{d-2})^{eff}$ the slope of $\ln(P(N_d))$ will be different above and below $N_d^0$ and this difference in slope tells us by which amount we should correct our initial guess.) The condition that the entropy of manifolds is exponentially bounded reads

$$\frac{f'(N_d^0, \kappa_{d-2})}{f(N_d^0, \kappa_{d-2})} \to 0 \quad \text{for} \quad N_d^0 \to \infty. \tag{4.227}$$

By repeating the fine-tuning process described above for different values of $N_d^0$ one can test if this condition is fulfilled. In particular one can investigate the nature of the (hopefully) subleading correction to the exponential bound. We note that if for $d > 2$ we have a $\gamma_{str}$ like in 2 dimensions i.e. $f(N_d, \kappa_{d-2}) = N_d^{\gamma(\kappa_{d-2})-3}$ the equation (4.225) reads

$$\kappa_d^c(\kappa_{d-2})^{eff} = \kappa_d^c(\kappa_{d-2}) + \frac{\gamma(\kappa_{d-2}) - 3}{N_d} \tag{4.228}$$

and a measurement of the distribution of the volume of our manifolds would hence allow us to determine $\gamma(\kappa_{d-2})$.

Fixing $\kappa_d$ to its critical value $\kappa_d^c(\kappa_{d-2})^{eff}$ and adding to the action the term $\triangle \kappa_d |N_d - N_d^0|$ is also what allows us to imitate a fixed volume simulation without violating the ergodicity of our algorithm. By adjusting $\triangle \kappa_d$ we can obtain an appropriate width of the distribution of volumes and by making measurements only on manifolds with volume $N_d^0$ we effectively study the fixed volume partition function. The value of $\triangle \kappa_d$ is typically only a few percent of $\kappa_d^c(\kappa_{d-2})^{eff}$. The strategy of fixing $\kappa_d$ at its critical value is fully compatible with our desire to search for a critical point where a continuum theory might exist since as explained in section 2.1 a necessary ingredient in our recipe for defining a continuum limit is to let $\kappa_d$ approach $\kappa_d^c$ from above. Let us point out again that the aim of the numerical simulations is to search the restricted coupling constant space $\kappa_d = \kappa_d^c(\kappa_{d-2}, \{g_i\})$ for a critical point where an appropriately defined susceptibility for quantum gravity diverges. Here the coupling constants $\{g_i\}$ might be coupling constants for matter fields or for higher derivative terms. We note that in order to pursue this aim one has to perform a fine tuning of $\kappa_d$ for each choice of the remaining coupling constants. In addition we have chosen to perform for each choice of coupling constants yet another fine tuning, namely we introduce a probability for performing each of the $(d+1)$ moves which we adjust so that in a given simulation all the $(d+1)$ different moves are performed approximately the same number of times. The purpose of this arrangement is to avoid bias towards special types of manifolds caused by a difference in acceptance rates for the different types of moves. Let us also mention that we restrict the $o(n_i)$'s in the following way

$$o(n_i) \geq d - i + 1, \qquad i \leq d - 2. \tag{4.229}$$

Of course we have $o(n_{d-1}) = 2$. To do the updating of the geometry we use a standard Metropolis algorithm. Possible matter fields living on the geometries are updated using well established techniques from fixed lattice simulations. However, special care must of course be taken for the updating of matter fields related to changes in the geometry. We will not enter into a discussion of these points here. A description of the line of action taken in the simulations that we will report on below can be found in references [112] and [113].

### 4.2.2 Observables

In quantum gravity we do not have the same type of observables as in conventional quantum field theory. Expectation values of operators which refer to specific points in the space time manifold have no meaning. Observables which do make sense are averages of integrated local operators. One such observable which we have made use of in our simulations is the average integrated scalar curvature per volume

$$\langle R \rangle = \frac{\int_{\mathcal{M}} d^d \xi \sqrt{g} R}{\int_{\mathcal{M}} d^d \xi \sqrt{g}} \longrightarrow c_d \frac{N_{d-2}}{N_d} - K_{d+1,2} \tag{4.230}$$

where the notation is as in section 2.1 (cf. to equation (2.7) and (2.10)). Another type of observable which is available in quantum gravity is the average correlation between local operators $\mathcal{O}(\xi_1)$ and $\mathcal{O}(\xi_2)$ defined as a function of geodesic distance, for example as

$$\langle G(r) \rangle = \langle \int_{\mathcal{M}} \int_{\mathcal{M}} d^d \xi_1 \sqrt{g(\xi_1)} d^d \xi_2 \sqrt{g(\xi_2)} \mathcal{O}(\xi_1) \mathcal{O}(\xi_2) \delta(d(\xi_1, \xi_2) - r) \rangle \tag{4.231}$$

where $d(\xi_1, \xi_2)$ is the geodesic distance between $\xi_1$ and $\xi_2$ for a given Riemannian manifold. A particularly simple example of such an observable is the average volume $\langle V(L) \rangle$ of the ball with radius $L$ defined as the volume of the set of points $\{\xi_i\}$ at geodesic distance $d(\xi_i, \xi_0) \leq L$ from some reference point, $\xi_0$. By means of $\langle V(L) \rangle$ we can define the Hausdorff dimension, $d_H$, describing the fractal structure of space time by

$$\langle V(L) \rangle \sim L^{d_H}. \tag{4.232}$$

In numerical simulations of simplicial gravity we obviously have the possibility of studying in detail the fractal structure of our universes. From a numerical point of view it is convenient in stead of working with the strict definition of geodesic distance on a simplicial manifold to work with the so-called $n_d$-distance between $d$-simplexes. We define the $n_d$-distance, $D$, between two $d$-simplexes in a $d$-dimensional simplicial manifold as the length of the shortest path between the two $d$-simplexes obtained by moving between centers of neighbouring $d$-simplexes. By $\langle D \rangle$ we mean the average value of $D$ for a given manifold averaged over our ensemble of manifolds. Hence $\langle D \rangle$ can be thought of as a typical radius of our universes. We have measured the average value of $D$ as well as its (average) distribution, $P(D)$. From these measurements we can in principle extract $d_H$ by fitting $P(D)$ to the functional behaviour $P(D) \sim D^{d_H - 1}$. Likewise we can in principle determine the so-called cosmological Hausdorff dimension $d_{ch}$ defined by [45]

$$\langle D \rangle \sim N_d^{1/d_{ch}}. \tag{4.233}$$

In practise it is hard to extract in a reliable way any of the Hausdorff dimensions. (We note that the two definitions are not necessarily equivalent.) However, $\langle D \rangle$ as well as $\langle R \rangle$ have proven very convenient when scanning the coupling constant space of the models of dynamically triangulated gravity.

To investigate further the nature of a possible phase transition we will need a susceptibility like observable. One candidate for such an observable is the curvature-curvature correlation function. In continuum notation it reads

$$\chi(\kappa_{d-2}) = \langle \int_{\mathcal{M}} \int_{\mathcal{M}} d^d\xi_1 d^d\xi_2 \sqrt{g(\xi_1)} R(\xi_1) \sqrt{g(\xi_2)} R(\xi_2) \rangle - \langle \int_{\mathcal{M}} d^d\xi \sqrt{g(\xi)} R(\xi) \rangle^2 \tag{4.234}$$

and it is easily seen that it is proportional to the second derivative of $-\log Z$ with respect to the inverse gravitational constant (cf. to equation (2.4)). In the context of dynamically triangulated gravity with $N_d$ fixed $\chi(\kappa_{d-2})$ turns into

$$\chi(\kappa_{d-2}, N_d) = \langle N_{d-2}^2 \rangle_{N_d} - \langle N_{d-2} \rangle_{N_d}^2. \tag{4.235}$$

What we hope to encounter is a point where $\chi(\kappa_{d-2}, N_d)/N_d$ diverges when $N_d \to \infty$. We have in addition to $\chi(\kappa_{d-2}, N_d)$ also studied another somewhat related quantity, namely $\langle R^2 \rangle - \langle R \rangle^2$ where

$$\langle R^2 \rangle = \frac{\int_{\mathcal{M}} d^d\xi \sqrt{g(\xi)} R^2(\xi)}{\int_{\mathcal{M}} d^d\xi \sqrt{g(\xi)}} \longrightarrow \langle \frac{1}{N_4} \sum_{n_{d-2}} o(n_{d-2}) \left[ \frac{c_d - o(n_{d-2})}{o(n_{d-2})} \right]^2 K_{d+1,2} \rangle. \tag{4.236}$$

This expression for the average squared curvature of a $d$-dimensional simplicial manifold comes about when one assigns to each of its $(d-2)$-simplexes a volume density $V_{n_{d-2}} = o(n_{d-2})/K_{d+1,2}$ and a curvature density $R_{n_{d-2}} = (c_d - o(n_{d-2}))/V_{n_{d-2}}$. We will discuss this construction in more detail in section 4.4.3. The correlator $\langle R^2 \rangle - \langle R \rangle^2$ has turned out to be a useful indicator for changes in the geometrical system.

In the case where matter is coupled to gravity one of course has additional observables associated with the matter fields. As for the geometry possible observables for the matter systems are restricted to global quantities and correlation functions defined in terms of geodesic distances. In the following we will only discuss aspects related to the behaviour of global observables of the matter systems. For a discussion of the problem of defining correlation functions as well as the problem of extracting from these critical indices we refer to [114].

## 4.3 Numerical Results for D=3

### 4.3.1 Pure Gravity

The justification for simulating 3-dimensional dynamically triangulated gravity is mainly that it is a natural intermediate step between the exactly solvable 2-dimensional case and the far more complicated 4-dimensional case. Classical 3-dimensional general relativity it trivial. However, the Einstein Hilbert action is in three as in four dimensions unbounded from below and non-renormalizable. Hence it might very well be that numerical simulations will reveal to us a theory with a non trivial structure. In particular coupling to matter might lead to complex interactions.

The first step when aiming at simulating three dimensional simplicial quantum gravity is of course to check that the partition function of the model makes sense. In reference [16] it was verified by numerical means that the number of triangulations of the three-sphere with a given volume is exponentially bounded by the volume. Having settled this question one knows that the line $\kappa_3 = \kappa_3^c(\kappa_1)$ described in section 2.1 and 4.2.1 exists and one can start looking for critical behaviour along this line. Several groups have pursued this idea [17, 115, 116, 117] and their results agree.

One finds that when moving along the line $\kappa_3 = \kappa_3^c(\kappa_1)$ one encounters at a certain value of $\kappa_1$ (between 3.5 and 4.0) a seemingly sudden change between two types of geometry. The two regions in the coupling constant space corresponding to the two types of geometry are denoted as the hot and the cold phase of 3-dimensional simplicial quantum gravity. For small values of $\kappa_1$ the geometrical system is in the hot phase. The hot phase is characterized by the space time manifolds being very crumpled. Typically there will be vertices of very high order and the number of vertices increases only very slowly with the volume. The average integrated curvature per volume $\langle R \rangle$ will be small and eventually for small enough values of $\kappa_1$ become negative. The average radius of the universes $\langle D \rangle$ will also be small and almost volume independent. In addition the distribution of $D$, $P(D)$ will be strongly peaked. The Hausdorff dimension defined in either way will hence be very large. Possibly it is even infinite. Presumably there is no $\gamma_{str}$ in this phase [16, 17]. The cold phase comes into existence when $\kappa_1$ gets large. In this phase the space time manifolds are very extended. Their average radius, $\langle D \rangle$, is large and grows almost linearly with the volume. Furthermore there are large fluctuations in $\langle D \rangle$ and $P(D)$ is a very broad distribution with an almost plateau like maximum. The Hausdorff dimension of the space time manifolds in the cold phase is probably close to one. The average curvature $\langle R \rangle$ is large and positive. It seems that in this phase a $\gamma_{str}$ does exist [17]. The cold phase has been interpreted as signaling the dominance of the conformal mode. Being regulated the Euclidean path integral can of course not diverge. However, the 1-dimensional space time manifolds with a rapidly changing connectivity are probably the lattice analogues of the conformal instability. For small values of $\kappa_1$ the conformal instability is suppressed by the large entropy of very compact universes. In connection with the transition between the hot and the cold phase pronounced hysteresis is observed. This is illustrated in figure 3 where we have shown the behaviour of $\langle R \rangle$ as our geometrical system goes through a thermal cycle.

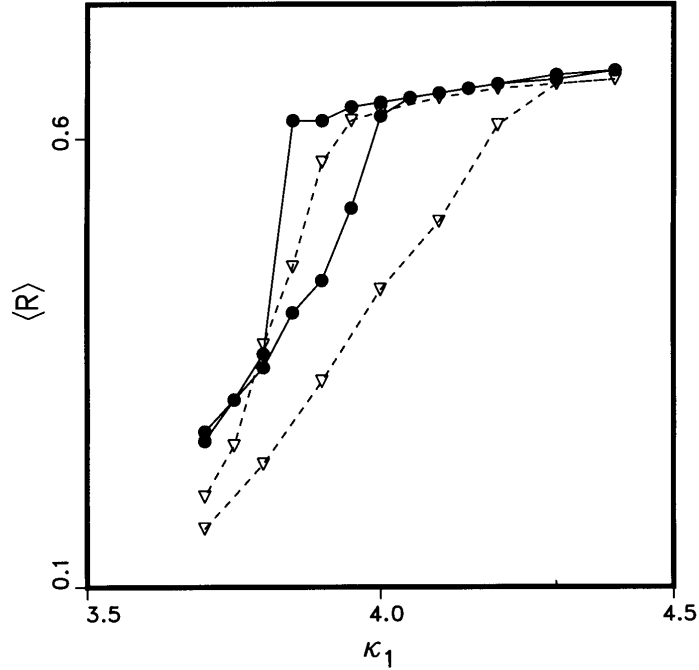

Figure 3. Hysteresis curves for 3-dimensional dynamically triangulated gravity. Triangles: pure gravity. Circles: gravity plus one Ising spin system when the coupling between matter and geometry is maximal, i.e. $\beta = \beta_c(\kappa_1)$. $N_3 = 10000$. From reference [119].

The fact that hysteresis is observed shows that the transition between the two phases is of first order. Hence we have no hope of defining along the lines of conventional statistical mechanics a continuum theory of gravity with a propagating massless particle. It seems as if the classical result that no gravitational waves can propagate in 3 dimensions survives quantization. The situation might be different if matter is coupled to gravity. We will consider this situation in the next section. Let us mention here that in the static triangulation approach to simplicial quantum gravity one finds for $d = 3$ a second order phase transition both in the case of pure gravity as well as in the case where a $R^2$ term is added to the Einstein Hilbert action [118]. So far no clear understanding of this rather fundamental difference between the two models of quantum gravity has been achieved.

### 4.3.2 Gravity plus Matter

Studying the interaction between space time and matter is of course interesting in its own right. For $d = 3$ we furthermore have the possibility that the presence of matter fields might cause the first order phase transition observed for pure gravity to change into a (from continuum physics point of view) more interesting second order phase transition.

So far only the case of 3-dimensional simplicial gravity coupled to an Ising spin system has been investigated. As mentioned in section (2.2), when coupling matter to simplicial gravity one can choose to place the matter field variables either on the vertices of the $d$-dimensional simplicial manifold or in the center of its $d$-simplexes. We have chosen the latter possibility [119]. In this way we are sure that the number of matter field variables grows with the volume of the universe. This is not necessarily true if the matter fields variables are placed on the vertices of the simplicial manifold (cf. to section 4.3.1). Our partition function reads

$$Z(\kappa_3, \kappa_1, \beta) = \sum_{N_3} e^{-\kappa_3 N_3} \sum_{T \sim S^3(N_3)} \frac{1}{C(T)} e^{\kappa_1 N_1} \sum_{\{\sigma\}} e^{\beta \sum_{\langle i,j \rangle} (\delta_{\sigma_i \sigma_j} - 1)} \quad (4.237)$$

where $\sigma_i \in \{-1, +1\}$ and where $\sum_{\langle i,j \rangle}$ as usual denotes the sum over neighbouring pairs of spins. We note that when the spins are placed in the center of the 3-simplexes each spin has 4 neighbours. By $\sum_{\{\sigma\}}$ we mean the sum over all spin configurations for a given manifold. As usual we are only interested in exploring the restricted coupling constant space $\kappa_3 = \kappa_3^c(\kappa_1, \beta)$. Hence we have effectively only two coupling constants $\kappa_1$ and $\beta$. What we aim at is mapping out the phase structure of the model in the $(\kappa_1, \beta)$ plane. To prepare for the presentation of the result it is convenient first to recall the characteristics of the Ising model on a regular lattice. A one dimensional Ising chain never becomes critical. The spins are always disordered. For $d = 2$ the Ising model has a second order phase transition from a disordered to an ordered phase at some value $\beta_c$ of $\beta$ and mean field theory indicates that a similar behaviour is seen for $d > 2$.

The phase diagram of the model (4.237) is shown in figure 4. The dotted line is the line $\kappa_1 = \kappa_1^c(\beta)$, i.e. it indicates the transition between the two phases of the geometrical system. (Due to hysteresis this line is not unambiguously defined. We shall explain shortly what we mean by $\kappa_1^c(\beta)$.) The almost horizontal line indicates a transition between a non magnetized and a magnetized phase for the Ising model. The location of the line $\beta = \beta_c(\kappa_1)$ is determined with the aid of the magnetization curves for the spin system (for $N_3 = 10000$). The transition between the two phases becomes sharper as the volume of the universe increases and we presumably face here a second order phase transition. We note that the dependence of $\beta_c$ on $\kappa_1$ is rather weak. In the elongated phase of the geometrical system we see no sign of a phase transition for the spin system. Presumably in the infinite volume limit the spin system has zero magnetization for all values of $\beta$. (We refer to reference [119] for details.) The absence of a phase transition for the spin system in the cold phase of gravity seems to show that the Hausdorff dimension of our space time manifolds is the dimension relevant for coupling to matter. Accordingly it is tempting to conjecture that for the phase transition of the spin system encountered in the hot phase of gravity, where the Hausdorff dimension is very large, the critical exponents take their mean field values. It would be interesting to test this conjecture but a reliable determination of the critical exponents of the matter system requires more extensive numerical studies. Figure 4 shows that there is a clear back reaction from the spins on the geometry. The spins affect the geometry by shifting $\kappa_1^c$ towards smaller values. The effect is largest when the spin system is itself critical and $\kappa_1^c$ takes its pure gravity value for $\beta \to 0$ and $\beta \to \infty$. This is what one would intuitively expect. When $\beta \to 0$ or $\beta \to \infty$ the Ising spins are independent and individually they can not influence the geometry. Only when there are correlations between the spin variables their presence might be of importance. It is interesting to note that the spin system seems to be counteracting its own criticality. It effectively pushes the geometry towards a lower Hausdorff dimension. A similar effect is seen in 2 dimensions when multiple spins are coupled to gravity [37].

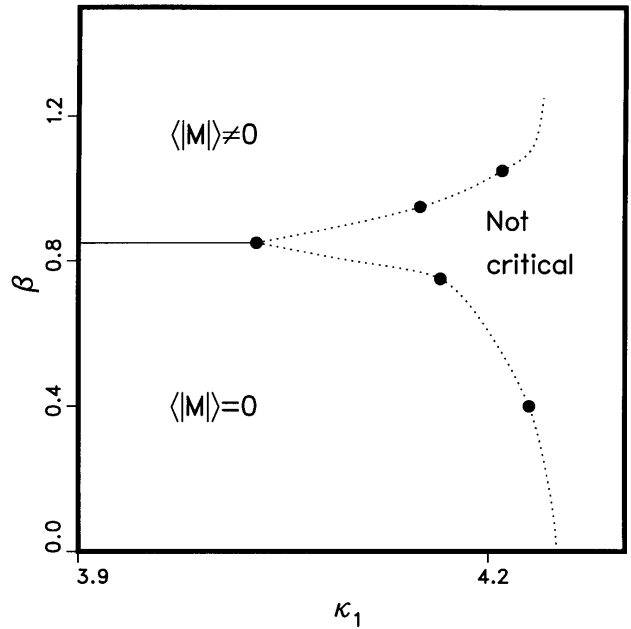

Figure 4: The phase diagram for 3-dimensional dynamically triangulated gravity coupled to one Ising spin system. Dotted line: the transition between the hot and the cold phase of the geometrical system. Full drawn line: transition between a disordered and an ordered phase of the spin system. Filled circles are results for $\kappa_1^c(\beta)$ determined for $N_3 = 10000$. From reference [119].

The back reaction from the spins does not change the order of the phase transition for the geometrical system, however. For all values of $\beta$ one observes hysteresis in connection with the transition between the hot and cold phase of the geometry. Although the hysteresis gets less pronounced as $\beta \to \beta_c$ it never completely disappears. In figure 3 we have shown the behaviour of $\langle R \rangle$ as the geometrical system for $\beta = \beta_c(\kappa_1)$ goes through a thermal cycle. When an Ising spin system was coupled to two-dimensional dynamically triangulated gravity one found that at the critical point of the spin system the exponent $\gamma_0$ describing the fractal structure of space time was changed from $-\frac{1}{2}$ to $-\frac{1}{3}$. In the 3-dimensional case the spin system can only become critical when the geometry is in its hot phase and as mentioned earlier in this phase presumably no $\gamma_{str}$ exists. Along the line $\kappa_1 = \kappa_1^c(\beta)$ we might have a $\gamma_{str}$ and it would be interesting to explore what happens to this index as well as the critical indices of the spin system exactly at $\kappa_1 = \kappa_1^c(\beta^c)$. Needless to say that this is a project which calls for a considerable amount of computer time. Let us finally comment on our definition of $\kappa_1^c$. If one starts out with values of $\kappa_1$ and $\beta$ corresponding to the hot phase of the geometrical system and changes $\kappa_1$ in small steps keeping $\beta$ fixed the value of $\langle R \rangle$ will trace out the lower branch of a hysteresis curve as in figure 3. For values of $\kappa_1$ well inside the hot phase the trajectory is unique and we define the critical value of $\kappa_1$ as the value of $\kappa_1$ for which this trajectory when extrapolated intersects the parts of the hysteresis curves corresponding to the cold phase of the geometrical system. The filled circles in figure 4 indicate values of $\kappa_1^c$ obtained according to this philosophy using universes of size $N_3 = 10000$.

Since hysteresis occurs even at $\beta = \beta_c$ coupling an Ising spin system to 3-dimensional gravity does not improve our possibilities of obtaining from the lattice approach a continuum theory of quantum gravity in three dimensions. One could of course try to couple more spin fields or other types of matter fields to the geometrical system but this line of action has so far not been pursued. Let us mention here that 3-dimensional dynamically triangulated gravity coupled to an Ising spin system has also been investigated in references [120] and [121]. In reference [120] the spins were placed on the vertices of the simplicial manifolds while in reference [121] both the case of spins living on the vertices and the case of spins living in the center of the $d$-simplexes were explored, but only for values of $\kappa_1$ corresponding to the hot phase of the geometrical system. In neither reference the possible influence of the spin system on the phase transition of the geometrical system was investigated.

## 4.4 Numerical Results for D=4

### 4.4.1 Pure Gravity

Numerical investigations have shown that the number of triangulations of $S^4$ with a given volume is exponentially bounded by the volume. Hence the partition function of 4-dimensional dynamically triangulated gravity is well defined in a certain region of the coupling constant space, $\kappa_4 > \kappa_4^c(\kappa_2)$. As already explained, when aiming at defining a continuum version of the model one must search the line $\kappa_4 = \kappa_4^c(\kappa_2)$ for critical points. Such investigations have been carried out by several groups [18, 19, 20, 21] and their results are in agreement.

When one explores the restricted coupling constant space $\kappa_4 = \kappa_4^c(\kappa_2)$ one finds that the geometrical system can exist in two phases. For small values of $\kappa_2$ the geometrical system is in a hot phase and for large values of $\kappa_2$ the system is in a cold phase. The hot and the cold phase of 4-dimensional simplicial gravity have the same characteristica as the hot and the cold phase of 3-dimensional simplicial gravity (cf. to section 4.3.1). However, the nature of the transition between the two phases is different in 3 and 4 dimensions. In 4 dimensions no hysteresis is observed. This opens the possibility that the phase transition might be of second order and hence that it might be possible to define in the vicinity of the critical point a continuum theory with macroscopic correlations. In figure 5 we have shown the behaviour of the susceptibility $\chi(\kappa_2)/N_4$ as the system passes from one phase to the other. The susceptibility clearly has a peak and the position of this peak coincides with the value of $\kappa_2$ for which the change in geometry is seen to take place. The height of the peak does increase with the volume of the universe. However, our data does not allow us to determine whether we face a divergence or not. In connection with the the transition between the hot and the cold phase we also observe a peak in the correlator $\langle R^2 \rangle - \langle R \rangle^2$. Neither for this correlator have we been able to analyse in detail the volume dependence. It seems unlikely that the transition should be of first order but it is not excluded that it could be of order higher than two.

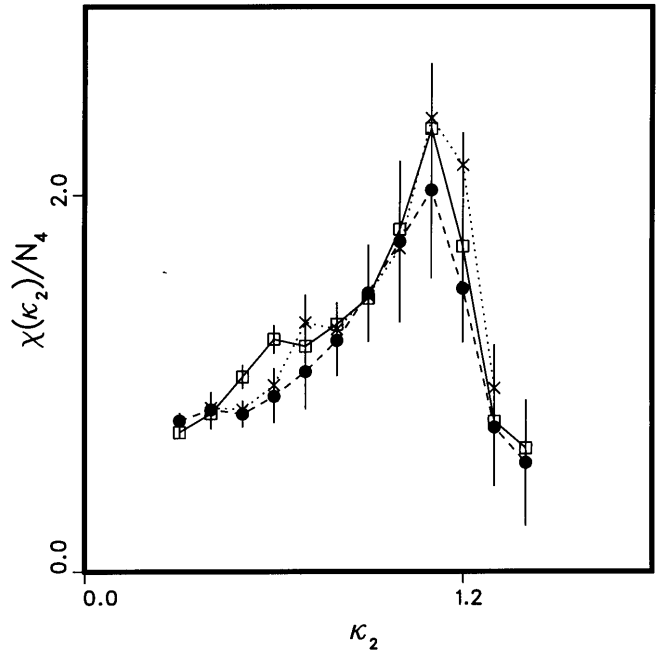

Figure 5. Four-dimensional dynamically triangulated gravity. The behaviour of the susceptibility $\chi(\kappa_2)/N_4$ as the geometrical system undergoes the transition from the hot to the cold phase. $N_4 = 4000(\bullet)$, $N_4 = 9000(\square)$ and $N_4 = 16000(\times)$. From reference [123].

Even if we succeed in arriving at a convincing argument that the transition is of second order there remains a number of unclarified points. A natural part of a prescription for defining a continuum limit is to introduce a scaling parameter, $a$, to be considered as the lattice spacing, and to send $a$ to zero and $N_4$ to infinity while keeping fixed the physical volume of space time $a^4 N_4$. This implies the following scaling for $\langle R \rangle$

$$\langle R_{continuum} \rangle = \langle R_{lattice} \rangle \frac{1}{a^2}. \qquad (4.238)$$

Hence it seems that we are led to the conclusion that the curvature of our continuum space time manifolds can only be finite if the lattice curvature scales to zero at the critical point. However, one finds that the value of $\langle R_{lattice} \rangle$ lies relatively far above zero at the critical point. (The same is true in 3 dimensions (cf. to figure 3).) It is still not completely excluded that $\langle R_{lattice} \rangle(\kappa_2^c) \neq 0$ is a finite volume effect but we consider it unlikely. We are convinced that $\langle R_{lattice} \rangle(\kappa_2^c) \neq 0$ is not due to the missing tessellation of flat space for $d > 2$ since already when averaged over only the $K_{d+1,2}$ $(d-2)$-simplexes belonging to a single $d$-simplex the curvature can easily take the value zero. One point of view to take is that the pure gravity theory is too naive and that $\langle R_{lattice} \rangle(\kappa_2^c)$ will actually be zero for a more realistic theory of gravity including for instance matter fields of higher derivative terms. Another point of view would be to give up the interpretation of $\langle R_{continuum} \rangle$ as the actual physical curvature of our continuum space time manifold. Lacking an alternative interpretation we have chosen to investigate a class of more elaborate models of dynamically triangulated gravity in 4 dimensions. These investigations are reviewed in the following two sections. There is another observable whose value at the critical point has important consequences for a possible continuum limit, namely the Hausdorff dimension. As mentioned earlier in the hot phase of dynamically triangulated gravity the Hausdorff dimension is very large, possibly infinite, while in the cold phase it is close to one. Since the transition from large to small values of the Hausdorff dimension is continuous one might find a Hausdorff dimension close to four at the critical point. However, we will refrain from making any claims abut the value of $d_H$ or $d_{ch}$ at the phase transition point since we do not believe that the data allows for a reliable estimate.

Four-dimensional simplicial quantum gravity has been studied also in the static triangulation approach [122]. In this approach one finds a first order phase transition for the model describing pure gravity. However, the nature of the transition changes when a higher derivative term is added to the action. We will return to this point in section 4.4.3.

Let us finally mention that in reference [13] the effect of choosing a measure different from the uniform one in dynamically triangulated 4-dimensional gravity was investigated. It was found that a change of the measure changes the location of the critical point. The abruptness of the transition between the hot and the cold phase was in general weakened when the measure was chosen different from uniform. The value of $\langle R \rangle$ at the critical point did not change significantly. A future aim of these investigations would be a detailed comparison of appropriately defined critical exponents.

### 4.4.2 Gravity plus Matter

Apart from the study of the interaction between matter and space time being interesting in its own right we also have the possibility that the introduction of matter fields in 4-dimensional dynamically triangulated gravity will improve the scaling of the curvature at the critical point (cf. to section 4.4.1). We have studied 4-dimensional simplicial quantum gravity coupled to an Ising spin system as well as 4-dimensional simplicial quantum gravity coupled to various numbers of Gaussian scalar fields. In all cases the matter field variables were placed in the center of our $d$-simplexes. Here we will only briefly review the results. For details we refer to [113].

Our partition function for the system consisting of one Ising spin system coupled to 4-dimensional dynamically triangulated gravity is completely analogous to (4.237) and the phase diagram of the model looks precisely as in the three-dimensional case (cf. to figure 4). The presence of the spin variables causes the critical value of $\kappa_2$ to be shifted towards smaller values and the shift is largest when the spin system is critical. The nature of the transition between the two phases of the geometrical system seems to be the same at all points on the line $\kappa_2 = \kappa_2^c(\beta)$. There is a small increase of the height of the peak of the correlator $\langle R^2 \rangle - \langle R \rangle^2$ as $\beta \to \beta_c$ but otherwise the data is compatible with the only effect of the matter field being a shift in $\kappa_2^c$. In particular no significant change of the value of $\langle R \rangle(\kappa_2^c)$ away from its pure gravity value is seen for any value of $\beta$.

Our partition function when we couple $n_g$ Gaussian fields, $\phi^1, \ldots \phi^{n_g}$, to our geometrical system reads

$$Z(\kappa_4, \kappa_2) = \sum_{N_4} e^{-\kappa_4 N_4} \sum_{T \sim S^4(N_4)} \frac{1}{C(T)} e^{\kappa_2 N_2} \int \prod_{i,\alpha} \frac{d\phi_i^\alpha}{\sqrt{2\pi}} \left[ \prod_{\alpha=1}^{n_g} \delta\left(\sum_i \phi_i^\alpha\right) \right] e^{\frac{1}{2} \sum_{<i,j>,\alpha} \left(\phi_i^\alpha - \phi_j^\alpha\right)^2}$$
(4.239)

where $i$ and $j$ label the 4-simplexes of our manifolds and where the $\delta$-functions are introduced in order to eliminate the translational mode. We note that coupling a Gaussian field to simplicial gravity does not require the introduction of an additional coupling constant. A coupling constant in front of the Gaussian action can be absorbed in a redefinition of the cosmological constant by an appropriate rescaling of the scalar field. This of course reflects the fact that a Gaussian scalar field is automatically critical in the infinite volume limit. As in the case of the Ising spin system we see no drastic effects when Gaussian fields are coupled to the geometry. The critical value of $\kappa_2^c$ is again shifted towards smaller values but the nature of the phase transition of the geometrical system seems unaltered. The height of the peak of the correlator $\langle R^2 \rangle - \langle R \rangle^2$ is somewhat larger than for pure gravity, though. There are some indications of an enhancement of the interaction between matter and space time when the number of matter fields is increased. We have considered up till four Gaussian fields. The larger the number of Gaussian fields, the larger is the shift in $\kappa_2^c$ and the more pronounced is the peak in $\langle R^2 \rangle - \langle R \rangle^2$. However the behaviour of $\langle R \rangle$ as we pass the transition in geometry seems unaltered. In particular the value of $\langle R \rangle$ at the critical points of the coupled systems is not significantly different from its value at the critical point of pure gravity. In the connection with the change in geometry for our space time manifolds we see a change in the expectation value $\langle (\phi^\alpha)^2 \rangle$. The value of $\langle (\phi^\alpha)^2 \rangle$ as well as the fluctuations of this quantity increase when we enter the elongated phase. As in the 3-dimensional case a detailed investigation of a possible change in critical indices for matter fields or geometry requires more extensive numerical studies.

### 4.4.3 Higher Derivative Regularization

So far the space time part of the action for our models of dynamically triangulated gravity has been given by the pure Einstein Hilbert action. In this section we will briefly describe the effects of adding to the Einstein Hilbert action a $R^2$ term. In the continuum formulation the introduction of a $R^2$ term (with positive coupling constant) would lead to a regularization of the path integral. In dynamically triangulated gravity the path integral involving only the Einstein Hilbert term is already regularized (for fixed space time volume). The purpose of introducing a $R^2$ term in the dynamically triangulated model is to test whether this extension of the coupling constant space allows us to reach other critical points which might prove especially well suited for defining a continuum theory of quantum gravity. What we have in mind is of course a unmistakable second order phase transition point where the curvature scales to zero.

To define simplicial analogues of quantities like $\int_{\mathcal{M}} d^d\xi \sqrt{g} R^p$, $p>1$, one can obviously not maintain the original idea of Regge that the curvature is $\delta$-function like having support only on the $(d-2)$-simplexes. A prescription for modifying the philosophy of Regge so that expressions like $\int_{\mathcal{M}} d^d\xi \sqrt{g} R^p$, $p>1$ make sense has been given by Hamber and Williams [10]. The idea consists in viewing the curvature $R_{n_{d-2}}$ normally associated with a given $(d-2)$-simplex, $n_{d-2}$, as being distributed in a $d$-dimensional volume around $n_{d-2}$. One assigns a volume element $V_d(n_{d-2})$ and a curvature density $R_{n_{d-2}}/V_d(n_{d-2})$ to each $(d-2)$-simplex, $n_{d-2}$, in a $d$-dimensional simplicial manifold and calculates integrals involving powers of the curvature according to

$$\int_{\mathcal{M}} d^d\xi \sqrt{g} R^p \longrightarrow \sum_{n_{d-2}} V_d(n_{d-2}) \left\{ \frac{R_{n_{d-2}}}{V_d(n_{d-2})} \right\}^p. \qquad (4.240)$$

For a $d$-dimensional simplicial manifold consisting of regular simplexes one has $R_{n_{d-2}} = (c_d - o(n_{d-2}))$ and $V_{n_{d-2}} = o(n_{d-2})/K_{d+1,2}$ where we note that $K_{d+1,2}$ is the number of $(d-2)$-dimensional sub-simplexes in a $d$-dimensional simplex. With these assignments and the prescription (4.240) for integrating over the space time manifold one gets

$$\int_{\mathcal{M}} d^d\xi \sqrt{g} R^2 \longrightarrow \sum_{n_{d-2}} o(n_{d-2}) \left\{ \frac{c_d - o(n_{d-2})}{o(n_{d-2})} \right\}^2 K_{d+1,2}. \qquad (4.241)$$

Furthermore the relations (2.7) and (2.10) are maintained. In the continuum formulation the addition of a $R^2$ term to the Einstein Hilbert action results in the favourization of space time manifolds with zero local curvature. This statement does not immediately translate to dynamically triangulated gravity. In dynamically triangulated gravity it is not possible to have zero local curvature for $d > 2$. The addition of a $R^2$ term like the one appearing on the right hand side of (4.241) does suppress manifolds with large numerical values of the curvature. However, the discretized $R^2$ term takes its minimum value at $R = -0.46$ not at $R = 0$ (where $R$ was defined in equation (4.230)). Studying dynamically triangulated gravity with the lattice $R^2$ term should not be considered as studying Einstein Hilbert gravity with exactly the term $\int_{\mathcal{M}} d^d\xi \sqrt{g} R^2$ added but as exploring a certain class of models with higher derivative terms. Let us briefly review what has been learnt from such studies. The details can be found in reference [123]. In this work we consider 4-dimensional simplicial gravity with the following action

$$S[T] = \kappa_4 N_4 - \kappa_2 N_2 + \frac{h}{c_4^2} \sum_{n_2} o(n_2) \left\{ \frac{c_4 - o(n_2)}{o(n_2)} \right\}^2 . \qquad (4.242)$$

The restricted coupling constant space of the model $\kappa_4 = \kappa_4^c(\kappa_2, h)$ is scanned keeping fixed the value of $h$ and looking for critical behaviour when varying $\kappa_2$. The maximal value of $h$ considered was $h = 20$. It is in practise impossible to increase $h$ further because the acceptance rates in our Monte Carlo updatings become too small. For all values of $h$ considered we encounter at a certain value $\kappa_2^c$ of $\kappa_2$ a transition between a hot and a cold phase of the geometrical system and in all cases the transition in geometry is accompanied by a peak in the susceptibility $\chi(\kappa_2)/N_4$ the height of which tends to grow somewhat with volume. For not too large values of $h$ the nature of the transition can not be distinguished from the nature of the transition encountered for $h = 0$. The value of $\langle R \rangle$ at the critical point does decrease with $h$ but the change is very small. For $h > 12$ the peak in the correlator $\langle R^2 \rangle - \langle R \rangle^2$ disappears. This might indicate that some new kind of transition takes place. However, $\langle R \rangle (\kappa_2^c, h)$ still decreases very slowly with $h$. Actually the data does not even allow us to conclude that $\langle R \rangle (\kappa_2^c, h) \to 0$ as $h \to \infty$. Hence adding a higher derivative term to the action does not cure the problems encountered for pure gravity.

Four-dimensional simplicial gravity with a $R^2$ term has also been studied in the static triangulation approach [122]. Here one finds that the first order phase transition encountered for pure gravity changes into a second order one when the value of $h$ gets sufficiently large. Furthermore the value of $\langle R \rangle$ at the critical point is always zero. This situation is very different from the one encountered for the dynamical triangulations. As in the 3-dimensional case the reason for this difference remains unclarified.

# References

[1] K.S. Stelle, Phys. Rev. D16 (1977) 953

[2] S. Weinberg, In *General Relativity, An Einstein Centenary Survey*, edited by S.W. Hawking and W. Israel, Cambridge University Press, Cambridge 1979

[3] S. Hawking, In *General Relativity, An Einstein Centenary Survey*, edited by S.W. Hawking and W. Israel, Cambridge University Press, Cambridge 1979

[4] G.W. Gibbons, S.W. Hawking and M.J. Perry, Nucl. Phys. B138 (1978) 141

[5] J. Greensite, Nucl. Phys. B361 (1991) 729

[6] A. Salam, In *Quantum Gravity: An Oxford Symposium*, edited by C.J. Isham, R. Penrose and D. Sciama, Oxford University Press 1975.

[7] C.J. Isham In *Quantum Gravity II: A Second Oxford Symposium*, edited by C.J. Isham, R. Penrose and D. Sciama, Oxford University Press 1981

[8] D.M.Y. Sommerville, Proc. R. Soc. London, A115 (1927) 103

[9] T. Regge, Nuovo Cimento 19 (1961) 558

[10] H. Hamber and R. Williams, Nucl. Phys. B248 (1984) 392

[11] H. Hamber, In *Critical Phenomena, Random Systems, Gauge Theories*, Proceedings of the Les Houches Summer School 1984, edited by K. Osterwalder and R. Stora, North-Holland 1986

[12] F. David, *Simplicial Quantum Gravity and Random Surfaces*, to appear in Proceedings of the Les Houches Summer School 1992

[13] B. Brügmann, Phys. Rev. D47 (1993) 3330 and Nucl. Phys. B (Proc. Suppl.) 30 (1993) 760

[14] E. Brezin, C. Itzykson, G. Parisi and J.B. Zuber, Comm. Math. Phys. 59 (1978) 35

[15] D. Bessis, C. Itzykson and J.B. Zuber, Adv. Appl. Math. 1 (1980) 109

[16] J. Ambjørn and S. Varsted, Phys. Lett. B266 (1991) 285

[17] D. Boulatov and A. Krzywicki, Mod. Phys. Lett. A6 (1991) 3003

[18] M.E. Agishstein and A.A. Migdal, Mod. Phys. Lett. A7 (1992) 1039

[19] M.E. Agishstein and A.A. Migdal, Nucl. Phys. B385 (1992) 395

[20] J. Ambjørn and J. Jurkiewicz, Phys. Lett. B278 (1992) 42

[21] S. Varsted, UCSD-PhTh-92-03 and PhD thesis, The Niels Bohr Institute, 1992

[22] J. Ambjørn, S. Jain, J. Jurkiewicz and C.F. Kristjansen, Phys. Lett. B305 (1993) 208

[23] J. Ambjørn, J. Jurkiewicz and C.F. Kristjansen, work in progress

[24] J. Ambjørn, B. Durhuus and J. Frölich, Nucl. Phys. B257 (1985) 433

[25] F. David, Nucl. Phys. B257 (1985) 543

[26] V.A. Kazakov, I.K. Kostov and A.A. Migdal, Phys. Lett. B157 (1985) 295

[27] F. David, Nucl. Phys. B257 (1985) 45

[28] V.A. Kazakov, Phys. Lett. A119 (1986) 140

[29] D.V. Boulatov and V.A. Kazakov, Phys. Lett. B186 (1987) 379

[30] V.A. Kazakov, Mod. Phys. Lett. A4 (1989) 2125

[31] M. Staudaucher, Nucl. Phys. B336 (1990) 349

[32] E. Brezin, M.R. Douglas, V.A. Kazakov and H. Shenker, Phys. Lett. B237 (1990) 43

[33] M.R. Douglas, In *Random Surfaces and Quantum Gravity*, Proceedings of the Cargese Workshop 1990, edited by O. Alvarez, E. Marinari and P. Windey, Plenum Press, New York, 1991

[34] J.M. Daul, V.A. Kazakov and I.K. Kostov, *Rational Theories of 2D Gravity from the Two Matrix Model*, CERN-TH-6834/93

[35] C.F. Baille and D.A. Johnston, Mod. Phys. Lett. A7 (1992) 1519 and Phys. Lett. B286 (1992) 44

[36] S.M. Catterall, J.B. Kogut and R.L. Renken, Phys. Lett. B292 (1992) 277 and Nucl. Phys. B (Proc. Suppl) 30 (1993) 775

[37] J. Ambjørn, B. Durhuus, T. Jónsson and G. Thorleifsson, Nucl. Phys. B398 (1993) 568 and G. Thorleifsson, Nucl. Phys. B (Proc. Suppl.) 30 (1993) 787

[38] M.R. Douglas and S.H. Shenker, Nucl. Phys. B335 (1990) 635

[39] E. Brezin and V.A. Kazakov, Phys. Lett. B236 (1990) 144

[40] D.J. Gross and A.A. Migdal, Phys. Rev. Lett. 64 (1990) 717, Nucl. Phys. B340 (1990) 333

[41] V.G. Knizhnik, A.M. Polyakov and A.B. Zamolodchikov, Mod. Phys. Lett. A3 (1988) 819

[42] F. David, Mod. Phys. Lett. A3 (1988) 1651

[43] J. Distler and H. Kawai, Nucl. Phys. B321 (1989) 509

[44] M. Gross and H. Hamber, Nucl. Phys. B364 (1991) 703

[45] J. Ambjørn, B. Durhuus and T. Jónsson, Mod. Phys. Lett. A6 (1991) 1133

[46] G. t'Hooft, Nucl. Phys. B72 (1974) 61

[47] J. Ambjørn, L. Chekhov, C.F. Kristjansen and Yu. Makeenko, Nucl. Phys. B404 (1993) 127

[48] J. Ambjørn, L. Chekhov and Yu. Makeenko, Mod. Phys. Lett. B282 (1992) 341

[49] J. Ambjørn, C.F. Kristjansen and Yu. Makeenko, Mod. Phys. Lett A7 (1992) 3187

[50] Yu. Makeenko, Mod. Phys. Lett. (Brief Reviews) A6 (1991) 1901 and references therein

[51] S.R. Wadia, Phys. Rev. D24 (1981) 970

[52] A.A. Migdal, Phys. Rep. 102 (1983) 199

[53] F. David, Mod. Phys. Lett. A5 (1990)1019

[54] J. Ambjørn and Yu. Makeenko, Mod. Phys. Lett. A5 (1990) 1753

[55] J. Ambjørn, J. Jurkiewicz and Yu. Makeenko, Phys. Lett. B251 (1990) 517

[56] O. Alvarez and P. Windey, Nucl. Phys. B348 (1991) 490

[57] M. Douglas, Phys. Lett. B238 (1990) 176

[58] C. Bachas and P.M.S. Petropoulos, Phys. Lett. B247 (1990) 363

[59] M. Fukuma, H. Kawai and R. Nakayama, Int. J. Mod. Phys. A6 (1991) 1385

[60] R. Dijkgraaf, E. Verlinde and H. Verlinde, Nucl. Phys. B348 (1991) 435

[61] Yu. Makeenko, A. Marshakov, A. Mironov and A. Morozov, Nucl. Phys. B356 (1991) 574.

[62] M.L. Kontsevich, Funk. Anal. & Prilozh. 25 (1991) 50 (in Russian), Comm. Math. Phys. 147 (1992) 1

[63] S. Kharchev, A. Marshakov, A. Mironov and A. Morozov, Nucl. Phys. B397 (1993) 339

[64] J. Ambjørn and C.F. Kristjansen, *From 1-Matrix Model to Kontsevich Model*, NBI-HE-93-30, to appear in Mod. Phys. Lett. A

[65] I. Kostov, Nucl. Phys. B (Proc. Suppl. ) 10A (1989) 295

[66] L. Chekhov and Yu. Makeenko, Phys. Lett. B278 (1992) 271.

[67] D.J. Gross and M.J. Newman, Phys. Lett. B266 (1991) 291

[68] C. Itzykson and J.-B. Zuber, Int. J. Mod. Phys. A7 (1992) 5661.

[69] E. Witten, *On the Kontsevich Model and other Models of Two–Dimensional Gravity*, IASSNS-HEP-91/24 (June, 1991).

[70] P. Deligne and M. Mumford, *The Irreducibility of the Space of Curves of given Genus*, Publ. I.H.E.S. 45 (1969) 75 and M. Mumford, *Towards an Enumerative Geometry of the Moduli Space of Curves.*, in *Arithmetic and Geometry*, Vol II, Birkhäuser, 1983

[71] L. Chekhov, *Matrix model for Discretized Moduli Space*, SMI-92-04

[72] L. Chekhov, *Matrix Models: A Way to Quantum Moduli Spaces*, PAR-LPTHE-93-22

[73] M.L. Mehta, Comm. Math. Phys. 79 (1981) 327

[74] C. Itzykson and J.B. Zuber, J. Math. Phys. 21 (1980) 411

[75] E. Gava and K.S. Narain, Phys. Lett. B263 (1991) 213

[76] J. Alfaro, Phys. Rev. D47 (1993) 4714

[77] M. Staudacher, Phys. Lett. B305 (1993) 332

[78] M.I. Dobroliubov, Yu. Makeenko and G.W. Semenoff, *Correlators of The Kazakov Migdal Model*, ITEP-YM-3-93

[79] J. Ambjørn, C.F. Kristjansen and Yu. Makeenko, work in progress

[80] F. David, Nucl. Phys. B348 (1991) 507

[81] P. Ginsparg and J. Zinn-Justin, Phys.Lett. B255 (1991) 189

[82] P.G. Silvestrov and A.S. Yelkhovsky, Phys. Lett. B251 (1990) 525

[83] S. Dalley, C. Johnson and T. Morris, Nucl. Phys. B368 (1992) 625

[84] S. Dalley, C. Johnson and T. Morris, Nucl. Phys. B368 (1992) 655

[85] J. Greensite and M. Halpern, Nucl. Phys. B242 (1984) 167

[86] E. Marinari and G. Parisi, Phys. Lett. B240 (1990) 375

[87] E. Marinari and G. Parisi, Phys. Lett. B247 (1990) 537

[88] M. Karliner and A.A. Migdal, Mod. Phys. Lett. A5 (1990) 2565

[89] J. Ambjørn, J. Greensite and S. Varsted, Phys. Lett. B249 (1990) 411

[90] J. Ambjørn and J. Greensite, Phys. Lett. B254 (1991) 66

[91] J. Ambjørn, C.V. Johnson and T.R. Morris, Nucl. Phys. B374 (1992) 496

[92] J. Gonzales and M. Vozmediano, Phys. Lett. B258 (1991) 55

[93] J. Miramontes, J. Guillen and M. Vozmediano, Phys. Lett. B253 (1991) 38

[94] J.L. Miramontes and J.S. Guillen, Int. J. Mod. Phys. A7 (1992) 6457

[95] J.L. Miramontes and J.S. Guillen, Nucl. Phys. B (Proc. Suppl.) 25A (1992) 195

[96] P.G. Silvestrov, Int. J. Mod. Phys. A7 (1992) 7401

[97] M. Carreau, E. Fahri, S. Gutmann and P.F. Mende, Annals of Physics, 204 (1990) 186

[98] J. Ambjørn and C.F. Kristjansen, Int. J. Mod. Phys. A8 (1993) 1259

[99] Jörgens and Rellich, *Eigenwerttheorie gewöhnlicher Differentialgleihungen* Springer-Verlag (1976)

[100] N. Sasakura, Mod. Phys. Lett. A6 (1991) 2613

[101] N. Godfrey and M. Gross, Phys. Rev. D43 (1991) 1749

[102] M. Gross, Nucl. Phys. B (Proc. Suppl.) 25A (1992) 144

[103] D.V. Boulatov, Mod. Phys. Lett. A7 (1992) 1629

[104] D.V. Boulatov, Int. J. Mod. Phys. A8 (1993) 3139

[105] V.G. Turaev and O.Y. Viro, Topology 31 (1992) 865

[106] T. Kohno, Topology 31 (1992) 203

[107] H. Ooguri, *Schwinger Dyson Equation in three dimensional simplicial quantum gravity*, preprint HUTP-92/A051

[108] H. Ooguri, Mod. Phys. Lett. A7 (1992) 2799

[109] J.W. Alexander, Ann. Math. 31 (1930) 292

[110] M. Gross and S. Varsted, Nucl. Phys. B378 (1992) 367

[111] B. Baumann and B. Berg, Nucl. Phys. B285 (1987) 391

[112] J. Ambjørn, Z. Burda, J. Jurkiewicz and C.F. Kristjansen, Nucl. Phys. B (Proc. Suppl.) 30 (1993) 771

[113] J. Ambjørn, Z. Burda, J. Jurkiewicz and C.F. Kristjansen, *4d quantum gravity coupled to matter*, preprint NBI-HE-93-3, to appear in Phys. Rev. D

[114] J. Ambjørn, Z. Burda, J. Jurkiewicz and C.F. Kristjansen, Act. Phys. Pol. B 23 (1992) 991

[115] M.E. Agishstein and A.A Migdal, Mod. Phys. Lett. A6 (1991) 1863

[116] J. Ambjørn and S. Varsted, Nucl. Phys. B373 (1992) 557

[117] J. Ambjørn, D. Boulatov, A. Krzywicki and S. Varsted, Phys. Lett. B276 (1992) 432

[118] H.W Hamber and R.M. Williams, Phys. Rev. D47 (1993) 510

[119] J. Ambjørn, Z. Burda, J. Jurkiewicz and C.F. Kristjansen, Phys. Lett. B297 (1992) 253

[120] S. Catterall, J. Kogut and R. Renken, Nucl. Phys. B389 (1993) 601

[121] C. Baille, Phys. Rev. D46 (1992) 2480

[122] H. Hamber, *Phases of Simplicial Gravity in Four Dimensions: Estimates for the Critical Exponents*, UCI-Th-92-29

[123] J. Ambjørn, J. Jurkiewicz and C.F. Kristjansen, Nucl. Phys. B393 (1993) 601